\documentclass[a4paper,11pt]{article}
\pdfoutput=1 

\usepackage{jheppub} 
\usepackage{tikz,tikz-cd}                     
\usetikzlibrary{shapes.geometric}
\usepackage{subfiles}
\usepackage{subcaption}
\usepackage[T1]{fontenc} 
\usepackage[utf8]{inputenc}
\usepackage{mathtools}
\usepackage{tensor}
\usepackage{mathabx}
\usepackage{textcomp}
\usepackage{comment}
\usepackage[makeroom]{cancel}
\usetikzlibrary{decorations.markings,positioning}
\usetikzlibrary{decorations.markings}
\usetikzlibrary{arrows}

\tikzset{->-/.style={decoration={
  markings,
  mark=at position .5 with {\arrow{>}}},postaction={decorate}}}

\tikzset{
smallnode/.style={circle, draw, very thick, minimum size=2mm},
smallcircle/.style={circle, fill, scale=0.5},
roundnode/.style={circle, draw, very thick, minimum size=10mm},
squarednode/.style={rectangle, draw, very thick, minimum size=10mm},
roundSO/.style={circle, draw, fill=gray!100, very thick, minimum size=10mm},
roundUSp/.style={circle, draw, fill=black, very thick, minimum size=10mm, text=white},
squareSO/.style={rectangle, draw, fill=gray!100, very thick, minimum size=10mm},
squareUSp/.style={rectangle, draw, fill=black, very thick, minimum size=10mm, text=white},
c1/.style={circle, draw, very thick, minimum size=1mm},
c2/.style={circle, draw, fill=black!100, very thick, minimum size=1mm}
}
\tikzset{every loop/.style={}}
\tikzset{
    arrowMe/.style={
        postaction=decorate,
        decoration={
            markings,
            mark=at position .5 with {\arrow[thick]{#1}}
        }
    }
}

\newtheorem{theorem}{Theorem}

\newcommand{\iso}{\cong}

\newcommand{\HS}{\mathrm{HS}}
\newcommand{\HWG}{\mathrm{HWG}}
\newcommand{\HL}{\mathrm{HL}}

\newcommand{\iu}{{i\mkern1mu}}
\newcommand{\Qsd}{\mathsf{Q}}

\DeclareMathOperator{\rN1}{r}
\DeclareMathOperator{\tr}{tr}
\DeclareMathOperator{\Sym}{Sym}
\DeclareMathOperator{\Tr}{Tr}

\DeclareMathOperator{\PE}{\mathrm{PE}}
\DeclareMathOperator{\PLog}{\mathrm{PLog}}

\title{\boldmath
\center The Coulomb and Higgs Branches of \\$\mathcal{N}=1$ Theories of Class $\mathcal{S}_k$}

\preprint{DESY 19-222}

\author{Thomas Bourton$^{a \spadesuit}$,}
\author{Alessandro Pini$^{a,b,\clubsuit}$ and}
\author{Elli Pomoni$^{a,^\heartsuit}$}

\affiliation{
$^a$ DESY Theory Group, Notkestra{\ss}e 85, 22607 Hamburg, Germany  \\
$^b$   I.N.F.N.- Sezione di Torino, Via Pietro Giuria 1, 10125 Torino, Italy
}

\emailAdd{$^\spadesuit$tbourton@gmail.com}
\emailAdd{$^\clubsuit$alessandro.pini@to.infn.it}
\emailAdd{$^\heartsuit$elli.pomoni@desy.de}

\abstract{

\bigskip

Even though for generic $\mathcal{N}=1$ theories it is not possible to separate distinct branches of supersymmetric vacua, in this paper we study a special class of $\mathcal{N}=1$ SCFTs, these of Class $\mathcal{S}_k$ for which it is possible to define Coulomb and Higgs branches precisely as for the  $\mathcal{N}=2$ theories of Class $\mathcal{S}$ from which they descend.
We study the BPS operators that parameterise these branches of vacua using the different limits of the superconformal index as well as the Coulomb and Higgs branch Hilbert Series.
Finally, with the tools we have developed, we provide a check that six dimensional $(1,1)$ Little String theory can be deconstructed from a toroidal quiver in class $\mathcal{S}_k$.
}

\begin{document} 
\maketitle
\flushbottom

\section{Introduction}
Since the seminal work of Gaiotto \cite{Gaiotto:2009we}, a lot of milage has been gained in the study of four dimensional superconformal field theories (4d SCFTs) when compactifications of six dimensional SCFTs (living on M5 branes) on various Riemann surfaces are considered.
Focusing on the so called Gaiotto's class $\mathcal{S}$ of  $\mathcal{N}=2$ SCFTs we have learned a lot about the 4d gauge theory thinking about simpler theories in two dimensions; a 2d CFT in the case of the partition function on $\mathbb{S}^4$ \cite{Alday:2009aq} or a 2d TQFT in the case of the superconformal index \cite{Gadde:2009kb,Gadde:2011ik}.
For generic $\mathcal{N}=1$ SCFTs the six dimensional  (M-theory) approach
 \cite{Bah:2012dg} has been proven to be far more complex. However,
there is a  class $\mathcal{S}_k$  of $\mathcal{N}=1$ SCFTs,   obtained by compactifications on Riemann surfaces of the 6d SCFTs living on M5 branes probing a $\mathbb{Z}_k$ singularity  \cite{Gaiotto:2015usa},
 for which there is hope that similar results with $\mathcal{N}=2$ are attainable \cite{Coman:2015bqq, Mitev:2017jqj, Bourton:2017pee,Razamat:2018zus}.\footnote{There is a larger class of $\mathcal{N}=1$ theories which are obtained as $\Gamma$ orbifolds of $\mathcal{N}=2$ theories in class $\mathcal{S}$, known as $\mathcal{S}_{\Gamma}$ \cite{Heckman:2016xdl,Kim:2018lfo,Kim:2018bpg,Kim:2017toz,Bah:2017gph,Razamat:2020bix} for which there is also hope to extend our  results \cite{Coman:2015bqq, Mitev:2017jqj, Bourton:2017pee} to.
Theories in class $\mathcal{S}_{\Gamma}$ are  obtained by compactifications on Riemann surfaces of the 6d (1,0) SCFTs living on M5 branes probing a $\Gamma \in ADE$ singularity.}

In the study of  $\mathcal{N}=2$ SCFTs it has been proven fruitful to concentrate in one of the two distinct branches of supersymmetric vacua;
the Coulomb branch or the Higgs branch. These vacua are parameterised by vacuum expectation values of certain local gauge invariant operators with very special properties.
In the case of the Coulomb branch, the so called Coulomb branch operators obey chiral  1/2 BPS shortening
conditions and form a closed chiral ring.
On the other hand, the Higgs branch operators are annihilated by two of the supercharges
$\mathcal{Q}$ and two of the $\widetilde{\mathcal{Q}}$ and they also form a closed ring of operators with very special mathematical structure.

The study of the Coulomb branch of $\mathcal{N}=2$ theories led to several important developments related to non-perturbative properties of the corresponding gauge theories. In the seminal works \cite{Seiberg:1994rs,Seiberg:1994aj} Seiberg and Witten showed that the low-energy effective action of $\mathcal{N}=2$ gauge theory can be completely specified by an holomorphic function containing both perturbative and non-perturbative contributions. Nekrasov gave a micropic derivation of the Seiberg-Witten results computing the instanton partition function  \cite{Nekrasov:2002qd} which finally was used by Pestun on the supersymmetric localization on $\mathbb{S}^4$ \cite{Pestun:2007rz}.
On the other hand, it is also long known that the Higgs branch of a $\mathcal{N}=2$ theory is endowed with a HyperK\"{a}hler structure that fixes the metric \cite{Hitchin:1986ea, Antoniadis:1996ra, hyper}. 

Correlation functions of the Coulomb branch operators are non-trivial functions of the marginal 
coupling constants of the SCFT. Nonetheless, they can be determined from a simple deformation of Pestun's matrix model via \cite{Baggio:2014ioa,Baggio:2014sna,Baggio:2015vxa,Gerchkovitz:2016gxx}\footnote{Or can also be obtained using data from the SW curve for non-Lagrangian cases \cite{Grassi:2019txd}.}. On the other hand, correlation functions of Higgs branch  operators are independent of the coupling constants  \cite{Beem:2013sza,Niarchos:2018mvl} and thus fixed, up to multiplet recombination, by a tree level calculation, which can be mapped to a 2d chiral algebra computation \cite{Beem:2013sza}.

The purpose of this paper is to show that theories in class $\mathcal{S}_k$, precisely because they are orbifold daughters of  $\mathcal{N}=2$ SCFTs in class $\mathcal{S}$ also have  similar branches of supersymmetric vacua as their  $\mathcal{N}=2$ mother theories.
They enjoy, as opposed to generic $\mathcal{N}=1$ theories, a second $U(1)$ (anomaly free \cite{Gaiotto:2015usa}) global symmetry on the top of their R-symmetry, which allows for a well defined Coulomb and Higgs branch.

The content of this paper is as follows. In Section \ref{sec:quivers} we review some basic facts about the theories of class $\mathcal{S}_k$ that we are going to study in this paper, while taking the opportunity to introduce notation.
In Section \ref{sec:modulispace} we introduce the Hilbert series (HS) which will be our first tool for counting operators in chiral rings.
Our second tool, the Superconformal index (SCI) and some interesting limits of it are introduced in Section \ref{sec:sci}.
In Section \ref{sec:genuszero} we study some examples of genus zero theories\footnote{Throughout the paper we will use the jargon ``genus $g$ theories'' by which we mean that these are theories in class  $\mathcal{S}_k$ obtained by compactifications on Riemann surfaces of genus $g$ of the 6d (1,0) SCFTs living on M5 branes probing a $\mathbb{Z}_k$ singularity.} with a Lagrangian description. We separately compute the Coulomb branch SCI, the Coulomb branch Hilbert series and show that they are equal. What is more, we compute the Hall Littlewood limit of the superconformal index, as well as the Higgs branch Hilbert series and also find that they are equal, precisely as for the mother  $\mathcal{N}=2$ SCFTs of class $\mathcal{S}$   \cite{Gadde:2011uv}.
Succeeding that, in Section \ref{sec:hsg1}, we study some examples of genus one theories and we compute their Higgs branch HS as well as their Hall Littlewood index. We take the opportunity to analyse how the 6d $(1,1)$ little string theory (LST) \cite{ArkaniHamed:2001ie} can be deconstructed using a toroidal class $\mathcal{S}_k$ theory, in the past also referred to as the ``{\it Conformal Moose}''.
In Section \ref{sec:trinion}, we consider strongly interacting class $\mathcal{S}_k$ theories, without a Lagrangian description, arising as S-dual descriptions (pants decompositions) of the $A_1$ four punctured sphere in class $\mathcal{S}_2$. We compute the Coulomb and Hall Littlewood indices for these theories and show that they can be written in a simple closed form. Finally, we end with some conclusions in Section \ref{sec:conclusions}.

In Appendix \ref{app:ag} we provide a short review of the so called \textit{algebra-geometry correspondence}, whilst in Appendix \ref{sec:appB} we collect the mathematical identities that we used through the paper. Finally important facts about  $\mathcal{N}=1$ superconformal representation theory, short and semi-short multiplets, and the
Superconformal Index 
are reviewed in Appendices \ref{app:SCAreps} and \ref{sec:appD}.

\section{\boldmath 4d $\mathcal{N}=1$ quiver gauge theories in Class $\mathcal{S}_k$}\label{sec:quivers}

In this Section we review some background material for  quiver gauge theories in Class $\mathcal{S}_k$ introduced in \cite{Gaiotto:2015usa}. Experts  can skip reading this Section. We will mostly follow the notation of \cite{Razamat:2018zus}. From an M-theory perspective these theories arise considering a stack of $N$ M5-branes probing a $A_{k-1}$ singularity. In the following our analysis will be based on the corresponding type IIA brane configuration reported in Table \ref{tab:sk}. 
\begin{table}[]
\center{
\begin{tabular}{|c||c|c|c|c|c|c|c|c|c|c|c|}
\hline
& $x^0$  & $x^1$  & $x^2$ & $x^3$ & $x^4$ & $x^5$ & $x^6$ & $x^7$ & $x^8$ & $x^9$ & $(x^{10})$\\
\hline
\hline
$M$ NS5 & \textminus & \textminus & \textminus & \textminus & \textminus & \textminus & . & . & . & . & .    \\
\hline
$N$ D4 & \textminus & \textminus  & \textminus  & \textminus & . & . & \textminus & . & . & . & \textminus  \\
\hline
$A_{k-1}$ orbifold & .  & . & .  & . & $*$ & $*$  & . &  $*$ &$*$ & . & .\\
\hline
\end{tabular}
}
\caption{\it Type IIA brane configuration for theories of Class $\mathcal{S}_k$.\label{tab:sk}}
\end{table}

For $k=1$ we are describing  $\mathcal{N}=2$ theories of Class $\mathcal{S}$ which arise through the compactification of the 6d $(2,0)$ SCFT living on the stack of M5-branes over a 2d Riemann surface of genus $g$ decorated with punctures. The R-symmetry group of these theories is $U(2) \simeq SU(2)_{R_{\mathcal{N}=2}} \times U(1)_{r_{\mathcal{N}=2}}$. Looking at Table \ref{tab:sk} we observe that the $ \mathfrak{su}(2)_{R_{\mathcal{N}=2}} \simeq \mathfrak{so}(3)_{R_{\mathcal{N}=2}}$ subalgebra of the R-symmetry group is described by rotations along $\{x^7,x^8,x^9\}$ directions. This symmetry is broken to a $U(1)_R$ by the orbifold projection. The unbroken $U(1)_R$ corresponds to the Cartan generator of the $SU(2)_{R_{\mathcal{N}=2}}$. On the other hand, the  $U(1)_{r_{\mathcal{N}=2}} \simeq SO(2)$ subgroup is described by rotations along $\{x^4$,$x^5\}$ directions and it is left unbroken by the orbifold.

\begin{figure}[b]
\center{
\begin{tikzpicture}[scale=0.3]


\node[squarednode] (1) {};
\node[roundnode] (2) [right=of 1] {};
\node[roundnode] (3) [right=of 2] {};
\node[roundnode] (4) [right=of 3] {};
\node[squarednode] (5) [right=of 4] {};


\draw[-,arrowMe=>] (1) to [bend left] (2) {};
\draw[-,arrowMe=>] (2) to [bend left] (3) {};
\draw[-,arrowMe=>] (3) to [bend left] (4) {};
\draw[-,arrowMe=>] (4) to [bend left] (5) {};

\draw[-,arrowMe=<] (1) to [bend right] (2) {};
\draw[-,arrowMe=<] (2) to [bend right] (3) {};
\draw[-,arrowMe=<] (3) to [bend right] (4) {};
\draw[-,arrowMe=<] (4) to [bend right] (5) {};

\draw[->] (2) to [loop above] (2) {};
\draw[->] (3) to [loop above] (3) {};
\draw[->] (4) to [loop above] (4) {};

\node (c1) [below left =0.25cm and 0.5 cm of 1] {};
\node (c2) [right =10cm of c1]  {};
\draw (c1)--(c2);

\node (C0) [below = 0.7cm of 1] {$n=0$};
\node (C1) [right = 0.85cm of C0] {$n=1$};
\node (C2) [right = 0.85cm of C1] {$n=2$};
\node (C3) [right = 0.85cm of C2] {$n=3$};
\node (C4) [right = 0.85cm of C3] {$n=4$};

\end{tikzpicture}
} 
\caption{\it Linear quiver for Class $\mathcal{S}$ theories (with $\ell-2=4$ and arbitrary $N$). The index $n=1,..., \ell-3$ labels the different gauge nodes of the quiver.\label{fig:quiverS}}
\end{figure}
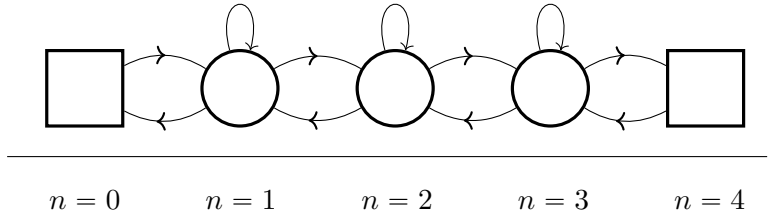

For $k =1$, the Class $\mathcal{S}$ theories obtained from the string theory setup in Table \ref{tab:sk}
are known as \textit{linear quiver} gauge theories, with a quiver diagram like the one in Figure \ref{fig:quiverS}. The circular nodes of the quiver correspond to $SU(N)$ gauge groups while the square boxes correspond to $SU(N)$ flavour groups. We can glue together by gauging the two extremal flavour nodes of the linear quiver, this way we obtain a \textit{circular} or \textit{elliptic} Class $\mathcal{S}$ theory. An example of the latter is given in Figure \ref{fig:quivercircular}.

\begin{figure}
\begin{center}
\includegraphics[scale=0.70]{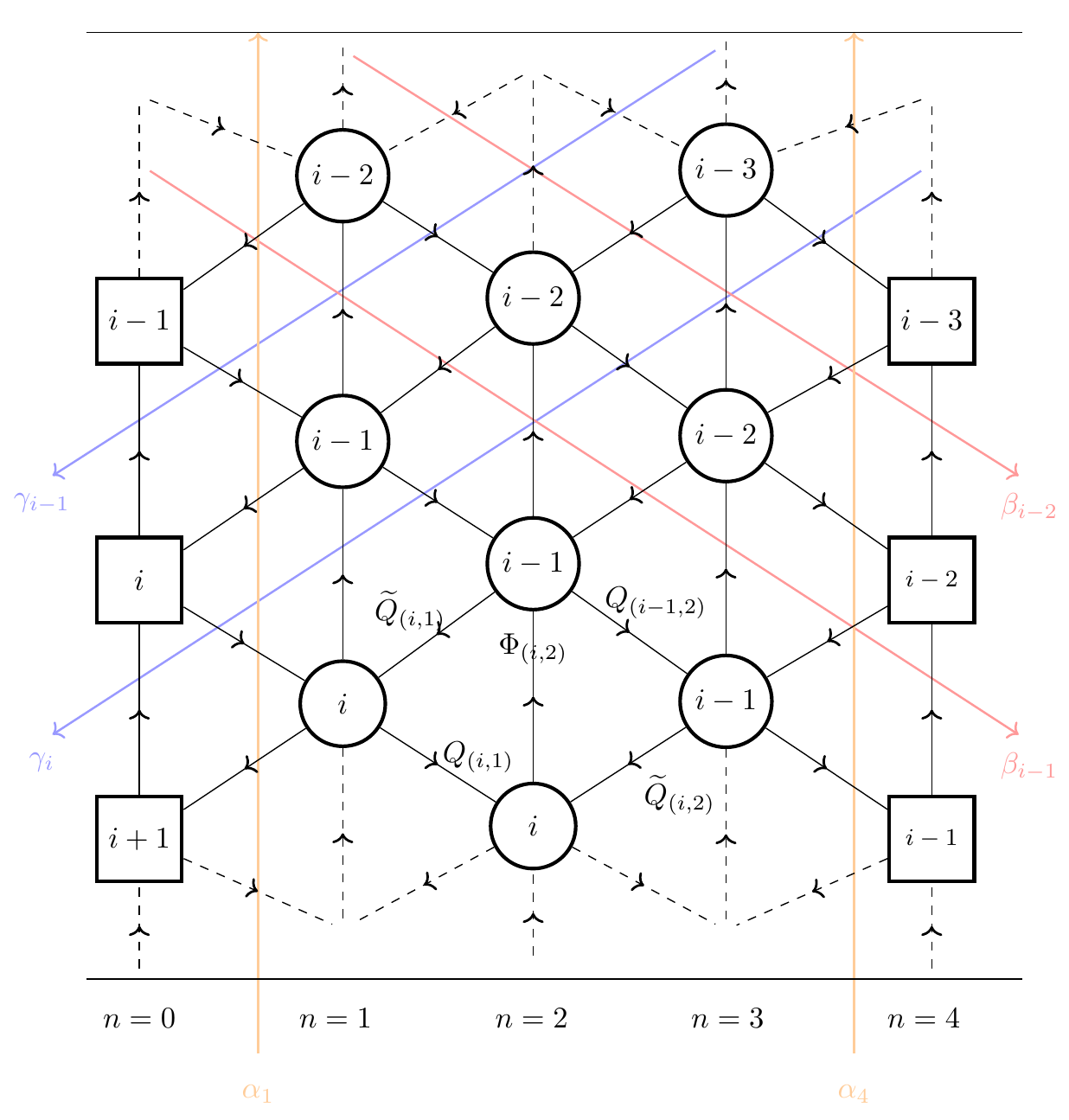}
\end{center}
\caption{\it Quiver for a Class $\mathcal{S}_k$ theory. The gauge nodes are labelled by $(i,n)$, where $i=1,...,k$ is the $\mathbb{Z}_k$ orbifold index while $n=1,...,\ell-3$ is the label from the $\mathcal{N}=2$ mother theory. In this example we set $\ell-2=4$. \label{fig:quiverSK}}
\end{figure}

The 4d Class $\mathcal{S}_k$ theories are realised as twisted compactifications  of the 6d $\mathcal{N}=(1,0)$ SCFT on the worldvolume theory on $N$ M5-branes probing the transverse $A_{k-1}$ singularity which has $\mathfrak{u}(1)_t\oplus\mathfrak{su}(k)_{\beta}\oplus\mathfrak{su}(k)_{\gamma}$ global symmetry. The compactifications typically preserve only a Cartan subalgebra $\mathfrak{u}(1)_t\oplus\mathfrak{u}(1)^{\oplus k-1}_{\beta}\oplus\mathfrak{u}(1)^{\oplus k-1}_{\gamma}$ which are `intrinsic' global symmetries carried by all class $\mathcal{S}_k$ theories. Moreover, there is also the $\mathcal{N}=1$ $\mathfrak{u}(1)_r$ R-symmetry. The Cartan generators of these symmetries are given by a linear combination of the Cartans $\{R_{\mathcal{N}=2},r_{\mathcal{N}=2}\}$ of the $\mathfrak{su}(2)_R \oplus \mathfrak{u}(1)_r$ $\mathcal{N}=2$ R-symmetry algebra 
\begin{equation}
r = \frac{2}{3} \left( 2R_{\mathcal{N}=2}-r_{\mathcal{N}=2} \right) \,,\quad q_t=R_{\mathcal{N}=2}+r_{\mathcal{N}=2}\,.
\end{equation}
where $q_t$ denotes the Cartan of the $\mathfrak{u}(1)_t$ symmetry. We report an example of such a quiver gauge theory in Figure \ref{fig:quiverSK}. 
We summarize the transformations of the field content for this theory in Table \ref{tab:fieldsSK}.

For $k \geq 2$, the orbifold breaks half of the initial  supersymmetry giving rise to 4d $\mathcal{N}=1$. 
States of the $\mathcal{N}=2$ mother theories of Class $\mathcal{S}$ live in unitary representations of the $\mathfrak{su}(2,2|2)$ superconformal algebra. 
They are labelled by the $(E,j_1,j_2,R_{\mathcal{N}=2},r_{\mathcal{N}=2})$ quantum numbers which denote representations under the maximal bosonic subalgebra
\begin{equation}
\mathfrak{u}(1)_{E} \oplus \mathfrak{su}(2)_1 \oplus \mathfrak{su}(2)_2 \oplus \mathfrak{su}(2)_{R_{\mathcal{N}=2}} \oplus \mathfrak{u}(1)_{r_{\mathcal{N}=2}} \subset \mathfrak{su}(2,2|2) \,.
\end{equation}
The orbifold action breaks the $\mathfrak{su}(2,2|2)$ superalgebra down to $\mathfrak{su}(2,2|1) \oplus \mathfrak{u}(1)_{q_t}$. Representations of this algebra are labelled by the Cartans $(E,j_1,j_2,r ; q_t)$ of the maximal bosonic subalgebra
\begin{equation}
\mathfrak{u}(1)_{E} \oplus \mathfrak{su}(2)_1 \oplus \mathfrak{su}(2)_2 \oplus \mathfrak{u}(1)_r \oplus \mathfrak{u}(1)_{q_t} \subset \mathfrak{su}(2,2|1) \oplus \mathfrak{u}(1)_{q_t} \,  .
\end{equation}
From the initial eight $\mathcal{N}=2$ supercharges, only  the ones with  $q_t=0$ charge survive after the orbifold projection,
\begin{equation}
\mathcal{Q}_{2} \, : \, \left(\frac{1}{2},\pm \frac{1}{2},0, -\frac{1}{2} , +\frac{1}{2}\right)  \,,\quad \widetilde{\mathcal{Q}}_1  \, : \,   \left(\frac{1}{2},0 ,\pm \frac{1}{2},  \frac{1}{2} , - \frac{1}{2}\right)    \, ,
\end{equation}
to give the four supercharges of the $\mathfrak{su}(2,2|1)$ superalgebra
\begin{equation}
\mathcal{Q} \, : \,  \left(\frac{1}{2},\pm \frac{1}{2},0, -1 ; 0\right)  \,,\quad \widetilde{\mathcal{Q}} \, : \,   \left( \frac{1}{2},0,\pm \frac{1}{2}, 1 ; 0\right)   \, .
\end{equation}
The  $(j_1,j_2,r ; q_t)$ quantum numbers of the field content is presented in Section \ref{sec:sci} in Tables \ref{tab:lettersft} and \ref{tab:lettersphi}. While the complete list of supercharges is reported in Table \ref{tab:supercharges}.

\begin{table}[]
{\renewcommand{\arraystretch}{1.2}
\center{
\begin{tabular}{|c||c|c|c|c|c|c|c|c|}
\hline
& $SU(N)_{(i,n-1)}$  & $SU(N)_{(i,n)}$  & $SU(N)_{(i-1,n)}$ & $U(1)_t$ & $U(1)_{\alpha_n}$ & $U(1)_{\beta_{i+1-n}}$ & $U(1)_{\gamma_i}$ \\
\hline
\hline
$V_{(i,n)}$ & $\mathbf{1}$ & Adj & $\mathbf{1}$ & 0 & 0 & 0 & 0\\
\hline
$\Phi_{(i,n)}$ & $\mathbf{1}$ & $\mathbf{N}$ & $\overline{\mathbf{N}}$ & -1 & 0 & -1 & +1\\
\hline
$Q_{(i,n-1)}$ & $\mathbf{N}$ & $\overline{\mathbf{N}}$ & $\mathbf{1}$ & +1/2 & -1 & +1 & 0\\
\hline
$\widetilde{Q}_{(i,n-1)}$ & $\overline{\mathbf{N}}$ & $\mathbf{1}$ & $\mathbf{N}$ & +1/2 & +1 & 0 & -1\\
\hline 
\end{tabular}
}
}
\caption{\it Symmetry transformations of the fields of a class $\mathcal{S}_k$ theory. \label{tab:fieldsSK}}
\end{table}
The punctures carry additional data associated to inserting a variety of defect operators in the  worldvolume of the 6d $\mathcal{N}=(1,0)$ SCFT, which are localised  at the punctures  \cite{Gaiotto:2015usa,Hassler:2017arf,Heckman:2016xdl} of the Riemann surface, and spacetime filling in $\{x^0,x^1,x^2,x^3\}$. In this paper we will focus only on maximal and minimal punctures which are well understood. A complete classification of puctures is yet missing \cite{Heckman:2016xdl,Hassler:2017arf}.

Maximal punctures are labelled with a  ``colour'' $c\in \{1,2,\dots,k\}$ quantum number, a sign $\sigma=\pm1$ and an  ``orientation'' $o=l/r$. We will label these maximal punctures with the notation $s_{c}^{o,\sigma}$. Maximal punctures also carry an associated $\mathfrak{su}(N)^{\oplus k}$ flavour symmetry algebra. Minimal punctures carry a $\mathfrak{u}(1)$ symmetry under which the baryonic operators of the form $\det Q_i$ and $\det \widetilde{Q}_i$
are charged, while mesonic operators of the form $\mathcal{M}_{i,i} = \tilde{Q}_i Q_{i}$  are uncharged.

\begin{figure}
\centering
\begin{subfigure}{.5\textwidth}
\centering
\begin{tikzpicture}[square/.style={regular polygon,regular polygon sides=4},thick,inner sep=0.1em,scale=0.7]
    \node (L1) at (0,0) [square,draw,minimum size=1.8cm]{$i-1$};
    \node (R1) at (4,0) [square,draw,minimum size=1.8cm]{$i-1$};
    \node (L2) at (0,-3) [square,draw,minimum size=1.8cm]{$i$};
    \node (R2) at (4,-3) [square,draw,minimum size=1.8cm]{$i$};
    \node (L3) at (0,-6) [square,draw,minimum size=1.8cm]{$i+1$};
    \node (R3) at (4,-6) [square,draw,minimum size=1.8cm]{$i+1$};

    \draw [->] (L1.0) -- (R1.180) node[midway,above] {$Q_{i-1}$};
    \draw [->] (L2.0) -- (R2.180) node[midway,above] {$Q_{i}$};
    \draw [->] (L3.0) -- (R3.180) node[midway,above] {$Q_{i+1}$};
    
    \draw [->] (R1.225) -- (L2.45) node[midway,above,sloped] {$\widetilde{Q}_{i}$};
    \draw [->] (R2.225) -- (L3.45) node[midway,above,sloped] {$\widetilde{Q}_{i+1}$};
     \draw  (R3.225) -- (2,-7.5) node[near end,above,sloped] {$\widetilde{Q}_{i+2}$};
      \draw [->] (2,1.5) -- (L1.45) node[near start,above,sloped] {$\widetilde{Q}_{i-1}$};
     
 \end{tikzpicture}
 \end{subfigure}%
 \begin{subfigure}{.5\textwidth}
 \centering
  \begin{tikzpicture}[thick,scale=1.7]
  \draw (0,0) ellipse (1.3cm and 1.3cm);
  \draw [black,fill=black] (0,0.6) circle (2pt) node [below=2pt]{$\alpha$};
  \draw [black] (-0.6,-0.3) circle (6pt) node [below=8pt]{$\mathbf{z}_l$} node{$l$};
  \draw [black] (0.6,-0.3) circle (6pt) node [below=8pt]{$\mathbf{z}_r$} node {$r$};

\end{tikzpicture}
 \end{subfigure}%
  \caption{\it On the left we depict the quiver diagram for the free trinion theory of bifundamental chiral multiplets (arising as the orbifold of a bifundamental hypermultiplet). On the right we depict the associated sphere with one minimal puncture and two maximal punctures $s_{c}^{l,+}$ and $s_{c+1}^{r,+}$ (the fugacities $\mathbf{z}_l$ and $\mathbf{z}_r$ label the global symmetry groups associated with the two maximal punctures).}
  \label{fig:freetrinion}
\end{figure}
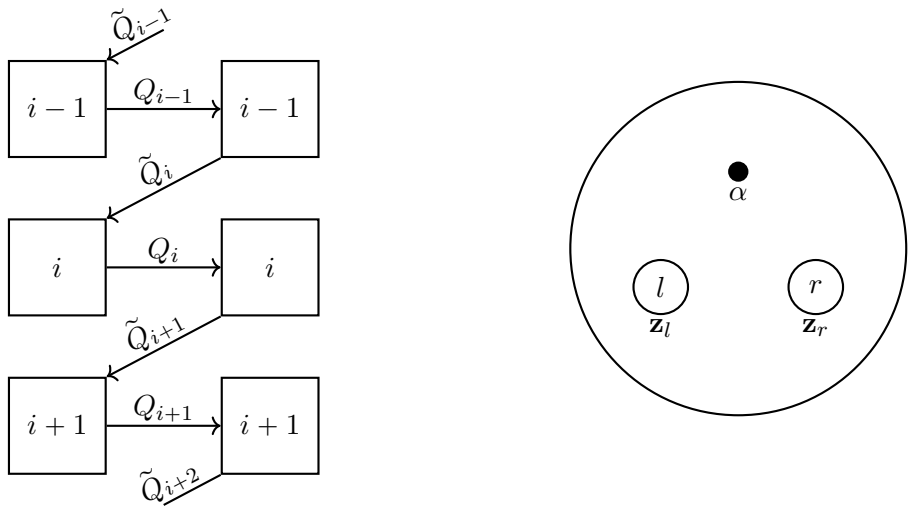

The analogous of the Class $\mathcal{S}$ linear quiver gauge theory, in the context of Class $\mathcal{S}_k$, are theories associated to spheres with $\ell-2$ minimal punctures and two maximal punctures $s_{c_l}^{l,+}$ \& $s_{c_r}^{r,+}$  with $c_r=(c_l+\ell-2)\bmod k$, and $\frac{1}{2k}$ unit of flux for $U(1)_t$. As we will discuss these theories will be the basic building blocks for more involved theories, for this reason we refer to these as ``core'' theories. 
These theories admits a weakly coupled Lagrangian description associated to a pair of pants decomposition of the Riemann surface into a chain of $n=0,1,\dots,\ell-3$ spheres with one minimal puncture and two maximal punctures $s_{c_l+n}^{l,+}$ \& $s_{c_l+n+1}^{r,+}$. These three punctured spheres correspond to quiver theories of bifundamental chiral multiplets called the \textit{free trinion} which is pictured in Figure \ref{fig:freetrinion}. See Table \ref{tab:letters} for the corresponding charges under symmetries. 

\begin{figure}
\centering
\begin{subfigure}{.5\textwidth}
\centering
\begin{tikzpicture}[thick,inner sep=0.1em,scale=0.7]
    \node (L1) at (0,0) [circle,draw,minimum size=1.3cm]{$i-1$};
    \node (L2) at (0,-3) [circle,draw,minimum size=1.3cm]{$i$};
    \node (L3) at (0,-6) [circle,draw,minimum size=1.3cm]{$i+1$};
    
    \draw [->] (0,-7.5) -- (L3.270) node[near start,midway,left] {$\Phi_{i+2}$};    
    \draw [->] (L3.90) -- (L2.270) node[midway,left] {$\Phi_{i+1}$};
    \draw [->] (L2.90) -- (L1.270) node[midway,left] {$\Phi_{i}$};
    \draw [-] (L1.90) -- (0,1.5) node[near end,left] {$\Phi_{i-1}$};
     
 \end{tikzpicture}
 \end{subfigure}%
 \begin{subfigure}{.5\textwidth}
 \centering
  \begin{tikzpicture}[thick,scale=1.7]
  \draw (0,0) ellipse (1.3cm and 1.3cm);
  \draw [black] (-0.7,0) circle (6pt) node [below=8pt]{$\mathbf{z}$} node{$l$};
  \draw [black] (0.7,0) circle (6pt) node [below=8pt]{$\mathbf{z}$} node {$r$};

\end{tikzpicture}
 \end{subfigure}%
  \caption{\it Quiver associated to $\Phi$-gluing.}
  \label{fig:tube}
\end{figure}
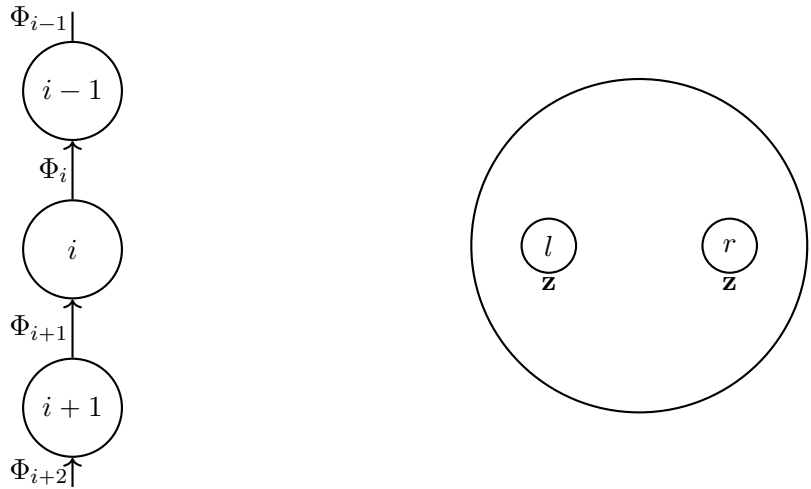

The maximal punctures of equal colour, opposite orientation and equal sign of the $n^{\text{th}}$ and $n+1^{\text{th}}$ three punctured spheres are then glued with tubes associated to spheres with two maximal punctures $s_{c_l+n+1}^{l,+}$ \& $s_{c_l+n+1}^{r,+}$. This gluing corresponds to gauging the diagonal $\mathfrak{su}(N)_{\mathbf{z}}^{\oplus k}\subset\mathfrak{su}(N)_{\mathbf{z}_l}^{\oplus k}\oplus \mathfrak{su}(N)_{\mathbf{z}_r}^{\oplus k}$ of the two maximal punctures with free $\mathcal{N}=1$ vector multiplets and $k$ bifundamental chiral multiplets $\Phi_i$.
This type of gluing is called $\Phi$-gluing \footnote{For this type of theories it exists another type of gluing, called $S$-gluing, that was originally introduced in \cite{Hanany:2015pfa}.} and is pictured in Figure \ref{fig:tube}. The representations under the global symmetries are summarised in Table \ref{tab:lettersphi}. 

From these theories it is possible to construct theories associated to tori with $\ell$ minimal punctures by $\Phi$-gluing the `open' maximal punctures $s_{c_l}^{l,+}$ \& $s_{c_r}^{r,+}$, these are $\mathbb{Z}_k\times\mathbb{Z}_{\ell}$ orbifold theories of $\mathcal{N}=4$ SYM. Unless $c_r-c_l=0\bmod k$ (or equivilantly $\ell-2\bmod k=0$) this procedure breaks the $\mathfrak{u}(1)^{\oplus k-1}_{\beta}\oplus\mathfrak{u}(1)^{\oplus k-1}_{\gamma}$ symmetry. The quiver diagram of such theories can be found in Figure \ref{fig:Skquivergenus1}. We will discuss theories of this type in Section \ref{sec:hsg1}.

Finally, a general Lagrangian theory in Class $\mathcal{S}_k$, made using the above ingredients, associated to a genus $g$ Riemann surface with $\ell$ punctures has superpotential
\begin{equation}\label{eqn:Sksuperpotential}
W_{\mathcal{S}_k}=\sum_{n=1}^{3g-3+\ell}\sum_{i=1}^k\tr\left( \widetilde{Q}_{(i,n-1)}Q_{(i,n-1)}\Phi_{(i,n)}-Q_{(i-1,n)}\widetilde{Q}_{(i,n)}\Phi_{(i,n)}\right)\,.
\end{equation}
\bigskip

\section{The Moduli Space of Supersymmetric Vacua}
\label{sec:modulispace}
In order to appreciate how special class $\mathcal{S}_k$ theories are, in the first part of this Section, we review the notion of moduli space of vacua valid for a $4d$ $\mathcal{N}=1$ theory.

Theories with $\mathcal{N}=1$ supersymmetry can  have a moduli space of supersymmetric vacua $\mathcal{M}$ which is given by the space of solutions to the F-term and D-term constraints modulo gauge transformations by gauge group $G$. 
Setting the F-terms and the D-terms to zero and modding out by $G$ is equivalent to dropping the D-term constraints and modding out by complexified gauge transformations $G^{\mathbb{C}}$ \cite{Luty:1995sd}. Therefore the moduli space can be described by the following quotient between the \textit{master space} $\mathcal{F}$ \cite{1993alg.geom..4001B,1993alg.geom..6005B} and the complexified gauge group
\begin{equation}\label{eqn:modspace}
\mathcal{M} \simeq \mathcal{F}/G^{\mathbb{C}}\,,\quad \mathcal{F}=\left\{(v_1,v_2,\dots)\middle|F_1=F_2=\dots=0\right\}\,,
\end{equation}
here $v_i$ denote the scalar vevs, $F_i=\partial W/\partial v_i$ are the F-term constraints and $W$ is the superpotential. From a geometrical point of view $\mathcal{F}$ is a complex algebraic variety, therefore it can be characterised in terms of a quotient ring $\mathcal{R}/\mathcal{I}$ where $\mathcal{R}=\mathbb{C}[v_1,v_2,\dots]$ is the polynomial ring in the $v_i$ and $\mathcal{I} = \langle F_1,F_2,\dots\rangle$ 
is the the F-terms ideal\footnote{ Given a polynomial ring $\mathbb{C}[x_1,...,x_n]$ and some polynomials $f_1,...,f_s$ $\in$ $\mathbb{C}[x_1,...,x_n]$  we define
$
\langle f_1,...,f_s \rangle := \sum_{i=1}^{s}h_if_i \ \ \textrm{with} \ \ h_1,...,h_s \in \mathbb{C}[x_1,...,x_n] \, \ , 
$
and we call $\langle f_1,...,f_n \rangle$ the \textit{ideal generated} by $f_1,...,f_n$.}. 
On the other hand, the moduli space is given by the spectrum  $\mathcal{M} = spec \{ \mathcal{A} \}$ of some algebra $\mathcal{A}$.
In Appendix \ref{app:ag} we have spelled out a lot of useful details regarding this \textit{algebra-geometry dictionary}.   

In generic $\mathcal{N}=1$ gauge theories, the superpotential can receive quantum corrections which can be perturbative and non-perturbative, which, however, are constrained by holomorphicity arguments \cite{Intriligator:1995au}. 
The same is true for the moduli space of vacua $\mathcal{M}\not\iso\mathcal{M}_{\text{Classical}}$.  In the quantum level it is generically different from the classical one since the metric receives corrections and singularities may appear (when  massive particles becomes massless). Generally, the problem of obtaining the quantum moduli space can be a difficult one.

As long as we have supersymmetry, there exist further tools which allows us to study
 the algebra  $\mathcal{A}$, the spectum of which gives the moduli space  $\mathcal{M} = spec \{ \mathcal{A} \}$. These are the Hilbert Series (which we will introduce in this Section) and the superconformal index (which we will introduce in the next Section).

\subsection{Higgs and Coulomb Branches for Class $\mathcal{S}_k$ Theories and the Hilbert Series}
\label{sec:hc}

Before plunging in Class  $\mathcal{S}_k$ a lightning review of Class  $\mathcal{S}$ is in order.
For  $\mathcal{N}=2$ SCFTs  the Coulomb branch is parameterised by the so called Coulomb branch operators which obey chiral  1/2 BPS shortening
conditions $\mathcal{Q}^{\mathcal{I}}_\alpha \mathcal{O}_{C} = 0$ $\forall$ $\mathcal{I},\alpha$ ($\mathcal{E}_r$ in \cite{Dolan:2002zh} notation) with
$E=r_{\mathcal{N}=2}$  and form a closed chiral ring.
On the other hand, the Higgs branch operators obey the shortening
condition $E=2R_{\mathcal{N}=2}$ ($\hat{\mathcal{B}}_R$  in \cite{Dolan:2002zh} notation) which means that they are annihilated by one
$\mathcal{Q}$ and one $\widetilde{\mathcal{Q}}$ and they also form a closed subsector of operators with interesting mathematical structure.
These two shortening conditions $E=r_{\mathcal{N}=2}$ and $E=2R_{\mathcal{N}=2}$ define subsectors of
operators which are closed (do not mix with other under
renormalization) and have ring structure.

The full moduli space for theories in class $\mathcal{S}_k$ are very rich and share many similarities with $\mathcal{N}=2$ theories, in particular they possess Coulomb, Higgs and mixed phases. The Coulomb moduli spaces have been studied in \cite{Coman:2015bqq,Razamat:2018zus,curves}. Here we will mainly focus on two truncations of the full moduli space which, in $\mathcal{N}=2$ nomenclature, we will refer to as \textit{Higgs} and \textit{Coulomb} branches. As we will see these truncations are well defined and will be useful.

\subsubsection{The Higgs Branch}
\label{sec:hcbc}
Let us first present the definition for the Higgs branch for theories of Class $\mathcal{S}_k$. This definition is valid for theories either with a Lagrangian description or those related to theories with a Lagrangian description by dualities and can be made by restricting to scalar operators that have
\begin{equation}\label{eqn:hbdefcharges}
 E=2R_{\mathcal{N}=2}=r+\frac{2q_t}{3}\,,\quad r_{\mathcal{N}=2}=\frac{2q_t}{3}-\frac{r}{2}=0 \, ,
 \end{equation}
 where $q_t$ and $r$ denote the charges of the $\mathfrak{u}(1)_t$ and $\mathcal{N}=1$ $\mathfrak{u}(1)_{r}$ R-symmetry respectively. Provided that the R-symmetry is not broken (which is true for any $\mathcal{N}=1$ SCFT) and that $\mathfrak{u}(1)_t$ is also anomaly free (shown in  \cite{Gaiotto:2015usa}) we can always decompose a ring $\mathcal{R}$ under the $\mathfrak{u}(1)_{r}\oplus \mathfrak{u}(1)_{t}$ grading. 
 The two conditions on \eqref{eqn:hbdefcharges} combine to 
  \begin{equation}
  \label{eqn:hbdefTotal}
  E=\frac{3}{2}r=2q_t\, ,
 \end{equation}
 with the first equality being the  BPS condition   $\overline{\mathcal{B}}_{r,(j_1,0)}$ in the notation \cite{Dolan:2002zh}  (or   $L\overline{B}_1[j_1,0]_E^{(r)}$ in the the notation of \cite{Cordova:2016emh}) as summarised in Table \ref{tab:shorts}. These types of multiplets are {\it isolated short multiplets}  from
all other types of multiplets (both the continuum of long and  other  short multiplets)  by a gap.
 What is special about the Higgs branch for theories of Class $\mathcal{S}_k$, as opposed to general $\mathcal{N}=1$ SCFTs,  is that the R-symmetry of the $\mathcal{N}=1$ superconformal algebra is further related to the charge $q_t$ of the extra flavour $U(1)_t$ as in \eqref{eqn:hbdefTotal}.

For our basic ``core'' theories 
this definition coincides with turning on generic diagonal vevs for scalars in $Q$ and $\widetilde{Q}$ chiral multiplets 
while setting to zero vevs for the scalars in the $\Phi$  chiral multiplets. 
Those choices of vevs completely breaks the $SU(N)$ gauge symmetry at each node. Hence, we refer to this sub-branch of
$\mathcal{M}$ as the \textit{Higgs branch}. 

From the point of view of the $(i,n)^{\text{th}}$ gauge node the theory is SQCD with $N_f=3N_c$. It is known that the quantum moduli space of SQCD with $N_f\geq N_c+1$ coincides with the classical one \cite{Seiberg:1994bz}. Therefore we have
\begin{equation}\label{eqn:HB}
HB=HB_{\text{Classical}}= HB_{\text{Quantum}}=\mathcal{F}_H/G^{\mathbb{C}}\,,\quad \mathcal{F}_H=\left\{Q_{(i,n)},\widetilde{Q}_{(i,n)}\,\middle|\,F_{(j,m)}=0\right\}\,.
\end{equation}
where $Q_{(i,n)},\widetilde{Q}_{(i,n)}$ are the scalar vevs of the theory which have $r_{\mathcal{N}=2}=0$. The only non-trivial F-terms in a Lagrangian theory using the building blocks we described in Section \ref{sec:quivers} associated to a Riemann surface of genus $g$ with $\ell$ punctures on this branch are therefore just $F_{(i,n)}=\partial W_{\mathcal{S}_k}/\partial{\Phi_{(i,n)}}$ where $W_{\mathcal{S}_k}$ is given in \eqref{eqn:Sksuperpotential}. $\mathcal{F}_H$ can be characterised by the quotient ring $\mathcal{R}_H/\mathcal{I}_H$ where
\begin{equation}
\mathcal{R}_H=\mathbb{C}[Q_{(i,n)},\widetilde{Q}_{(i,n)}]\,,\quad\mathcal{I}_H=\langle F_{(1,1)},\dots,F_{(k,1)},\dots,F_{(1,3g-3+\ell)},\dots,F_{(k,3g-3+\ell)}\rangle\,.
\end{equation}

We are now in a position to define the Higgs branch Hilbert series for our class $\mathcal{S}_k$ theories. The total independent grading on the ring $\mathcal{R}_H$ can be paramterised by a fugacity $\tau$ for the generator $E=2q_t=\frac{3}{2}r$, from the conditions \eqref{eqn:hbdefcharges}, as well as the fugacities for $U(1)^{k-1}_{\beta}\times U(1)^{k-1}_{\gamma}$ and any other global symmetries. The Higgs branch Hilbert series is then defined to be \cite{Feng:2007ur,Benvenuti:2006qr} 
\begin{equation}\label{eqn:HS}
\HS(\tau;HB):=\Tr_{HB}\tau^Ee^{-\mathcal{J}}\,,
\end{equation}
where $\Tr_{HB}$ denotes the trace over the space of operators parametrising the Higgs branch \eqref{eqn:HB} and $e^{-\mathcal{J}}$ collectively denotes the fugacities for the remaining intrinsic $U(1)^{k-1}_{\gamma}\times  U(1)^{k-1}_{\beta}$ and any other global symmetries. See also \cite{Gray:2008yu,Hanany:2008kn} for a detailed analysis of the Hilbert series for $\mathcal{N}=1$ SQCD. 
For gauge theories \eqref{eqn:HS} takes the general form of an integral over the gauge group $G$ of the Hilbert series of the master space $\mathcal{F}_H$ which, by abuse of notation, we often simply denote as $\mathcal{F}_H$ 
\begin{equation}
\HS(\tau;HB)=\oint d\mu_G(\mathbf{z})\mathcal{F}_H(\tau,\mathbf{z},\dots)\,,
\end{equation}
where $d\mu_G(\mathbf{z})$ denotes the Haar measure of $G$. 

Examining Table \ref{tab:shorts}, where short representations of the $\mathcal{N}=1$ superconformal algebra are listed, it is clear that the Hilbert series \eqref{eqn:HS} counts the top components of $\overline{\mathcal{D}}_{(0,0)}$ and $\overline{\mathcal{B}}_{r,(0,0)}$ multiplets of the $\mathcal{N}=1$ superconformal algebra. These multiplets have $E=\frac{3}{2}r$ and $j_1=j_2=0$. We can therefore see that, for $k=1$, this definition does indeed coincide with the usual Higgs branch definition for $\mathcal{N}=2$ theories.

\subsubsection{The Coulomb Branch}\label{sec:CB}
We can also make a similar definition for the Coulomb branch. Namely, analogously to $\mathcal{N}=2$ theories, we may define a consistent truncation of the moduli space by restricting those operators which have 
\begin{equation} 
\label{eqn:cbdefcharges}
E=-r_{\mathcal{N}=2}=-\frac{2q_t}{3}+\frac{r}{2}\,,\quad R_{\mathcal{N}=2}=\frac{r}{2}+\frac{q_t}{3}=0 \, .
\end{equation} 
The two conditions in \eqref{eqn:cbdefcharges} combine to 
  \begin{equation}
  \label{eqn:cbdefTotal}
  E=\frac{3}{2}r= - q_t\, ,
 \end{equation}
 with the first equality being the  BPS condition of {\it isolated short multiplets};  $\overline{\mathcal{B}}_{r,(j_1,0)}$ in the notation \cite{Dolan:2002zh}  (or   $L\overline{B}_1[j_1,0]_E^{(r)}$ in the the notation of \cite{Cordova:2016emh}), summarised in Table \ref{tab:shorts}. 
 What is special about the Coulomb branch for theories of Class $\mathcal{S}_k$, as opposed to general $\mathcal{N}=1$ SCFTs,  is that the R-symmetry of the $\mathcal{N}=1$ superconformal algebra is further related to the charge $q_t$ of the extra flavour $U(1)_t$ as in \eqref{eqn:cbdefTotal}.

For Lagrangian theories this coincides with setting the scalars arising from the free trinions to zero while giving the scalar in the $\Phi$ chiral multiplets (defined in Table \ref{tab:fieldsSK}) 
generic diagonal vevs. On this branch the gauge symmetry is broken down to the stabiliser subgroup
 of $SU(N)^{k(3g-3+\ell)}$ with respect to the vevs $\Phi_{(i,n)}$ which is given by $U(1)^{(k-1+\delta_{k,1})(3g-3+\ell)}$ where $\ell$ is the number of punctures. We have that
\begin{equation}\label{eqn:CB}
CB=CB_{\text{Classical}}= CB_{\text{Quantum}}=\mathcal{F}_C/G^{\mathbb{C}}\,,\quad \mathcal{F}_C=\left\{\Phi_{(i,n)}\right\}\,,
\end{equation}
where $\Phi_{(i,n)}$ are the scalar vevs of the theory that have $R_{\mathcal{N}=2}=0$. On this branch all of the F-terms in a Lagrangian theory are trivial because they are all proportional to either a $Q$ or $\widetilde{Q}$. Therefore $\mathcal{F}_C$ is simply associated to the freely generated ring $\mathcal{R}_C=\mathbb{C}[\Phi_{(i,n)}]$.
Similarly, we may also define the Hilbert series for the Coulomb branch
\begin{equation}\label{eqn:HSCB}
\HS(T;CB):=\Tr_{CB}T^Ee^{-\mathcal{J}}\,,
\end{equation}
where $\Tr_{CB}$ denotes the trace over the space of operators parametrising \eqref{eqn:CB}. In other words, the space of scalar operators of the theory satisfying \eqref{eqn:cbdefcharges} $(E=\frac{3}{2}r=-q_t)$. Because the F-terms are trivial, for Lagrangian theories the F-flat Hilbert series \eqref{eqn:HSCB} may be computed by multiplying the contribution coming from each $\Phi$-multiplet and integrating over the gauge group. One extra simplification that arises is the fact that, because $Q=\widetilde{Q}=0$, the $(i,n)^{\text{th}}$ node in the quiver is not coupled to the $(j,m\neq n)^{\text{th}}$. Therefore \eqref{eqn:HSCB} reduces to a product of factors associated to the $\Phi$-gluing of colour $c$ and positive sign $\sigma=+$
\begin{equation}
\label{eq:hscb}
\HS(T;CB)=\prod_{\Phi^+_c}h_{\Phi^+_c}\,,\quad h_{\Phi^+_c}=\oint d\mu\,\mathcal{F}_{\Phi^+_c}\,,\quad \mathcal{F}_{\Phi^+_{c}}=\prod_{i=1}^k\frac{(1-T)^{\delta_{1,k}}}{\prod_{A,B=1}^N\left(1-T\frac{\gamma_i}{\beta_{i+c-2}}\frac{z_{i,A}}{z_{i-1,B}}\right)}\,,
\end{equation} 
the $z$'s denote fugacities for the product gauge group, which are integrated over using the invariant measure $d\mu$. The integrals $h_{\Phi_c^+}$ were computed in \cite{Razamat:2018zus}
\begin{equation}\label{eqn:hscbhfn}
h_{\Phi^+_c}=\PE\left[\sum_{i=1}^k\frac{\gamma^N_i}{\beta^N_{i+c-2}}T^N+\sum_{A=1}^{N-1}T^{Ak}-\delta_{1,k}T\right]\,,
\end{equation}
where the symbol $\PE$ denotes the Pleythistic Exponetial defined in \eqref{eqn:PE}. As we will discuss in Section \ref{sec:sci} the factor $\mathcal{F}_{\Phi^+_c}$, appearing in equation (\ref{eq:hscb}), coincides with the Coulomb limit of the Superconformal index of the $\Phi$-gluing factor \eqref{eqn:CoulombBuild}. Therefore we reach the conclusion that for these Lagrangian theories $\HS(T;CB)=I^{\text{C}}$, where $I^{\text{C}}$ denotes the Coulomb branch Superconformal index (defined in \eqref{eqn:CoulombInd}).

\section{The Superconformal Index and some Interesting Limits}
\label{sec:sci}
\begin{table}[t]
\centering
\renewcommand{\arraystretch}{1.2}
\begin{tabular}{ |c||c|c|c|c|c|c|c| } 
 \hline
 $\mathcal{Q}$ & $j_1$ & $j_2$ & $R_{\mathcal{N}=2}$ & $r_{\mathcal{N}=2}$ & $r$ & $q_t$&$\delta=2\{\mathcal{Q},\mathcal{S}\}$ \\[2pt] 
  \hline\hline
 $\mathcal{Q}_{1\pm}$ & $\pm\frac{1}{2}$ & $0$ & $+\frac{1}{2}$ & $+\frac{1}{2}$ & $+\frac{1}{3}$ & $+1$ &$\delta_{1\pm}=E\pm2j_1-\frac{r}{2}-\frac{4q_t}{3}$\\[2pt] \hline
$\mathcal{Q}_{\pm}=\mathcal{Q}_{2\pm}$ & $\pm\frac{1}{2}$ & $0$ & $-\frac{1}{2}$ & $+\frac{1}{2}$ & $-1$ & $0$&$\delta_{\pm}=\delta_{2\pm}=E\pm2j_1+\frac{3r}{2}$ \\ [2pt]\hline
$\widetilde{\mathcal{Q}}_{\dot\pm}=\widetilde{\mathcal{Q}}_{1\dot\pm}$ & $0$ & $\pm\frac{1}{2}$ & $+\frac{1}{2}$ & $-\frac{1}{2}$ & $+1$ & $0$&$\widetilde{\delta}_{\dot-}=\widetilde{\delta}_{1\dot\pm}=E\pm2j_2-\frac{3}{2}r$ \\ [2pt]\hline
$\widetilde{\mathcal{Q}}_{2\dot\pm}$ & $0$ & $\pm\frac{1}{2}$ & $-\frac{1}{2}$ & $-\frac{1}{2}$ & $-\frac{1}{3}$ & $-1$&$\widetilde{\delta}_{2\dot\pm}=E\pm2j_2+\frac{r}{2}+\frac{4q_t}{3}$ \\ [2pt]\hline
\end{tabular}
\caption{\it Supercharges of the $\mathfrak{su}(2,2|2)$ superalgebra and its $\mathfrak{su}(2,2|1)$ subsuperalgebra. The orbifold breaks $\mathfrak{su}(2,2|2)\xrightarrow[]{}\mathfrak{su}(2,2|1)\oplus\mathfrak{u}(1)_t$ and projects out any supercharge with $q_t\neq0$. In radial quantisation $\mathcal{S}=\mathcal{Q}^{\dagger}$.}
\label{tab:supercharges}
\end{table}
The right-handed $\mathcal{N}=1$ Superconformal index computed with respect to $\widetilde{\mathcal{Q}}_{\dot-}$ is given by \cite{Kinney:2005ej,Romelsberger:2005eg} (see Table \ref{tab:supercharges} for the quantum numbers of the supercharges and \cite{Rastelli:2016tbz} for a review).
\begin{equation}
\begin{aligned}
\label{eqn:SCI}
I\left(t,p,q,\dots\right)=&\Tr(-1)^Fp^{j_1+j_2+\frac{r}{2}-\frac{2q_t}{3}}q^{-j_1+j_2+\frac{r}{2}-\frac{2q_t}{3}}t^{q_{t}}e^{-\mathcal{J}}e^{-\beta\widetilde{\delta}_{\dot-}}\\
=&\Tr(-1)^F\sigma^{\frac{1}{2}\delta_{1+}}\rho^{\frac{1}{2}\delta_{1-}}\tau^{\frac{1}{2}\widetilde{\delta}_{2\dot+}}e^{-\mathcal{J}}e^{-\beta'\widetilde{\delta}_{\dot-}}
\end{aligned}
\end{equation}
where $\Tr$ denotes the trace over the Hilbert space on $\mathbb{S}^3$ in the radial quantisation,  $(E,j_1,j_2,r)$ denote the Cartans of the maximal compact bosonic subalgebra $\mathfrak{u}(1)_E\oplus\mathfrak{su}(2)_1\oplus\mathfrak{su}(2)_2\oplus \mathfrak{u}(1)_{r}\subset\mathfrak{su}(2,2|1)$ and $q_t$ denotes the generator for the `intrinsic' global $\mathfrak{u}(1)_t$ symmetry.
What is more, $e^{-\mathcal{J}}$ collectively denotes the fugacities for the remaining intrinsic $U(1)^{k-1}_{\gamma}\times  U(1)^{k-1}_{\beta}$ and any other global symmetries.
Finally, the different $\delta_{1\pm},\delta_{2\pm},\widetilde{\delta}_{1\dot\pm},\widetilde{\delta}_{2\dot\pm}$ are given by
\begin{gather}
\label{eqn:deltas} 
\delta_{1\pm}=E\pm2j_1-\frac{r}{2}-\frac{4q_t}{3}\,,\quad \widetilde{\delta}_{2\dot\pm}=E\pm2j_2+\frac{r}{2}+\frac{4q_t}{3}\,,\\
\label{eqn:newparam}
p=\tau\sigma\,,\quad q=\tau\rho\,,\quad t=\tau^2\,.
\end{gather}
The superconformal index \eqref{eqn:SCI} receives contributions only from those states satisfying
\begin{equation}\label{eqn:BPS}
\widetilde{\delta}_{\dot-}=\widetilde{\delta}_{1\dot-}=2\{\widetilde{\mathcal{Q}}_{\dot-},\widetilde{\mathcal{S}}^{\dot-}\}=E-2j_2-\frac{3}{2}r=0\,.
\end{equation}
Special attention should be paid to the fugactity $t$. When $k=1$ the combinations 
\begin{equation}\label{eqn:Neq2Neq1embed}
R_{\mathcal{N}=2}=\frac{r}{2}+\frac{q_t}{3}\,,\quad r_{\mathcal{N}=2}=\frac{2q_t}{3}-\frac{r}{2}\,,
\end{equation} 
and \eqref{eqn:deltas} are elements of the enhanced $\mathfrak{su}(2,2|2)$ superconformal algebra, see Table \ref{tab:supercharges}. When $k\geq2$ there is generically no $\mathcal{N}=2$ enhancement and $q_{t}$ generates a global $U(1)_{t}$ symmetry of the corresponding theory.

Note that if the lowest component of a chiral superfield is given by $f$ then the fermion which also contributes to the index has 
\begin{equation}
\delta_{1\pm}[\mathcal{\widetilde{Q}}_{\dot+}\overline{f}]=2-\delta_{1\pm}[f]\,,\quad \widetilde{\delta}_{2\dot+}[\mathcal{\widetilde{Q}}_{\dot+}\overline{f}]=4-\widetilde{\delta}_{2\dot+}[f]\,.
\end{equation}
The importance of the above relations is due to the fact  that, for any chiral superfield $f$, the condition that each state contributing to the index has $\delta_{1\pm},\widetilde{\delta}_{2\dot+}\geq0$ is equivalent to
\begin{equation}
0\leq\delta_{1\pm}[f]\leq2\,,\quad 0\leq\widetilde{\delta}_{2\dot+}[f]\leq4\,.
\end{equation}
We will employ the above inequalities in Section \ref{sec:trinion} in order to ensure that the limits of the SCI are well defined for theories without a Lagrangian description. We report in the last columns of  Table \ref{tab:lettersft} and Table \ref{tab:lettersphi}  the values of the different $\delta,\widetilde{\delta}$ for the field content of the free trinion and $\Phi$-gluing. 

We reviewed the construction of the basic Lagrangian theories in class $\mathcal{S}_k$ in Section \ref{sec:quivers}. The letters of the free trinion of Figure \ref{fig:freetrinion} that contribute to the index are listed in Table \ref{tab:lettersft}.
\begin{table}
\renewcommand{\arraystretch}{1.3}
\centering
\begin{tabular}{|c||c|c|c|c|c|c|c|c|} 
\hline
 &  $j_1$ & $j_2$ & $r$ & $q_t$& Index&$\delta_{1+}$&$\delta_{1-}$&$\widetilde{\delta}_{2\dot+}$\\ 
 \hline\hline
  $Q_i$ & $0$ & $0$ & $\frac{2}{3}$ &$\frac{1}{2}$&$\sqrt{t}\frac{\beta_{i+c-1}}{\alpha}\chi_{l,i}\overline{\chi}_{r,i}$&$0$&$0$&$2$\\[2pt]\hline
 $\overline{\psi}_{\dot+i}$ & $0$ & $+\frac{1}{2}$ & $\frac{1}{3}$ & $-\frac{1}{2}$& $-\frac{pq}{\sqrt{t}}\frac{\alpha}{\beta_{i+c-1}}\chi_{r,i}\overline{\chi}_{l,i}$&$2$&$2$&$2$\\ [2pt]
 \hline
\hline
  $\widetilde{Q}_i$ & $0$ & $0$ & $\frac{2}{3}$ &$\frac{1}{2}$ &$\sqrt{t}\frac{\alpha}{\gamma_i}\chi_{r,i-1}\overline{\chi}_{l,i}$&$0$&$0$&$2$\\[2pt] 
 \hline
 $\overline{\widetilde{\psi}}_{\dot+i}$ & $0$ & $+\frac{1}{2}$ & $\frac{1}{3}$ &$-\frac{1}{2}$ & $-\frac{pq}{\sqrt{t}}\frac{\gamma_i}{\alpha}\chi_{l,i}\overline{\chi}_{r,i-1}$&$2$&$2$&$2$\\ [2pt]
 \hline
\hline
    $\partial_{\pm\dot+}$ &  $\pm\frac{1}{2}$ & $\frac{1}{2}$ & $0$ &$0$&  $p$, $q$&$0$, $2$&$2$, $0$&$2$, $2$\\[2pt] 
 \hline
\end{tabular}
\caption{\textit{Letters satisfying the BPS condition \eqref{eqn:BPS} for the free trinion theory associated to a sphere with one minimal puncture (with associated $U(1)$ valued fugacity $\alpha$) and two maximal punctures $s_{c}^{l,+}$ and $s_{c+1}^{r,+}$. Here $\chi_{o,i}\equiv\chi_{(1,0,\dots,0)}(\mathbf{z}_{o,i})$, $\overline{\chi}_{o,i}\equiv\chi_{(0,0,\dots,1)}(\mathbf{z}_{o,i})$ are shorthand for the characters of the fundamental and anti-fundamental representations of $SU(N)$, defined in \eqref{eqn:SUNChar}. We importantly note that $\delta_{1\pm},\widetilde{\delta}_{2\dot+}\geq0$.}}
\label{tab:lettersft}
\end{table}
The free trinion, that corresponds to an orbifolded $\mathcal{N}=2$ hypermultiplet, contributes to the index a factor
\begin{equation}\label{eqn:ftindex}
\begin{aligned}
&I_{s_{c}^{l,+},s_{c+1}^{r,+}}=\prod_{i=1}^k\prod_{A,B=1}^N\Gamma_e\left(\frac{\sqrt{t}\beta_{i+c-1}}{\alpha}\frac{z_{l,i,B}}{z_{r,i,A}}\right)\Gamma_e\left(\frac{\sqrt{t}\alpha}{\gamma_i}\frac{z_{r,i-1,B}}{z_{l,i,A}}\right)\,,
\end{aligned}
\end{equation}
where $\Gamma_e(z)$ denotes the Elliptic Gamma function, defined in \eqref{eqn:EllGamma} and with $\prod_{i=1}^k\gamma_i=\prod_{i=1}^k\beta_i=1$. Moreover, as discussed in Section \ref{sec:quivers}, we observe that the colour label of the two maximal punctures is sfhited by one.

We will also sometimes adopt the notation $I_{s_{c}^{l,+},s_{c+1}^{r,+}}=I\indices{_{\mathbf{z}_l\alpha}^{\mathbf{z}_r}}$, leaving implicit the colour of the punctures. The $\Phi$-gluing of two maximal punctures of the same sign $\sigma=+$, of colour $c$ and opposite orientation contributes
\begin{equation}\label{eqn:tubeindex}
I_{\Phi^+_c}=\prod_{i=1}^k\frac{\kappa\prod_{A,B=1}^N\Gamma_e\left(\frac{pq}{t}\frac{\gamma_i}{\beta_{i+c-2}}\frac{z_{i,B}}{z_{i-1,A}}\right)}{\Gamma_e\left(\frac{pq}{t}\right)^{\delta_{k,1}}\Delta(\mathbf{z}_{i})\prod_{A\neq B}\Gamma_e\left(\frac{z_{i,A}}{z_{i,B}}\right)}\,.
\end{equation}
\begin{table}
\renewcommand{\arraystretch}{1.2}
\centering
\begin{tabular}{|c||c|c|c|c|c|c|c|c|} 
\hline
 &  $j_1$ & $j_2$ & $r$ & $q_t$& Index&$\delta_{1+}$&$\delta_{1-}$&$\widetilde{\delta}_{2\dot+}$\\ 
 \hline\hline
  $\Phi_i$ & $0$ & $0$ & $\frac{2}{3}$ &$-1$& $\frac{pq}{t}\frac{\gamma_i}{\beta_{i+c-2}}\chi_i\overline{\chi}_{i-1}$&$2$&$2$&$0$\\ [2pt]
 \hline
     $\overline{\lambda}_{\dot{+}i}$ &  $0$ & $+\frac{1}{2}$ & $+\frac{1}{3}$ &$+1$ & $-t\frac{\beta_{i+c-2}}{\gamma_i}\chi_{i-1}\overline{\chi}_i$&$0$&$0$&$4$\\ [2pt]\hline\hline
   $\lambda_{\pm i}$ & $\pm\frac{1}{2}$ & $0$ & $1$ &$0$& $-p\chi^{\text{adj.}}_i$, $-q\chi^{\text{adj.}}_i$&$0$, $2$&$2$, $0$&$2$, $2$\\ [2pt]
 \hline
  $\widetilde{F}_{\dot+\dot+i}$ &$0$ & $1$ &$0$& $0$ & $pq\chi^{\text{adj.}}_i$&$2$&$2$&$4$\\ [2pt]
 \hline
    $\partial\lambda_i=0$ & $0$ & $+\frac{1}{2}$ & $1$ &$0$ & $pq\chi^{\text{adj.}}_i$&$2$&$2$&$4$\\ [2pt]
 \hline\hline
    $\partial_{\pm\dot+}$ &  $\pm\frac{1}{2}$ & $\frac{1}{2}$ & $0$ &$0$&  $p$, $q$&$0$, $2$&$2$, $0$&$2$, $2$\\[2pt] 
 \hline
\end{tabular}
\caption{\textit{Letters satisfying the BPS condition \eqref{eqn:BPS} of the free $\mathcal{N}=1$ theory corresponding to a tube which implements the $\Phi$-gluing of two punctures of equal colour, opposite orientation and sign $\sigma=+$. For the vector multiplet piece we must take into account the equation of motion $\partial\lambda=\partial_{+\dot+}\lambda_{-}+\partial_{-\dot+}\lambda_{+}=0$. Here $\chi_i\equiv\chi_{(1,0,\dots,0)}(\mathbf{z}_i)$, $\overline{\chi}_i\equiv\chi_{(0,0,\dots,1)}(\mathbf{z}_i)$ and $\chi^{\text{adj.}}_i\equiv\chi_{(1,0,\dots,1)}(\mathbf{z}_i)$ are shorthand for the characters of the fundamental, anti-fundamental and adjoint representations of $SU(N)$, defined in \eqref{eqn:SUNChar}. We importantly note that $\delta_{1\pm},\widetilde{\delta}_{2\dot+}\geq0$.}}
\label{tab:lettersphi}
\end{table}
Physically the above corresponds to the contribution of an orbifolded $\mathcal{N}=2$ vector multiplet. It contains an $\mathcal{N}=1$ bifundamental chiral field $\Phi$ and $\mathcal{N}=1$ vector $V$ in the adjoint, as can be seen in Table \ref{tab:lettersphi}. The factors $\kappa$ and $\Delta$ are defined in \eqref{eqn:veccont}.\footnote{The factor $\delta_{k,1}$ is to account for the fact that for $k=1$ $\Phi$ sits in the adjoint of $\mathfrak{su}(N)$ $\text{adj.}\iso \mathbf{N}\otimes\overline{\mathbf{N}}-1$, while for $k>1$ $\Phi$ sits in bifundamenal representations.} Note that in the above, and throughout, products and sums over $i$ shall always be taken modulo $k$, i.e. $i+k\sim i$ unless otherwise stated.
For instance, we denote the $\Phi$-gluing of two three punctured spheres to obtain the theory associated to a sphere with two minimal and two maximal punctures at the level of the index by \footnote{The theory obtained in this way corresponds to SQCD$_n$, that is to say the orbifold of $\mathcal{N}=2$ SQCD.}
\begin{equation}\label{eqn:4puncsphereindex}
I_{s_{c}^{l,+},s_{c+2}^{r,+}}=\oint\prod_{i=1}^kd\mu_i\,I_{s_{c}^{l,+},s_{c+1}^{r,+}}I_{\Phi_{c+1}^+}I_{s_{c+1}^{l,+},s_{c+2}^{r,+}}\equiv I\indices{_{\mathbf{u}\alpha\delta}^{\mathbf{v}}}\,,
\end{equation}
we observe that the colour label of the second maximal puncture has been shifted by two. Let's now consider more in details the expression \eqref{eqn:4puncsphereindex}, starting from it we can understand some general properties of the theories taken into consideration. In order to do this let's consider the simplest case: $k=N=c+1=2$ and let's expand \eqref{eqn:4puncsphereindex} in terms of $\mathcal{N}=1$ index equivalence classes\footnote{These equivalence classes have been defined in Appendix \ref{app:SCAreps}. In particular, for the equivalence class $I_{[\widetilde{r},j_1]_{\pm}}$, see equation \eqref{eqn:indequiv}.} $I_{[\widetilde{r},j_1]_{\pm}}$  \cite{Gadde:2009dj,Beem:2012yn,Evtikhiev:2017heo} as
\begin{equation}
\begin{aligned}
I\indices{_{\mathbf{u}\alpha\delta}^{\mathbf{v}}}=&1+\frac{1}{t'^2}\left[1+\frac{\beta^2}{\gamma^2}+\frac{\gamma^2}{\beta^2}\right]I_{[-\frac{2}{3},0]_-}+\sum_{i=1}^2t'\left[\beta_i^2\left(\frac{1}{\alpha^2}+\frac{1}{\delta^2}\right)+\gamma_i^2\left(\alpha^2+\delta^2\right)\right.\\
&+\beta_i\gamma_i\chi_1(u_i)\chi_1(u_{i+1})+\beta_i\gamma_i\chi_1(v_i)\chi_1(v_{i+1})+\left(\frac{\alpha}{\delta}+\frac{\delta}{\alpha}\right)\frac{\beta_i}{\gamma_i}\chi_1(u_i)\chi_1(v_{i+1})\\
&\left.+\left(\alpha\delta+\frac{1}{\alpha\delta}\right)\chi_1(u_i)\chi_1(v_i)\right]I_{[-\frac{2}{3},0]_-}+\sum_{i=1}^2\left[\left(\frac{\alpha\delta}{\beta_i\gamma_i}+\frac{\beta_i\gamma_i}{\alpha\delta}\right)\chi_1(u_i)\chi_1(v_{i+1})\right.\\
&\left.+\chi_2(u_i)+\chi_2(v_i)+4\vphantom{\frac{\alpha\delta}{\beta_i\gamma_i}}\right]I_{[0,0]_+}+\sum_{i=1}^2\left[\left(\frac{\beta_i^3}{\alpha\delta\gamma_i}+\frac{\alpha\delta\beta_i}{\gamma_i^3}\right)\chi_1(u_i)\chi_1(v_{i+1})\right.\\
&\left.+\frac{\gamma_i^2}{\beta_i^2}\left(\chi_2(u_{i+1})+\chi_2(v_i)+1\right)\right]I_{[0,0]_-}+\mathcal{O}((pq)^{4/3})
\end{aligned}
\end{equation}
where $\beta=\beta_1=\beta_2^{-1}$ and $\gamma=\gamma_1=\gamma_2^{-1}$ and we defined $t'=t/(pq)^{2/3}$ so that the expansion is made using the free R-symmetry. The equivalence class $[\frac{1}{3},\frac{1}{2}]_-$,  which has only a single representative $\hat{\mathcal{C}}_{(\frac{1}{2},0)}$ containing a spin $3/2$ current and contributes to the index a factor proportional to $+(pq)^{2/3}(p+q)/(1-p)(1-q)$, is absent. There are only two possible explantions of this fact. The first is that  the theory has no enhancement to $\mathcal{N}=2$ supersymmetry. The second is that the theory contains a number of $\overline{\mathcal{B}}_{\frac{7}{3}(\frac{1}{2},0)}$ multiplets (which is the single representative of the $[\frac{1}{3},\frac{1}{2}]_+$ equivalence class), that cancel the contribution arising from the equivalence class $[\frac{1}{3},\frac{1}{2}]_{-}$ in such a way that we do not see it in the index expansion. Note that all of the equivalence classes in the above contain only a single representative. In particular one can replace $I_{[-\frac{2}{3},0]_-}=I_{\overline{\mathcal{B}}_{\frac{4}{3},(0,0)}}$, $I_{[0,0]_+}=I_{\hat{\mathcal{C}}_{(0,0)}}$ and $I_{[0,0]_-}=I_{\overline{\mathcal{B}}_{2,(0,0)}}$. The \textit{net degeneracy} \cite{Beem:2012yn}, defined in \eqref{eqn:netdegen}, of the $[0,0]_{\pm}$ equivalence classes counts
\begin{equation}
\begin{aligned}
\text{ND}[0,0]=&\#[0,0]_+-\#[0,0]_-=\#\overline{\mathcal{B}}_{2,(0,0)}-\#\hat{\mathcal{C}}_{(0,0)}\\
=&\#\text{marginal operators}-\#\text{conserved currents}\\
=&30-36\,.
\end{aligned}
\end{equation}
We observe that, when $k=1$, the index \eqref{eqn:SCI} admits various interesting limits involving the three fugacities $p,q,t$ (or $\rho,\sigma,\tau$) in which the index receives contribution only from states annihilated by two or more $\mathcal{N}=2$ Poincar\'e supercharges (one of them, of course, always being $\widetilde{\mathcal{Q}}_{\dot-}$) \cite{Gadde:2011uv}.

On the other hand when $k\geq 2$ the situation is different. We notice that the index of a generic $\mathcal{N}=1$ SCFT admits no non-trivial limits in which the states contributing to it are annihilated by more than one supercharge since 
$\widetilde{\delta}_{\dot-}[\mathcal{Q}_{\alpha}]\neq 0$ and $\widetilde{\delta}_{\dot-}[\widetilde{\mathcal{Q}}_{\dot+}]\neq0$.
However, when the $\mathcal{N}=1$ SCFT has flavour symmetry, we may consider taking limits also involving the flavour fugacities. For generic theories there is no guarantee that such limits are well defined. Moreover, the index in certain limits; although not leading to extra superconformal shortening, can often admit drastic simplifications. Similar ideas have also been deployed in studying `non-generic' $\mathcal{N}=1$ SCFTs in e.g. \cite{Beem:2012yn,Gaiotto:2015usa,Rastelli:2016tbz}.

\subsection{The Hall-Littlewood Limit of the Index}
We can study the limit which, for $\mathcal{N}=2$ theories, is equivalent to the so-called Hall-Littlewood limit of the index \cite{Gadde:2011uv}.\footnote{This limit has also been considered for $\mathcal{N}=1$ theories in \cite{Spiridonov:2009za,Rastelli:2016tbz}. Geometrically it corresponds to collapsing the $\mathbb{S}^3$ to a point.}
For $\mathcal{N}=1$ Class $\mathcal{S}_k$ theories it was first considered  in \cite{Gaiotto:2015usa}, and it reads
\begin{equation}
\sigma\to0\,,\quad \rho\to0\,,\quad \tau\,\,\text{fixed,}
\end{equation} 
or, equivalently, $p,q\to0$ with $t$ held fixed. From \eqref{eqn:ftindex} and \eqref{eqn:tubeindex} we see that the Hall-Littlewood limit of the indices for the Lagrangian building blocks is well defined. The existence of this limit is equivalent to the fact that each letter contributing to the basic building building blocks have $\delta_{1\pm}\geq0$ (see Table \ref{tab:lettersft} and Table \ref{tab:lettersphi}), as is the case for all $\mathcal{N}=2$ theories.  We can therefore write
\begin{equation}\label{eqn:HL}
\HL(\tau,\dots):=\lim_{\sigma,\rho\to0}I=\Tr_{\HL}(-1)^F\tau^{2q_t}e^{-\beta\widetilde{\delta}_{\dot-}}e^{-\mathcal{J}}\,,
\end{equation}
here $\Tr_{\HL}$ denotes the restriction of $\Tr$ to the states satisfying $\delta_{1\pm}=0$, i.e. 
$2q_t=\frac{3}{2}r+3j_2=E+j_2$ and $j_1=0$. The indices for the building blocks \eqref{eqn:ftindex} and \eqref{eqn:tubeindex} become
\begin{gather}
\HL_{s_{c}^{l,+},s_{c+1}^{r,+}}=\prod_{i=1}^k\prod_{A,B=1}^N\frac{1}{\left(1-\tau\frac{\beta_{i+c-1}}{\alpha}\frac{z_{l,i,B}}{z_{r,i,A}}\right)\left(1-\tau\frac{\alpha}{\gamma_i}\frac{z_{r,i-1,B}}{z_{l,i,A}}\right)}\label{eqn:HLFT}\\
\HL_{\Phi_c^+}=\frac{\prod_{i=1}^k\prod_{A,B=1}^N\left(1-\tau^2\frac{\beta_{i+c-2}}{\gamma_i}\frac{z_{i-1,A}}{z_{i,B}}\right)}{\left(1-\tau^2\right)^{\delta_{k,1}}}\label{eqn:HLSglue}\,.
\end{gather}

As pointed out in \cite{Gadde:2011uv,Hanany:2012dm}, for Class $\mathcal{S}$ theories at genus $g=0$, the Hall-Littlewood limit of the index coincides with the Hilbert series of the Higgs branch. For Lagrangian theories this can be explicitly proved and can be argued to extend to theories related to Lagrangian theories by S-duality \cite{Gaiotto:2012uq}. To the best of the author's knowledge a full proof that extends to all class $\mathcal{S}$ theories is currently lacking. In Section \ref{sec:genuszero} we will demonstrate that the same property also holds for Lagrangian theories made using $\Phi$-gluing in class $\mathcal{S}_k$ at genus $g=0$.

\subsection{The Coulomb Limit of the Index}
The Coulomb limit of the index for $\mathcal{N}=2$ theories is given by
\begin{equation}
\tau\to0\,,\quad \rho,\sigma\,\,\,\text{fixed,}
\end{equation}
or, equivalently, $t,p,q\to0$ with $T:=pq/t=\sigma\rho$ and $V:=p/q=\sigma/\rho$ held fixed. An extensive study of this limit of the index was given in \cite{Razamat:2018zus}. For generic $\mathcal{N}=1$ theories we would have no reason to believe that this limit exists since $\widetilde{\delta}_{2\dot+}\geq0$ is no longer guaranteed. However, let us assume that it does. In this limit the index would take the form
\begin{equation}\label{eqn:CoulombInd}
I^{\text{C}}(T,V,\dots)=\Tr_{\text{C}}(-1)^F\sigma^{\frac{1}{2}\delta_{1+}}\rho^{\frac{1}{2}\delta_{1-}}e^{-\mathcal{J}}e^{-\beta'\widetilde{\delta}_{\dot-}}=\Tr_{\text{C}}(-1)^FT^{E+j_2}V^{j_1}e^{-\mathcal{J}}e^{-\beta'\widetilde{\delta}_{\dot-}}
\end{equation} 
Here $\Tr_{\text{C}}$ denotes the restriction of $\Tr$ to states with $\widetilde{\delta}_{2\dot+}=0$. Indeed, we can see at the level of the Lagrangian building blocks that the limit does exist and, moreover, is conjectured to exist for all theories in class $\mathcal{S}_k$ of type $A_{N-1}$ \cite{Razamat:2018zus}. The indices for the Lagrangian building blocks \eqref{eqn:ftindex} and \eqref{eqn:tubeindex} become
\begin{equation}\label{eqn:CoulombBuild}
I^{\text{C}}_{s_{c}^{l,+},s_{c+1}^{r,+}}=1\,,\quad I^{\text{C}}_{\Phi^+_c}=\prod_{i=1}^k\frac{\left(1-T\right)^{\delta_{k,1}}}{\prod_{A,B=1}^N\left(1-T\frac{\gamma_i}{\beta_{i+c-2}}\frac{z_{i,B}}{z_{i-1,A}}\right)}\,.
\end{equation}
For Lagrangian theories made with $\Phi$ gluing the interpretation of this limit of the index is clear. The Coulomb limit of the index is simply counting the possible gauge invariants that can be made from the bifundamental scalar fields in the chiral multiplets $\Phi$. These operators are the top components of the $1/2$-BPS multiplets $\overline{\mathcal{D}}^{(-1)}_{(0,0)}$ and $\overline{\mathcal{B}}^{(-\frac{2r}{3})}_{r,(0,0)}$ which simultaneously have $E=\frac{3}{2}r=-q_t$ and $j_1=j_2=0$ (see Appendix \ref{app:SCAreps}). Indeed we demonstrated in Section \ref{sec:CB} that, for those theories, $I^{\text{C}}$ can be given the interpretation of a Hilbert series constructed to count the above $1/2$-BPS multiplets on the Coulomb branch.

\subsection{The Schur \& Madonald Limits of the Index}
\label{subsec:mc}
For completeness of our discussion we can also define analogues of the Schur and Macdonald limits of the index of \cite{Gadde:2011uv}. 
The analogue of the Macdonald index
 for the specific case of class $\mathcal{S}_k$ theories, was originaly considered in Appendix B of \cite{Gaiotto:2015usa}.
It is obtained by taking $\sigma\to0$, which is well defined for Lagrangian theories because each letter has $\delta_{1+}\geq0$, while holding $\rho,\tau$ fixed.
\begin{equation}
I^M(q,t,\dots)=\Tr_M(-1)^Fq^{-j_1+j_2+\frac{r}{2}-\frac{2q_t}{3}}t^{q_{t}}e^{-\mathcal{J}}e^{-\beta\widetilde{\delta}_{\dot-}}\,,
\end{equation} 
where $\Tr_{M}$ denotes the restriction of $\Tr$ to states with $\delta_{1+}=0$.
The Schur index is defined by setting $\rho=\tau$, or $q=t$
\begin{equation}\label{eqn:schur}
\begin{aligned}
I^S(q,p,\dots)=&\Tr(-1)^F\sigma^{\frac{1}{2}\delta_{1+}}\rho^{\frac{1}{2}\left(\delta_{1-}+\widetilde{\delta}_{2\dot+}\right)}e^{-\mathcal{J}}e^{-\beta\widetilde{\delta}_{\dot-}}\\=&\Tr(-1)^Fp^{j_1+j_2+\frac{r}{2}-\frac{2q_t}{3}}q^{-j_1+j_2+\frac{r}{2}+\frac{q_t}{3}}e^{-\mathcal{J}}e^{-\beta\widetilde{\delta}_{\dot-}}
 \,.
 \end{aligned}
\end{equation} 
Note that, unlike for $\mathcal{N}=2$ theories, the Schur index \eqref{eqn:schur} is dependent on $\sigma=p/\sqrt{q}$. However, for Lagrangian theories we have $\delta_{1+}\geq0$ and therefore we can consider a further limit, which we will call the reduced Schur index
\begin{equation}\label{eqn:redschur}
I^{RS}(q,\dots):=\lim_{p\to0}I^S=\Tr_{M}(-1)^Fq^{E-\frac{r}{2}-\frac{q_t}{3}}e^{-\mathcal{J}}e^{-\beta\widetilde{\delta}_{\dot-}}\equiv I^M|_{q=t}\,.
\end{equation} 
In particular, for the index $I^{RS}$ the Lagrangian building blocks become
\begin{align}
&I^{RS}_{s_{c}^{l,+},s_{c+1}^{r,+}}=\prod_{i=1}^k\prod_{A,B=1}^N\frac{1}{\left(\sqrt{q}\frac{\beta_{i+c-1}}{\alpha}\frac{z_{l,i,B}}{z_{r,i,A}};q\right)\left(\sqrt{q}\frac{\alpha}{\gamma_i}\frac{z_{r,i-1,B}}{z_{l,i,A}};q\right)}\,,\\
&I_{\Phi^+_c}^{RS}=(q;q)^{k-\delta_{k,1}}\prod_{i=1}^k\prod_{A,B=1}^N\left(q\frac{\beta_{i+c-2}}{\gamma_i}\frac{z_{i-1,A}}{z_{i,B}};q\right)\left(q\frac{z_{i,A}}{z_{i,B}};q\right)\,.
\end{align}

For the $k=N=2$ theory associated to a sphere with two minimal punctures and two maximal punctures $s_{1}^{l,+}$, $s_{3}^{r,+}$ the reduced Schur index can be expanded as
\begin{equation}
\begin{aligned}\label{eqn:IRSexpand}
I^{RS}=&1+\Big[\chi_1(m^2)\left(a^2 [0,1,0;0,0,0]+\frac{1}{a^2}[0,0,0;0,1,0]\right)+[1,0,0;0,0,1]\\
&+[0,0,1;1,0,0] \Big]q+\Big[\frac{1}{a^2}\chi_1(m^2)\left([0,0,1;1,1,0]+[1,0,0;0,1,1]\right)\\
&+a^2\chi_1(m^2)\left([1,1,0;0,0,1] + [0,1,1;1,0,0]\right)+[0,1,0;0,1,0]\\
&+[1,0,1;1,0,1]+[2,0,0;0,0,2]+[0,0,2;2,0,0]\\
&+ \chi_2(m^2)\left(a^4[0,2,0;0,0,0]+\frac{1 }{a^4}[0,0,0;0,2,0]+[0,1,0;0,1,0]\right)\Big]q^2\\
&+\mathcal{O}(q^3)\,.
\end{aligned}
\end{equation} 
Note that here, we used the symmetry enhancement of this theory \eqref{eqn:4punsymenhance}, which we will discuss in more detail in Section \ref{sec:coreint}. Here $[d_1,d_2,d_3;d_1',d_2',d_3']$ denotes the character of the enhanced $SU(4)^2$ symmetry.

For theories with $\mathcal{N}\geq2$ supersymmetry the quantity $I^S$ (which, for $\mathcal{N}\geq2$ supersymmetry, equals $I^{RS}$) plays a pivotal role in the chiral algebra 2d/4d correspondence. In particular $I^S$ is identified with the vacuum character of the associated chiral algebra. 

The stress tensor of the associated chiral algebra is identified with the top component of the $\mathfrak{su}(2)_{R_{\mathcal{N}=2}}$ current, namely $j^{\mu}_{11}$. This current lives in the stress tensor multiplet $\hat{\mathcal{C}}_{0(0,0)}$ which contains conserved $\mathfrak{su}(2)_{R_{\mathcal{N}=2}}$ and $\mathfrak{u}(1)_{r_{\mathcal{N}=2}}$ currents $j^{\mu}_{(IJ)}$ and $j^{\mu}$ with $I,J=1,2$ $\mathfrak{su}(2)_{R_{\mathcal{N}=2}}$ indices. This current enters the Schur index $I^S$ with a factor $I^S_{\hat{\mathcal{C}}_{0(0,0)}}=q^2/(1-q)$.

\begin{figure}
\centering
\begin{tikzcd}[row sep=0.2cm, column sep=0.2cm]
&&\color{red}[0,0]^{(0)}_2\arrow[dl]\arrow[dr]&&\\
&\begin{tabular}{c}$\color{red}[\frac{1}{2},0]^{(-1)}_{\frac{5}{2}}$,\\
$\color{green}[\frac{1}{2},0]^{(\frac{1}{3})}_{\frac{5}{2}}$\end{tabular}\arrow[dl]\arrow[dr]&&\arrow[dl]\arrow[dr]\begin{tabular}{c}$\underbracket{\color{red}[0,\frac{1}{2}]^{(1)}_{\frac{5}{2}}\color{black}}_{\dot+}$,\\$\color{blue}[0,\frac{1}{2}]^{(-\frac{1}{3})}_{\frac{5}{2}}$\end{tabular}&\\
\color{green}[1,0]^{(-\frac{2}{3})}_{3}\arrow[dr]&&\arrow[dl]\arrow[dr]\begin{tabular}{c}
$\underbracket{\color{green}[\frac{1}{2},\frac{1}{2}]^{(\frac{4}{3})}_{3}\color{black}}_{\dot+},\color{blue}[\frac{1}{2},\frac{1}{2}]^{(-\frac{4}{3})}_{3}$, \\
$\color{red}[\frac{1}{2},\frac{1}{2}]^{(0)}_{3}\color{black},\color{cyan}[\frac{1}{2},\frac{1}{2}]^{(0)}_{3}$\end{tabular}&&\arrow[dl]\underbracket{\color{blue}[0,1]^{(\frac{2}{3})}_{3}\color{black}}_{\dot{+} \dot{+}}\\
&\arrow[dr]\begin{tabular}{c}$\color{cyan}[1,\frac{1}{2}]^{(-1)}_{\frac{7}{2}}$,\\
$\color{green}[1,\frac{1}{2}]^{(\frac{1}{3})}_{\frac{7}{2}}$\end{tabular}&&\begin{tabular}{c}$\underbracket{\color{cyan}[\frac{1}{2},1]^{(1)}_{\frac{7}{2}}\color{black}}_{\dot+\dot+}$,\\$\color{blue}[\frac{1}{2},1]^{(-\frac{1}{3})}_{\frac{7}{2}}$\end{tabular}\arrow[dl]&\\
&&\color{cyan}[1,1]^{(0)}_4&&
\end{tikzcd}
\caption{\textit{Branching of $\hat{\mathcal{C}}_{0(0,0)}\iso \color{red}\hat{\mathcal{C}}_{(0,0)}\color{black}\oplus \color{green}\hat{\mathcal{C}}_{\left(\frac{1}{2},0\right)}\color{black}\oplus  \color{blue}\hat{\mathcal{C}}_{\left(0,\frac{1}{2}\right)}\color{black}\oplus \color{cyan}\hat{\mathcal{C}}_{\left(\frac{1}{2},\frac{1}{2}\right)}$. Underlined are those states, with given $j_2$, which have $\widetilde{\delta}_{\dot-}=E-2j_2-\frac{3}{2}\rN1=0$ and thus can contribute to the right-handed index \eqref{eqn:SCI}.}}
\label{fig:Chat000}
\end{figure}
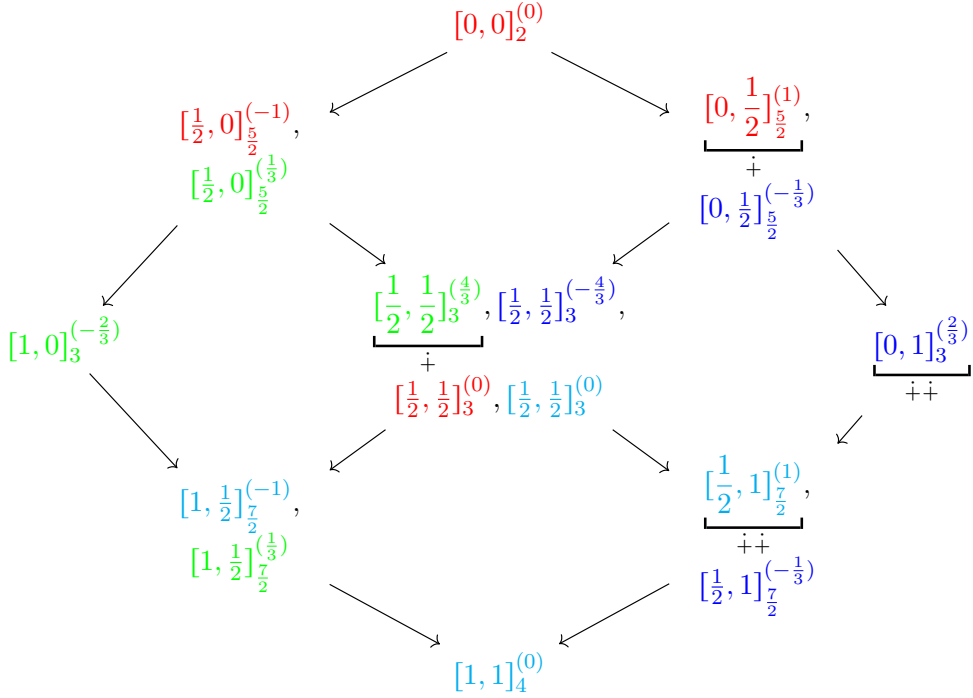

Under the decomposition $\mathfrak{su}(2,2|2)\to\mathfrak{su}(2,2|1)\oplus\mathfrak{u}(1)_t$ the $\mathcal{N}=2$ stress tensor multiplet decomposes as
\begin{gather}
\hat{\mathcal{C}}_{0(0,0)}\iso\, (\hat{\mathcal{C}}_{(0,0)},0)\oplus (\hat{\mathcal{C}}_{\left(\frac{1}{2},0\right)},+1)\oplus  (\hat{\mathcal{C}}_{\left(0,\frac{1}{2}\right)},-1)\oplus (\hat{\mathcal{C}}_{\left(\frac{1}{2},\frac{1}{2}\right)},0)\,,\\
I_{\hat{\mathcal{C}}_{0(0,0)}}=\frac{-pq}{(1-p)(1-q)}+\frac{t(p+q)}{(1-p)(1-q)}+\frac{t^{-1}p^2q^{2}}{(1-p)(1-q)}+\frac{-pq(p+q)}{(1-p)(1-q)}\,;
\end{gather} 
see Figure \ref{fig:Chat000}. 

Since, the stress tensor multiplet of the mother SCFT sits in trivial representations of any flavour symmetries (e.g. it sits in a trivial $SU(N_f)$ representation) under the $\mathbb{Z}_k$ orbifold the projection of the stress tensor should simply be $\hat{\mathcal{C}}_{0(0,0)}\xrightarrow{\mathbb{Z}_k}\hat{\mathcal{C}}_{(0,0)}\oplus\hat{\mathcal{C}}_{\left(\frac{1}{2},\frac{1}{2}\right)}$. Indeed the multiplets $\hat{\mathcal{C}}_{\left(\frac{1}{2},0\right)}$ and $\hat{\mathcal{C}}_{\left(0,\frac{1}{2}\right)}$ contain additional supersymmetry currents which would lead to enhanced $\mathcal{N}\geq2$ supersymmetry if present. 

We can identify in the decomposition the $\hat{\mathcal{C}}_{(0,0)}$ as the $\mathfrak{u}(1)_t$ flavour current multiplet while $\hat{\mathcal{C}}_{\left(\frac{1}{2},\frac{1}{2}\right)}$ is of course the $\mathcal{N}=1$ stress tensor multiplet whose lowest component is the $\mathfrak{u}(1)_r$ current. They are built from linear combinations of $j^{\mu}_{12}=j^{\mu}_{21}$ and $j^{\mu}$ in accordance with \eqref{eqn:Neq2Neq1embed}. 



\section{Genus Zero Theories}
\label{sec:genuszero}
In this Section we consider Class $\mathcal{S}_k$ theories which arise from compactifications of
6D SCFTs on  Riemann surfaces of
 genus $g=0$. We show that, for this subclass of theories, the Hall-Littlewood limit of the index coincides with the corresponding Higgs branch Hilbert series, as is the case for $\mathcal{N}=2$ Class $\mathcal{S}$ theories. In this Section, we provide the explicit  expression of the Higgs branch Hilbert series for some Lagrangian genus zero theories, i.e. the free trinion and the interacting SCFT associated to a sphere with two maximal and two minimal punctures. 
The study of 
 genus $g=0$ theories without a Lagrangian description is presented in Section \ref{sec:trinion}.

\subsection{Hilbert Series and the Hall-Littlewood Limit of the Index}\label{sec:HLeqHS}
We are now in a position to show that the Hall-Littlewood limit of the index coincides with the Higgs branch Hilbert series at genus $g=0$ for Lagrangian theories made using $\Phi$-gluing. For the theory corresponding to a sphere with $\ell-2$ minimal punctures and two maximal punctures the relevant F-terms for the Higgs branch are 
\begin{equation}\label{eqn:genuszeroFtermsHB}
F_{(i,n)}:=\frac{\partial W_{\mathcal{S}_k}}{\partial \Phi_{(i,n)}}= \widetilde{Q}_{(i,n-1)}Q_{(i,n-1)}-Q_{(i-1,n)}\widetilde{Q}_{(i,n)}=0\,,
\end{equation}
for $n=1,\dots,-3+\ell$ and $i+k\sim i=1,\dots, k$.
For genus $g=0$ these constitute $k(-3+\ell)$ independent constraints on the $Q_{(i,n)}$ and $\widetilde{Q}_{(i,n)}$. More precisely the ideal $\mathcal{I}_H$ comprised of the list of the $F_{(i,n)}$ forms a regular sequence in $\mathcal{R}_H=\mathbb{C}[Q_{(i,n)},\widetilde{Q}_{(i,n)}]$, see Appendix \ref{app:ag}. This means that the variety whose coordinate ring is given by the quotient ring $\mathcal{R}_H/\mathcal{I}_H$ is a complete intersection and we may apply letter counting techniques to compute the Hilbert series for the master space $\mathcal{F}_H$. 

The Hilbert series of the Higgs branch of the theory associated to the a sphere with $\ell-2$ minimal punctures and two maximal punctures $s_{1}^{l,+}$, $s_{\ell-1}^{r,+}$ is precisely given by
\begin{align}\label{eqn:HSgenericgzero}
&\HS(\tau,\dots;HB)= \oint\prod_{i=1}^k\prod_{n=2}^{\ell-2} d\mu_{(i,n)}\mathcal{F}_H(\tau,\dots)\\
&\mathcal{F}_H(\tau,\dots)=\frac{\PE\left[\sum_{n=1}^{\ell-2}\sum_{i=1}^k\sum_{A,B=1}^N\left(\tau\frac{\beta_{i+n-1}}{\alpha_n}\frac{z_{(i,n),B}}{z_{(i,n+1),A}}+\tau\frac{\alpha_n}{\gamma_i}\frac{z_{(i-1,n+1),B}}{z_{(i,n),A}}\right)\right]}{\PE\left[\sum_{n=2}^{\ell-2}\sum_{i=1}^k\sum_{A,B=1}^N\tau^2\frac{\beta_{i+n-2}}{\gamma_i}\frac{z_{(i-1,n),A}}{z_{(i,n),B}}+(3-\ell)\tau^2\delta_{k,1}\right]}
\end{align}
and we see that $\mathcal{F}_H=\prod_{n=1}^{\ell-2}\HL_{s_{n}^{l,+},s_{n+1}^{r,+}}\prod_{n=2}^{\ell-2}\HL_{\Phi_n^+}$ and therefore $\HS(\tau)\equiv\HL(\tau)$ for this class of genus zero theories. The contribution of the $\overline{\lambda}_{\dot+(i,n)}$ to the Hall-Littlewood index coming from $\Phi$-gluing  precisely plays the role of the F-term constraints \eqref{eqn:genuszeroFtermsHB} in the Higgs-branch Hilbert series. 

In other words, we see that for this class of theories $\Tr_{\HL}=\Tr_{HB}$. Indeed, one can see that if one further introduces the condition $j_2=0$ into those defining the Hall-Littlewood limit of the index $\widetilde{\delta}_{\dot-}=\delta_{1\pm}=0$ we have $E=2q_t=\frac{3}{2}r$, $j_1=j_2=0$ which are precisely the conditions defining the Higgs branch \eqref{eqn:HB}.
\subsection{The Free Trinion}
Let us consider the Hall-Littlewood index/ Hilbert series of the $\mathfrak{g}=A_{N-1}$ theory associated to a sphere with one minimal puncture with fugacity $\alpha$ and two maximal punctures $s_1^{l,+}$ and $s_2^{r,+}$ a.k.a. the free trinion. The expression for the Higgs branch Hilbert series was given in \eqref{eqn:HLFT} and it reads
\begin{equation}\label{eqn:HLFT2}
\HS_{s_{1}^{l,+},s_{2}^{r,+}}=\prod_{i=1}^k\prod_{A,B=1}^N\frac{1}{\left(1-\tau\frac{\beta_{i}}{\alpha}\frac{u_{i,A}}{v_{i,B}}\right)\left(1-\tau\frac{\alpha}{\gamma_i}\frac{v_{i-1,B}}{u_{i,A}}\right)}
\end{equation}
note that we set $z_l=u$ and $z_r=v$ with respect to \eqref{eqn:HLFT}.
We checked for various low values of $N$ in expansion around $\tau=0$ that the identity
\begin{equation}
\begin{aligned}
\prod_{A,B=1}^N\frac{1}{1-a\tau \frac{u_A}{v_B}}=&\sum_{\{n_1,\dots,n_N\}\geq0}(a\tau)^{\sum_{A=1}^NAn_A}\chi_{(n_1,n_2\dots,n_{N-1})}(\mathbf{u})\chi_{(n_{N-1},n_{N-2},\dots,n_1)}(\mathbf{v})\\
=&\sum_{l\geq0}\sum_{\lambda}(a\tau)^{Nl+\sum_{A=1}^{N-1}A(\lambda_A-\lambda_{A+1})}s_{\lambda}(\mathbf{u})s_{\overline{\lambda}}(\mathbf{v})
\end{aligned}
\end{equation}
holds, with $\prod_{A=1}^Nu_A=\prod_{A=1}^Nv_A=1$ and where the $SU(N)$ characters are defined in \eqref{eqn:SUNChar}. In the second line we rewrote the expression in terms of Schur polynomials, the relevant definitions and identities can be found in Appendix \ref{sec:appB}. Therefore, we can write \eqref{eqn:HLFT2} as
\begin{align}
&\HS_{s_{1}^{l,+},s_{2}^{r,+}}\begin{aligned}=&\sum_{l^{(j)},m^{(j)}\geq0}\sum_{\lambda^{(j)},\mu^{(j)}}\prod_{i=1}^k\frac{s_{\lambda^{(i)}}(\mathbf{u}_i)s_{\overline{\mu}^{(i)}}(\mathbf{u}_i)s_{\overline{\lambda}^{(i)}}(\mathbf{v}_i)s_{\mu^{(i)}}(\mathbf{v}_{i-1})}{\left(\frac{\beta_i\tau}{\alpha}\right)^{\sum_{A=1}^{N-1}A(\lambda_{A+1}^{(i)}-\lambda_A^{(i)})-Nl^{(i)}}\left(\frac{\alpha\tau}{\gamma_i}\right)^{\sum_{A=1}^{N-1}A(\mu_{A+1}^{(i)}-\mu_A^{(i)})-Nm^{(i)}}}
\end{aligned}\\
&\begin{aligned}=&\sum_{\lambda^{(j)},\mu^{(j)},\nu^{(j)},\eta^{(j)}}\prod_{i=1}^k\frac{c_{\lambda^{(i)}\overline{\mu}^{(i)}}^{\eta^{(i)}}c^{\nu^{(i)}}_{\mu^{(i+1)}\overline{\lambda}^{(i)}}s_{\eta^{(i)}}(\mathbf{u}_i)s_{\nu^{(i)}}(\mathbf{v}_i)}{\prod_{A=1}^{N-1}\left(\left(\frac{\beta_i\tau}{\alpha}\right)^{A(\lambda_{A+1}^{(i)}-\lambda_A^{(i)})}\left(\frac{\alpha\tau}{\gamma_i}\right)^{A(\mu_{A+1}^{(i)}-\mu_A^{(i)})}\right)}\PE\left[\frac{\alpha^N}{\gamma_i^N}\tau^N+\frac{\beta_i^N}{\alpha^N}\tau^N\right]
\end{aligned}
\end{align}
Specialising to the case $\mathfrak{g}=A_{1}$, where we explicitly know the Littlewood-Richardson coefficients $c_{\lambda\mu}^{\nu}$, we have
\begin{equation}
\HS_{s_{1}^{l,+},s_{2}^{r,+}}=\prod_{i=1}^k\sum_{n_i,n'_i\geq0}\sum_{d_i=|n_i-n'_i|}^{n_i+n'_i}\sum_{d'_i=|n_i-n'_{i+1}|}^{n_i+n'_{i+1}}\frac{\chi_{d_i}(u_i)\chi_{d'_i}(v_i)}{\left(1-\frac{\beta_i^2}{\alpha^2}\tau^2\right)\left(1-\frac{\alpha^2}{\gamma_i^2}\tau^2\right)}\left(\frac{\beta_i\tau}{\alpha}\right)^{n_i}\left(\frac{\alpha\tau}{\gamma_i}\right)^{n'_i}\,.
\end{equation}
\subsection{Core Interacting Theories}\label{sec:coreint}
Let us first consider the Hall-Littlewood index/ Higgs-branch Hilbert series for the interacting SCFT associated to a sphere with two minimal punctures and two maximal punctures $s_1^{l,+},s_{3}^{r,+}$. It is given by
\begin{align}
&\HS_{s^{l,+}_1,s^{r,+}_{3}}=\HL_{s^{l,+}_1,s^{r,+}_{3}}=\HL\indices{_{\mathbf{u}\delta\alpha}^{\mathbf{v}}}=\prod_{i=1}^k \oint d\mu_i\,\HL_{s^{l,+}_1,s^{r,+}_{2}}\HL_{\Phi_{2}^+}\HL_{s^{l,+}_{2},s^{r,+}_{3}}\\
&=\prod_{i=1}^k\oint d\mu_i\prod_{A,B=1}^N\frac{\left(1-\tau^2\frac{\beta_{i}}{\gamma_i}\frac{z_{i-1,A}}{z_{i,B}}\right)(1-\tau^2)^{-\delta_{k,1}}}{\left(1-\frac{\tau\beta_{i}}{\delta}\frac{u_{i,A}}{z_{i,B}}\right)\left(1-\frac{\tau\delta}{\gamma_i}\frac{z_{i-1,B}}{u_{i,A}}\right)\left(1-\frac{\tau\beta_{i+1}}{\alpha}\frac{z_{i,B}}{v_{i,A}}\right)\left(1-\frac{\tau\alpha}{\gamma_i}\frac{v_{i-1,A}}{z_{i,B}}\right)}\label{eqn:4puncHL}\,.
\end{align}
An important observation that will allow us to write down the Highest Weight Generating (HWG) function \cite{Hanany:2014dia} for the Hilbert series for this theory is the fact that there is, at the level of the Lagrangian, a symmetry enhancement 
\begin{equation}\label{eqn:4punsymenhance}
SU(N)^{2k}\times U(1)_{\gamma}^{k-1}\times U(1)_{\beta}^{k-1}\times U(1)_{\delta}\times U(1)_{\alpha}\to SU(2N)^{k}\times U(1)_{a}^{k-1} \times U(1)_{m}\,,
\end{equation}
which we can make manifest in \eqref{eqn:4puncHL} by writing
\begin{equation}
\mathbf{x}_i=\left(\sqrt{\frac{\gamma_i\beta_i}{\delta\alpha}}\mathbf{u}_i,\sqrt{\frac{\delta\alpha}{\gamma_i\beta_i}}\mathbf{v}_{i-1}\right)\,,\quad a_i=\sqrt{\frac{\beta_i}{\gamma_i}}\,,\quad m=\sqrt{\frac{\alpha}{\delta}}\,,
\end{equation}
corresponding to writing $q_i=(Q_i^L,\widetilde{Q}_{i-1}^R)$, $\widetilde{q}_i=(\widetilde{Q}_i^L,Q^R_{i-1})$ in the decomposition $\mathbf{2N}\to(\mathbf{N},\mathbf{1})_1\oplus(\mathbf{1},\mathbf{N})_{-1}$ under $SU(2N)\to SU(N)\times SU(N)\times U(1)$ and $Q_i^L,\widetilde{Q}_{i}^L$ are the bifundamental chiral multiplets of the first free trinion and $Q_i^R,\widetilde{Q}_{i}^R$ the second. Then the Hall-Littlewood index / Hilbert series can be written as
\begin{equation}\label{eqn:HSfull4punc}
\HS_{s^{l,+}_1,s^{r,+}_{3}}=\prod_{i=1}^k\oint\prod_{A=1}^{N-1}\frac{dz_{i,A}}{2\pi\iu z_{i,A}}\frac{\prod_{1\leq A<B\leq N}\left(1-\frac{z_{i,B}}{z_{i,A}}\right)\prod_{A,B=1}^N\left(1-\tau^2a_i^2\frac{z_{i-1,A}}{z_{i,B}}\right)}{(1-\tau^2)^{\delta_{k,1}}\prod_{A=1}^N\prod_{B=1}^{2N}\left(1-\tau a_im\frac{x_{i,B}}{z_{i,A}}\right)\left(1-\tau\frac{a_i}{m}\frac{z_{i-1,A}}{x_{i,B}}\right)}\,.
\end{equation}
The symmetry enhancement \eqref{eqn:4punsymenhance} allows us to conjecture the following expression for the Highest Weight Generating (HWG) function for the Hilbert series as
\begin{equation}
\HWG_{s^{l,+}_1,s^{r,+}_{3}}=\PE\left[\delta_{1,k}\tau^2+\sum_{i=1}^k\sum_{A=1}^{N-1}\mu^{(i)}_{A}\mu^{(i+1)}_{2N-A}a_i^Aa_{i+1}^A\tau^{2A}+\sum_{i=1}^k\mu^{(i)}_{N}a_i^{N}\left(m^{N}+\frac{1}{m^N}\right)\tau^N\right]\, \ ,
\end{equation}
where $\{\mu_{A}^{(i)}\}$ denotes a set of highest weights for the $i$-th flavour node.
The corresponding Hilbert series is given by
\begin{equation}\label{eqn:HSconject}
\begin{aligned}
& \HS_{s^{l,+}_1,s^{r,+}_{3}}=\sum_{\{n^{(i)}_{A},p^{(i)},l^{(i)}\geq0\}}\prod_{i=1}^k\left\{\left[n^{(i)}_{1},n^{(i)}_{2},\dots,n^{(i)}_{N-1},p^{(i)}+l^{(i)},n^{(i+1)}_{N-1},n^{(i+1)}_{N-2},\dots,n^{(i+1)}_{1}\right]_{\mathbf{x}_i}\right.\\
&\left.\times m^{Np^{(i)}-Nl^{(i)}}a_i^{Np^{(i)}+Nl^{(i)}+\sum_{A=1}^{N-1}An_{A}^{(i)}}a_{i+1}^{\sum_{A=1}^{N-1}An_{A}^{(i+1)}}\tau^{Np^{(i)}+Nl^{(i)}+2\sum_{A=1}^{N-1}An^{(i)}_{A}}\vphantom{\left[n^{(i-1)}_{N},\right]_x}\right\}\left(1-\tau^2\right)^{-\delta_{1,k}}\,.
\end{aligned}
\end{equation}
We checked in an expansion around $\tau=0$ that \eqref{eqn:HSconject} agrees with \eqref{eqn:HSfull4punc} for $(N,k)=\{(2,2),(2,3),(3,2)\}$. For theories associated to $\ell$-punctured spheres with two maximal punctures and $\ell-2>2$ minimal punctures there is no symmetry enhancement \eqref{eqn:4punsymenhance} and the number of monomials in the $\PLog$ of the HWG is not finite.

We can however perform the expansion of the integral \eqref{eqn:HSgenericgzero} around $\tau=0$. For example, for $k=N=2$, $\ell-2=3$ we have
\begin{equation}
\begin{aligned}
\HS=&1+ \left[\chi^L_{2} \chi^L_{1}\left(\frac{\beta }{\gamma }+\frac{\gamma }{\beta }\right) +\chi^R_{2}\chi^R_{1} \left(\beta  \gamma +\frac{1}{\beta  \gamma }\right)+\sum_{n=1}^{3}\sum_{i=1}^2\left(\frac{\beta^2_i}{\alpha _n^2}+\gamma_i^2\alpha _n^2\right)\right]\tau^2\\
&+\left[\left(\frac{\alpha
   _3 \alpha _2}{\alpha _1}+\frac{\alpha _1 \alpha _2}{\alpha _3}+\frac{\alpha _1 \alpha _3}{\alpha _2}+\frac{1}{\alpha _1 \alpha _3 \alpha _2}\right) \sum_{i=1}^2\beta_i\chi_i^L\chi_i^R\right.\\
   &\,\,+\left.\left(\alpha _2 \alpha _3 \alpha _1+\frac{\alpha _1}{\alpha _2 \alpha _3}+\frac{\alpha _3}{\alpha _2 \alpha _1}+\frac{\alpha _2}{\alpha _3 \alpha _1}\right) \sum_{i=1}^2\gamma_{i+1}\chi^L_i\chi_{i+1}^R\right]\tau ^3 +\mathcal{O}(\tau^4)
   \end{aligned}
\end{equation}
where $\beta_1=\frac{1}{\beta_2}=\beta$, $\gamma_1=\frac{1}{\gamma_2}=\gamma$ and $\chi^L_{i}=\chi_{1}(z_{(i\bmod k,1)})$, $\chi^R_{i}=\chi_{1}(z_{(i\bmod k,4)})$ denotes the characters of the fundamental representation of the corresponding $SU(2)$'s.

\section{Genus One Theories}
\label{sec:hsg1}
In this Section we consider genus one theories of Class $\mathcal{S}_k$.
 As is the case for $\mathcal{N}=2$ Class $\mathcal{S}$ theories,  for $g=1$ theories the Hall-Littlewood limit of the index does not coincide with the corresponding Higgs branch Hilbert series (which, as we discussed in Section \ref{sec:hcbc}, counts only ${\mathcal{\overline{B}}}$ and ${\mathcal{\overline{D}}}^{(\frac{1}{2})}_{(0,0)}$ multiplets), but it differs by
 $\mathcal{\overline{C}}, \hat{\mathcal{C}}$ and $\mathcal{D}^{(j+1)}$ type
 multiplets.

 The theories that we study are the $\mathcal{N}=1$ $\mathfrak{u}(N)^{\oplus\ell k}$ toroidal quiver gauge theories realised as the string length $l_s\to0$ limit of the worldvolume theory on a stack of $N$ D3-branes probing a transverse $\mathbb{C}^3/(\mathbb{Z}_{\ell}\times\mathbb{Z}_k)$ singularity where the quotient acts by
\begin{equation}\label{eqn:quotientactZkZl}
(z_1,z_2,z_3)\mapsto(\omega_kz_1,\omega_{\ell}z_2,\omega_{\ell}^{-1}\omega^{-1}_{k}z_3)\,,\quad \omega_k^k=\omega_{\ell}^{\ell}=1\,.
\end{equation}
These are orbifolds of $\mathcal{N}=4$ SYM. In the case where the $\mathfrak{u}(1)^{\oplus\ell k}\subset\mathfrak{u}(N)^{\oplus\ell k}$ is not gauged, these are precisely the theories in class $\mathcal{S}_k$ associated to Riemann surfaces of genus one with $\ell$ punctures. 
For $k=\ell \to \infty$ we will provide a check that the  six dimensional $(1,1)$ little string theory can be deconstructed from a toroidal quiver in class $\mathcal{S}_k$.

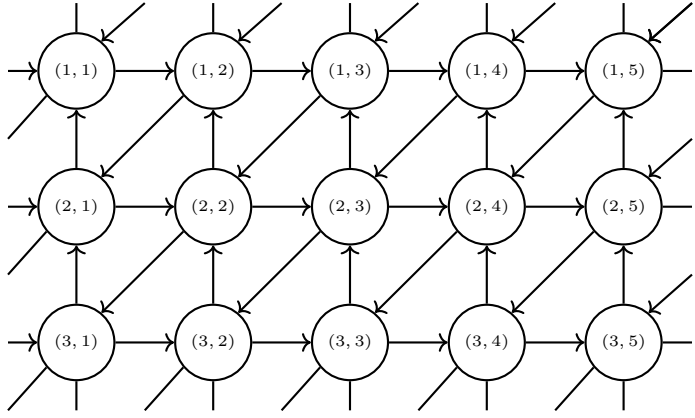
\begin{figure}
\centering
  \begin{tikzpicture}[square/.style={regular polygon,regular polygon sides=4},thick,inner sep=0.1em,scale=0.9]
    \node (G11) at (0,0)[circle,draw,minimum size=1cm]{\tiny $(1,1)$};
    \node (G21) at (2,0) [circle,draw,minimum size=1cm]{\tiny $(1,2)$};
    \node (G31) at (4,0) [circle,draw,minimum size=1cm]{\tiny $(1,3)$};
    \node (G41) at (6,0) [circle,draw,minimum size=1cm]{\tiny $(1,4)$};
    \node (G51) at (8,0)[circle,draw,minimum size=1cm]{\tiny $(1,5)$};
    \node (G12) at (0,-2)[circle,draw,minimum size=1cm]{\tiny $(2,1)$};
    \node (G22) at (2,-2) [circle,draw,minimum size=1cm]{\tiny $(2,2)$};
    \node (G32) at (4,-2) [circle,draw,minimum size=1cm]{\tiny $(2,3)$};
    \node (G42) at (6,-2)[circle,draw,minimum size=1cm]{\tiny $(2,4)$};
    \node (G52) at (8,-2)[circle,draw,minimum size=1cm]{\tiny $(2,5)$};
	\node (G13) at (0,-4)[circle,draw,minimum size=1cm]{\tiny $(3,1)$};
    \node (G23) at (2,-4) [circle,draw,minimum size=1cm]{\tiny $(3,2)$};
    \node (G33) at (4,-4) [circle,draw,minimum size=1cm]{\tiny $(3,3)$};
    \node (G43) at (6,-4)[circle,draw,minimum size=1cm]{\tiny $(3,4)$};
    \node (G53) at (8,-4)[circle,draw,minimum size=1cm]{\tiny $(3,5)$};
    
    \draw [<-] (G11.270) to (G12.90);
    \draw [<-] (G21.270) to (G22.90);
    \draw [<-] (G31.270) to (G32.90);
    \draw [<-] (G41.270) to (G42.90);
    \draw [<-] (G51.270) to (G52.90);
    \draw [<-] (G12.270) to (G13.90);
    \draw [<-] (G22.270) to (G23.90);
    \draw [<-] (G32.270) to (G33.90);
    \draw [<-] (G42.270) to (G43.90);
    \draw [<-] (G52.270) to (G53.90);
    \draw [-] (0,1) to (G11.90);
    \draw [-] (2,1) to (G21.90);
    \draw [-] (4,1) to (G31.90);
    \draw [-] (6,1) to (G41.90);
    \draw [-] (8,1) to (G51.90);
    \draw [-] (G13.270) to (0,-5);
    \draw [-] (G23.270) to (2,-5);
    \draw [-] (G33.270) to (4,-5);
    \draw [-] (G43.270) to (6,-5);
    \draw [-] (G53.270) to (8,-5);
    
    \draw [->] (-1,0) to (G11.180);
    \draw [->](G11.0) to (G21.180);
    \draw [->] (G21.0) to (G31.180);
    \draw [->] (G31.0) to (G41.180);
    \draw [->] (G41.0) to (G51.180);
    \draw [-] (G51.0) to (9,0);
    \draw [->] (-1,-2) to (G12.180);
    \draw [->] (G12.0) to (G22.180);
    \draw [->] (G22.0) to (G32.180);
    \draw [->] (G32.0) to (G42.180);
    \draw [->] (G42.0) to (G52.180);
    \draw [-] (G52.0) to (9,-2);
    \draw [->] (-1,-4) to (G13.180);
    \draw [->] (G13.0) to (G23.180);
    \draw [->] (G23.0) to (G33.180);
    \draw [->] (G33.0) to (G43.180);
    \draw [->] (G43.0) to (G53.180);
    \draw [-] (G53.0) to (9,-4);

    \draw [<-] (G11.50) to (1,1);
    \draw [<-] (G21.50) to (3,1);
    \draw [<-] (G31.50) to (5,1);
    \draw [<-] (G41.50) to (7,1);
    \draw [<-] (G51.50) to (9,1);

	\draw [<-] (G13.50) to (G22.220);
    \draw [<-] (G23.50) to (G32.220);
    \draw [<-] (G33.50) to (G42.220);
    \draw [<-] (G43.50) to (G52.220);
    \draw [<-] (G32.50) to (G41.220);
    \draw [<-] (G22.50) to (G31.220);
    \draw [<-] (G12.50) to (G21.220);
    \draw [<-] (G42.50) to (G51.220);
    
    \draw [-] (G13.220) to (-1,-5);
    \draw [-] (G12.220) to (-1,-3);
    \draw [-] (G11.220) to (-1,-1);
    \draw [-] (G23.220) to (1,-5);
    \draw [-] (G33.220) to (3,-5);
    \draw [-] (G43.220) to (5,-5);
    \draw [-] (G53.220) to (7,-5);
    \draw [<-] (G53.50) to (9,-3);
    \draw [<-] (G52.50) to (9,-1);
    \draw [<-] (G51.50) to (9,1);
    
  \end{tikzpicture}
  \caption{\it Section of the quiver diagram of the $\mathbb{Z}_k\times\mathbb{Z}_{\ell}$ orbifold theory of $\mathcal{N}=4$ SYM. Circular nodes denote $U(N)$ vector multiplets and directed arrows denote chiral multiplets. Horizontal lines between node $(i,n)$ and $(i,n+1)$ denote $Q_{(i,n)}$ fields. Vertical lines between node $(i,n)$ and $(i-1,n)$ denote $\Phi_{(i,n)}$ fields. Diagonal lines between $(i-1,n+1)$ and $(i,n)$ denote $\widetilde{Q}_{(i,n)}$ fields. The quiver should be periodically identified in both directions, such that it has the topology of a tessellation of the  torus.}
  \label{fig:Skquivergenus1}
\end{figure}

The action \eqref{eqn:quotientactZkZl} is an element of the $SO(6)_R$ R-symmetry group of $\mathcal{N}=4$ SYM. If we denote the $\mathcal{N}=4$ supercharges by $\mathcal{Q}_{\alpha}^{q_1q_2q_3}$ with $8q_1q_2q_3=-1$ and $\overline{\mathcal{Q}}_{\dot\alpha}^{q_1q_2q_3}$ with $8q_1q_2q_3=+1$ where the $q_i=\pm\frac{1}{2}$. The orbifold acts by
\begin{equation}
\mathcal{Q}_{\alpha}^{q_1q_2q_3}\mapsto \omega_k^{q_1-q_3}\omega_{\ell}^{q_2-q_3}\mathcal{Q}_{\alpha}^{q_1q_2q_3}\,,\quad
\overline{\mathcal{Q}}_{\dot\alpha}^{q_1q_2q_3}\mapsto \omega_k^{q_1-q_3}\omega_{\ell}^{q_2-q_3}\overline{\mathcal{Q}}_{\dot\alpha}^{q_1q_2q_3}\,.
\end{equation}
The surviving supercharges are those with $q_1=q_2=q_3$
\begin{equation}
\mathcal{Q}_{\alpha}^{---}=\mathcal{Q}_{\alpha}\,,\quad\overline{\mathcal{Q}}^{+++}_{\dot\alpha}=\overline{\mathcal{Q}}_{\dot\alpha}\,.
\end{equation}
The Cartans $q_1,q_2,q_3$ of $\mathfrak{so}(6)_R$ are related to the more natural $\mathcal{N}=1$ symmetries by
\begin{equation}\label{eqn:N4N1ides}
r=\frac{2}{3}(q_1+q_2+q_3)\,,\quad q_t=-q_1+\frac{1}{2}(q_2+q_3)\,,\quad q_b=-q_2+q_3\,,
\end{equation} 
there is also the overall $U(1)$ generated by $b$.

It will later be useful to have the Hilbert series for this quotient space. This is given by the \textit{Molien series} 
\cite{Feng:2007ur,Benvenuti:2006qr}
\begin{equation}
\begin{aligned}\label{eqn:moilen}
&\HS(\tau_1,\tau_2,\tau_3;\mathbb{C}^3/(\mathbb{Z}_{\ell}\times\mathbb{Z}_k))=M(\tau_1,\tau_2,\tau_3;\mathbb{C}^3/(\mathbb{Z}_{\ell}\times\mathbb{Z}_k))\\&=\frac{1}{|\mathbb{Z}_{\ell}||\mathbb{Z}_k|}\sum_{g\in\mathbb{Z}_{\ell}\atop{h\in\mathbb{Z}_k}}\frac{1}{(1-h\tau_1)(1-g\tau_2)(1-g^{-1}h^{-1}\tau_3)}=\frac{\sum_{r=0}^{LCM(\ell,k)-1}(\tau_1\tau_2\tau_3)^r\tau_1^{-\lfloor\frac{r}{k}\rfloor k}\tau_2^{-\lfloor\frac{r}{\ell}\rfloor\ell}}{(1-\tau_1^k)(1-\tau_2^{\ell})(1-\tau_3^{LCM(\ell,k)})}
\,,\end{aligned}
\end{equation}
where $\tau_1,\tau_2,\tau_3$ are related to the $SO(6)_R\supset U(1)^{3}\lefttorightarrow\mathbb{C}^3$ toric action generated by $q_1,q_2,q_3$. Moreover $LCM(\ell,k)$ denotes the Lowest Common Multiple of $\ell$ and $k$. For $\ell=k$ these varieties are complete intersections since 
\begin{equation}
M(\tau_1,\tau_2,\tau_3;\mathbb{C}^3/(\mathbb{Z}_{k}\times\mathbb{Z}_k))=\PE\left[\sum_{n=1}^3\tau_n^k+\tau_1\tau_2\tau_3-\tau_1^k\tau_2^k\tau_3^k\right]\,.
\end{equation}
This space can be realised as $\mathbb{C}^3/(\mathbb{Z}_{k}\times\mathbb{Z}_k)\hookrightarrow\mathbb{C}^4$ defined by the equation
\begin{equation}
w_1w_2w_3-w_4^k=0\,,
\end{equation}
with $w_i=z_i^k$, $w_4=z_1z_2z_3$.
As a warm up we will now review the computation for $k=1$.

\subsection{Class $\mathcal{S}$ $k=1$}
\begin{figure}
\begin{center}
\begin{tikzpicture}[scale=0.9]

\def\circledarrow#1#2#3{ 
\draw[#1,->] (#2) +(80:#3) arc(80:-260:#3);
}
 

\draw (0,3) circle [radius=0.5]  node (A) {$N$};
\draw (1.5*1.414,1.5*1.414)  circle [radius=0.5] node {$N$};
\draw (3,0)  circle [radius=0.5] node {$N$};
\draw (1.5*1.414,-1.5*1.414)  circle [radius=0.5] node {$N$};
\draw (0,-3)  circle [radius=0.5] node {$N$};
\draw (-1.5*1.414,-1.5*1.414)  circle [radius=0.5] node {$N$};
\draw (-3,0)  circle [radius=0.5] node {$N$};
\draw (-1.5*1.414,1.5*1.414)  circle [radius=0.5] node {$N$};


\draw [->,thick,domain=54:81,scale=3] plot ({1.05*cos(\x)}, {1.05*sin(\x)}) node[above right] {$\widetilde{Q}_1$};
\draw [->,thick,domain=81:54,scale=3] plot ({0.95*cos(\x)}, {0.95*sin(\x)}) node[below left] {$Q_1$};;

\draw [->,thick,domain=9:36,scale=3] plot ({1.05*cos(\x)}, {1.05*sin(\x)});
\draw [->,thick,domain=35:9,scale=3] plot ({0.95*cos(\x)}, {0.95*sin(\x)});

\draw [<-,thick,domain=-9:-36,scale=3] plot ({1.05*cos(\x)}, {1.05*sin(\x)});
\draw [<-,thick,domain=-36:-9,scale=3] plot ({0.95*cos(\x)}, {0.95*sin(\x)});

\draw [<-,thick,domain=-54:-80,scale=3] plot ({1.05*cos(\x)}, {1.05*sin(\x)});
\draw [<-,thick,domain=-80:-54,scale=3] plot ({0.95*cos(\x)}, {0.95*sin(\x)});

\draw [->,thick,domain=54:80,scale=-3] plot ({1.05*cos(\x)}, {1.05*sin(\x)});
\draw [->,thick,domain=80:54,scale=-3] plot ({0.95*cos(\x)}, {0.95*sin(\x)});

\draw [->,thick,domain=9:35,scale=-3] plot ({1.05*cos(\x)}, {1.05*sin(\x)});
\draw [->,thick,domain=35:9,scale=-3] plot ({0.95*cos(\x)}, {0.95*sin(\x)});

\draw [->,dashed,thick,domain=-35:-9,scale=-3] plot ({1.05*cos(\x)}, {1.05*sin(\x)});
\draw [->,dashed,thick,domain=-9:-35,scale=-3] plot ({0.95*cos(\x)}, {0.95*sin(\x)});

\draw [->,dashed,thick,domain=-81:-54,scale=-3] plot ({1.05*cos(\x)}, {1.05*sin(\x)});
\draw [->,dashed,thick,domain=-54:-81,scale=-3] plot ({0.95*cos(\x)}, {0.95*sin(\x)});

\node (A1) at (0,3.35) {};
\draw[->,thick] (A1) to [out=60,in=120,looseness=15] node[above] {$\Phi_1$} (A1);

\node (A2) at (3.35,0) {};
\draw[->,thick] (A2) to [out=-30,in=30,looseness=15] node[right] {} (A2);

\node (A3) at (0,-3.35) {};
\draw[->,thick] (A3) to [out=-60,in=-120,looseness=15] node[below] {} (A3);

\node (A4) at (-3.35,0) {};
\draw[->,thick] (A4) to [out=-150,in=-210,looseness=15] node[below] {} (A4);

\node (B1) at (0.28+1.5*1.414,0.28+1.5*1.414) {};
\draw[->,thick] (B1) to [out=15,in=75,looseness=15] node[above] {$\Phi_2$} (B1);

\node (B2) at (0.28+1.5*1.414,-0.28-1.5*1.414) {};
\draw[->,thick] (B2) to [out=-30,in=-90,looseness=15] node[right] {} (B2);

\node (B3) at (-0.28-1.5*1.414,-0.28-1.5*1.414) {};
\draw[->,thick] (B3) to [out=-105,in=-165,looseness=15] node[below right] {} (B3);

\node (B4) at (-0.28-1.5*1.414,0.28+1.5*1.414) {};
\draw[->,thick] (B4) to [out=165,in=105,looseness=15] node[above] {$\Phi_i$} (B4);

\node (text) {$\ell$ nodes};
\circledarrow{ultra thick, gray}{text}{1.8cm};

\end{tikzpicture}
\end{center}

\caption{\textit{The circular quiver with gauge group $U(N)^{\ell}$}.}\label{fig:quivercircular}
\end{figure}
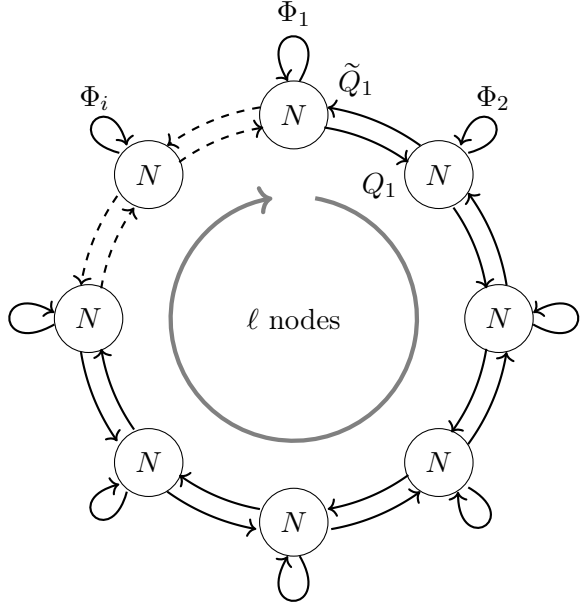

The quiver diagram for this theory is given in Figure \ref{fig:quivercircular}. In the following subsection we compute the Hilbert series of the mesonic moduli space $\mathcal{M}^{mes}$ \cite{Forcella:2008bb}.
\subsubsection{Hilbert Series}
. Let us begin with the case of $N=1$. The master space for $\mathcal{M}$ is then associated to $\mathcal{R}/\mathcal{I}$ with $\mathcal{R}=[Q_1,\dots,Q_{\ell},\widetilde{Q}_1,\dots,\widetilde{Q}_{\ell},\Phi_1,\dots,\Phi_{\ell}]$ and $\mathcal{I}=\langle F_{Q_1},\dots,F_{Q_{\ell}},F_{\widetilde{Q}_1},\dots,F_{\widetilde{Q}_{\ell}},F_{\Phi_1},\dots,F_{\Phi_{\ell}}\rangle$ with
\begin{equation}
F_{Q_{n}}=\widetilde{Q}_n(\Phi_{n+1}-\Phi_n)\,,\quad F_{\widetilde{Q}_n}=Q_n(\Phi_{n+1}-\Phi_n)\,,\quad F_{\Phi_n}=Q_{n-1}\widetilde{Q}_{n-1}-Q_n\widetilde{Q}_{n}\,,
\end{equation}
where $n\sim n+\ell$. We perform a primary decomposition of the above ideal and we select the prime ideal corresponding to a mesonic branch, the corresponding Hilbert series is 
\begin{equation}
\mathcal{F}^{mes}=\PE\left[\sum_{n=1}^{\ell}\left(\tau_1+\frac{z_n}{z_{n+1}}\tau_2+\frac{z_{n+1}}{z_{n}}\tau_3\right)-\sum_{n=1}^{\ell-1}\left(\tau_2\tau_3+\tau_1\right)\right]\,.
\end{equation}
The $\tau_i$ are the same as those in \eqref{eqn:moilen}. Explicitly, in terms of the parameters appearing in \eqref{eqn:SCI}
\begin{equation}
\tau_1=\frac{pq}{t}=T\,,\quad \tau_2=\frac{\sqrt{t}}{b}=\frac{\tau}{b}\,,\quad \tau_3=b\sqrt{t}=b\tau
\end{equation} 
where $b^{\ell}=\prod_{n=1}^{\ell}\alpha_n$ is the product of all of the fugacities for minimal punctures. So, the Hilbert series for $\mathcal{M}^{mes}$ is given by
\begin{equation}
\HS(\tau_1,\tau_2,\tau_3;\mathcal{M}^{mes})=\prod_{n=1}^{\ell}\oint_{\mid z_n \mid =1}\frac{dz_n}{2\pi\iu z_n}\mathcal{F}^{mes}=\frac{1-\tau_2^{\ell}\tau_3^{\ell}}{(1-\tau_1)(1-\tau_2\tau_3)(1-\tau_2^{\ell})(1-\tau_3^{\ell})}
\end{equation}
We observe that $\HS(\tau_1,\tau_2,\tau_3;\mathcal{M}^{mes})=M(\tau_1,\tau_2,\tau_3;\mathbb{C}\times\mathbb{C}^2/\mathbb{Z}_{\ell})$.
The Hilbert series for the Higgs branch can be easily obtained by considering the $\tau_1\to0$ limit
\begin{align}
&\HS(\tau_2,\tau_3;HB)=\lim_{\tau_1\to0}\HS(\tau_1,\tau_2,\tau_3;\mathcal{M}^{mes})=\PE\left[\tau_2\tau_3+\tau^{\ell}_2+\tau_3^{\ell}-\tau_2^{\ell}\tau_3^{\ell}\right]\label{eq:hsk1}
\end{align}

As we can see taking the PLog of (\ref{eq:hsk1}) the Higgs branch is generated by one generator $M=Q_1\widetilde{Q}_1=\dots=Q_{\ell}\widetilde{Q}_{\ell}$ of dimension $2$ and two generators $B=\prod_{n=1}^{\ell}Q_n$, $\widetilde{B}=\prod_{n=1}^{\ell}\widetilde{Q}_n$ of dimension $\ell$. They satisfy the following relation at order $2\ell$ 
\begin{equation}
\label{eq:rel}
M^{\ell}=B\widetilde{B} \, ,
\end{equation}
this corresponds to the variety $\mathbb{C}^2/\mathbb{Z}_{\ell}$. 
The Coulomb branch is simply a copy of $\mathbb{C}^{\ell}$ parametrised by $\{\Phi_1,\dots,\Phi_{\ell}\}$
\begin{align}
&\HS(\tau_1;CB)=\lim_{\tau_2,\tau_3\to0}\HS(\tau_1,\tau_2,\tau_3;\mathcal{M})=\PE\left[\ell\tau_1\right]
\,.
\end{align}
The corresponding expressions for the mesonic branch for general $N$,
can be obtained as the $N^{\text{th}}$ symmetric product of the $N=1$ case \cite{Feng:2007ur,Benvenuti:2006qr,Seiberg:1997ax,Gang:2011xp,Forcella:2008bb} and 
\begin{equation}
\mathcal{M}^{\textrm{mes}}=\Sym^N\left(\mathbb{C}\times\mathbb{C}^2/\mathbb{Z}_{\ell}\right)
\end{equation}
and the Hilbert series is given by
\begin{equation}
\HS\left(\tau_1,\tau_2,\tau_3;\Sym^N\left(\mathbb{C}\times\mathbb{C}^2/\mathbb{Z}_{\ell}\right)\right)=\frac{1}{N!}\left.\frac{\partial^N}{\partial\nu^N}\PE\left[\nu\HS\left(\tau_1,\tau_2,\tau_3;\mathbb{C}\times\mathbb{C}^2/\mathbb{Z}_{\ell}\right)\right]\right|_{\nu=0}\,.
\end{equation}

\subsubsection{Hall-Littlewood Index}
Let's now move to the computation of the Hall-Littlewood index of the theory in Figure \ref{fig:quivercircular}. Let's begin the case with $N=1$ and generic $\ell$ the computation can be explicitly done and we get
\begin{equation}
\label{eq:HL}
\textrm{HL}(\tau_2,\tau_3) =\prod_{n=1}^{\ell}\oint_{\mid z_n \mid =1 }\frac{dz_n}{2\pi\iu z_n}\PE\left[\sum_{n=1}^{\ell}\left(\alpha_n^{-1}\frac{z_n}{z_{n+1}}\tau+\alpha_n\frac{z_{n+1}}{z_{n}}\tau-\tau^2\right)\right]=\textrm{PE}[\tau_2^{\ell}+\tau_3^{\ell}-\tau_2^{\ell}\tau_3^{\ell}] \, \ ,
\end{equation}
therefore the ratio between the Higgs branch Hilbert series \eqref{eq:hsk1} and the above index reads
\begin{equation}
\frac{\textrm{HL}(\tau_2,\tau_3) }{\HS(\tau_2,\tau_3;HB)}= \textrm{PE}[-\tau_2\tau_3]=\textrm{PE}[-\tau^2] = \textrm{PE}[\HL_{\mathcal{D}_{0,(0,0)}}] \, ,
\end{equation}
here $\HL_{\mathcal{D}_{0,(0,0)}}$ denotes the Hall-Littlewood index of the free $\mathcal{N}=2$ vector multiplet \cite{Gadde:2011uv,Dolan:2002zh}.
We now want to compute the expression of the Hall-Littlewood index for a generic value of $N$. Since $\mathcal{D}_{0,(0,0)}$ is a free field multiplet it naturally decouples from the theory and we would like to conjecture that the Hall-Littlewood index for general $N$ can still be obtained as the coefficient $\textrm{HL}^{\ell}_{N}$ of the following expansion
\begin{equation}
\textrm{PE}[\nu \textrm{HL}_{N=1}^{\ell}(\tau_2,\tau_3)] = \sum_{N=1}^{\infty}\nu^N \textrm{HL}^{\ell}_{N}(\tau_2,\tau_3)\,,\quad  \textrm{HL}_{N=1}^{\ell}(\tau_2,\tau_3)=\textrm{PE}[\tau_2^{\ell}+\tau_3^{\ell}-\tau_2^{\ell}\tau_3^{\ell}]\,.
\end{equation}  
We verified this conjecture for various low values $N$ and $\ell$, that is to say for $(\ell,N)=\{(1,2),(2,2),(3,2),(1,3),(2,3),(1,4)\}$. 

\subsection{Class $\mathcal{S}_k$ $k\geq2$}
\begin{figure}
\centering
\begin{tikzpicture}[thick,inner sep=0.1em,scale=0.9]
    \node (L1) at (0,0) [circle,draw,minimum size=1.3cm]{$N$};
    \node (G1) at (4,0) [circle,draw,minimum size=1.3cm]{$N$};
    \node (G2) at (4,-4) [circle,draw,minimum size=1.3cm]{$N$};
    \node (L2) at (0,-4) [circle,draw,minimum size=1.3cm]{$N$};

    \draw [->] (L1.10) -- (G1.170) node[midway,above] {$ \scriptstyle{Q_2^L}$};
    \draw [->] (L2.10) -- (G2.170) node[midway,above] {$\scriptstyle{Q_1^L}$};
    \draw [<-] (L1.350) -- (G1.190) node[midway,below] {$ \scriptstyle{Q_2^R}$};
    \draw [<-] (L2.350) -- (G2.190) node[midway,below] {$\scriptstyle{Q_1^R}$};
    
    \draw [->] (G1.235) -- (L2.35) node[near start,below,sloped] {$\scriptstyle{\widetilde{Q}_1^L}$};
        \draw [<-] (G1.215) -- (L2.55) node[near end,above,sloped] {$\scriptstyle{\widetilde{Q}_2^R}$};
    \draw [->] (G2.125) -- (L1.325) node[near start,above,sloped] {$\scriptstyle{\widetilde{Q}_2^L}$};
        \draw [<-] (G2.145) -- (L1.305) node[near end,below,sloped] {$\scriptstyle{\widetilde{Q}_1^R}$};
    
     \draw [->] (G1.280) -- (G2.80) node[midway,right] {$\scriptstyle{\Phi_2^R}$};
     \draw [<-] (G1.260) -- (G2.100) node[midway,left] {$\scriptstyle{\Phi_1^R}$};
     \draw [->] (L1.280) -- (L2.80) node[midway,right] {$\scriptstyle{\Phi_2^L}$};
     \draw [<-] (L1.260) -- (L2.100) node[midway,left] {$\scriptstyle{\Phi_1^L}$};
     \node at (9,-2) [text width=5cm] {$Q_1^{L}:\frac{\beta_1}{\alpha_1}\tau$\quad $Q_1^R:\frac{\beta_2}{\alpha_2}\tau$\\
   $Q_2^{L}:\frac{\beta_2}{\alpha_1}\tau$\quad $Q_{2}^R:\frac{\beta_1}{\alpha_2}\tau$\\$\widetilde{Q}_1^L:\frac{\alpha_1}{\gamma_1}\tau$\quad $\widetilde{Q}_1^R:\frac{\alpha_2}{\gamma_1}\tau$ \\$\widetilde{Q}_2^L:\frac{\alpha_1}{\gamma_2}\tau$\quad $\widetilde{Q}_2^R:\frac{\alpha_2}{\gamma_2}\tau$\\$\Phi_i^L:\frac{\gamma_{i-1}}{\beta_{i}}\rho\sigma$\quad $\Phi_i^R:\frac{\gamma_{i-1}}{\beta_{i-1}}\rho\sigma$};
 \end{tikzpicture}
  \caption{\it Quiver diagram of the $k=2$ theory associated to a torus with $\ell=2$ minimal punctures.}
  \label{fig:c3z2z2}
\end{figure}
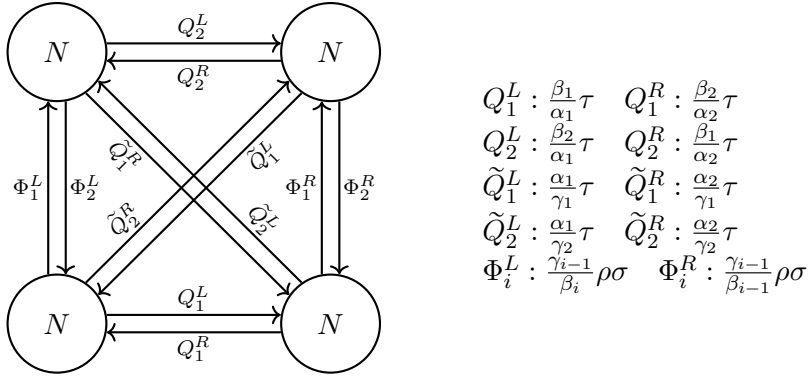

Let us move to the case of general values of $k,\ell$, while again focusing on $N=1$. 
Let's firstly consider $k=\ell=2$. The quiver diagram is given in Figure \ref{fig:c3z2z2}.

\subsubsection{Hilbert Series}
Using \textit{Macaulay2} \cite{M2}  we can compute the Hilbert series for $\mathcal{R}/\mathcal{I}$. The full moduli space $\mathcal{M}$ is then given as a projection onto gauge invariants
\begin{align}
&\HS(\tau_1,\tau_2,\tau_3;\mathcal{M})=\oint\prod_{A=1}^4\left(\frac{dz_A}{2\pi\iu z_A}\right)\mathcal{F}(\tau_1,\tau_2,\tau_3,\alpha,\delta,\gamma,\beta;z_A)\,,\\
&\begin{aligned}\mathcal{F}=&\frac{z_1z_2z_3z_4P(\tau_1,\tau_2,\tau_3,\alpha,\delta,\gamma,\beta;z_A)}{\left(\beta  \tau _1 z_1-\gamma  z_2\right) \left(\beta  z_1-\gamma  \tau _1 z_2\right) \left(\alpha  \beta  z_1-\tau _3 z_3\right) \left(\beta  \tau _3
   z_1-\delta  z_3\right) \left(\alpha z_2-\beta  \tau _3 z_4\right) \left(\tau _3 z_2-\beta  \delta  z_4\right)}\\
   &\frac{1}{ \left(\alpha  \tau _2 z_1-\gamma  z_4\right) \left(z_1-\gamma  \delta  \tau _2 z_4\right) \left(\alpha  \gamma  \tau _2
   z_2-z_3\right) \left(\gamma  z_2-\delta  \tau _2 z_3\right) \left(\tau _1 z_3-\beta  \gamma z_4\right) \left(z_3-\beta  \gamma  \tau _1 z_4\right)}
 \end{aligned}
\end{align}
with $P$ polynomial in the $z_A$ and $\tau_{1,2,3}$ and we have set $\gamma=\gamma_1$, $\beta=\beta_1$, $\alpha=\alpha_1$ and $\delta=\alpha_2$. After performing the integrals, we have
\begin{equation}
\begin{aligned}
\HS(\tau_1,\tau_2,\tau_3;\mathcal{M})=&\PE\left[2\tau^2_1+2\tau^2_2+2\tau^2_3\right]\left(\tau _2^2 \tau _3^4 \tau _1^4+\tau _2^2 \tau _1^4+\tau _2^4 \tau _3^2 \tau _1^4-4 \tau _2^2 \tau _3^2 \tau _1^4+\tau _3^2 \tau _1^4\right.\\
&-\tau _2^3 \tau _3^3 \tau _1^3+\tau _2 \tau _3^3 \tau _1^3+\tau
   _2^3 \tau _3 \tau _1^3-\tau _2 \tau _3 \tau _1^3+\tau _2^4 \tau _1^2+\tau _2^4 \tau _3^4 \tau _1^2-4 \tau _2^2 \tau _3^4 \tau _1^2\\
   &+\tau _3^4 \tau _1^2-3 \tau _2^2 \tau _1^2-4 \tau _2^4 \tau _3^2
   \tau _1^2+11 \tau _2^2 \tau _3^2 \tau _1^2-3 \tau _3^2 \tau _1^2+\tau _2^3 \tau _3^3 \tau _1-\tau _2 \tau _3^3 \tau _1\\
   &\left.-\tau _2^3 \tau _3 \tau _1+\tau _2 \tau _3 \tau _1+\tau _2^2 \tau _3^4+\tau _2^4
   \tau _3^2-3 \tau _2^2 \tau _3^2+1\right)
\end{aligned}
\end{equation}
Note that the final expression is independent of $\frac{\alpha}{\delta}$ and both $\gamma$ and $\beta$. This is because for the quivers with $U(N)$ gauge groups the $U(1)$ transformations generated by $q_{\alpha_n},q_{\gamma_i},q_{\beta_i}$ are isomorphic to gauge transformations, while for $SU(N)$ gauge groups they are global symmetries. 
As before the Higgs branch is reached by considering the $\tau_1\to0$ limit 
\begin{align}
&\begin{aligned}
\HS(\tau_2,\tau_3;HB)=&\lim_{\tau_1\to0}\HS(\tau_1,\tau_2,\tau_3;\mathcal{M})=\frac{1-3 \tau _3^2 \tau _2^2+\tau _3^2 \tau _2^4+\tau _3^4 \tau _2^2}{\left(1-\tau _2^2\right)^2 \left(1-\tau _3^2\right)^2}\label{eqn:HSU14}\\
=&\PE[\tau_2^2+\tau_3^2]+\PE[2\tau_2^2]+\PE[2\tau_3^2]-\PE[\tau_2^2]-\PE[\tau_3^2]\,,
\end{aligned}\\
&\PLog[\HS(\tau_2,\tau_3;HB)]=2 \tau _2^2+2
   \tau _3^2-3 \tau _3^2 \tau _2^2+\tau _3^2 \tau _2^4+\tau _3^4 \tau _2^2-3 \tau _3^4 \tau _2^4+\dots\,.
\end{align}
We notice that the Hilbert series splits into that for $\mathbb{C}^2/(\mathbb{Z}_2\times\mathbb{Z}_2)$ (the mesonic Higgs branch moduli space), and two copies of $(\mathbb{C}/\mathbb{Z}_2)^2$, minus the two common $\mathbb{C}/\mathbb{Z}_2$-line intersections.
From the Plethystic Logarithm we recognise the generators as $B_i=Q_i^LQ_i^R$, $\widetilde{B}_i=\widetilde{Q}^L_{i}\widetilde{Q}^R_{i+1}$, with fugacity $\tau_2^2$ and  $\tau_3^2$, respectively. At the next order we have the relations $B_1\widetilde{B}_1=B_1\widetilde{B}_2=B_2\widetilde{B}_1=B_2\widetilde{B}_2$ thanks to the F-terms $\widetilde{Q}_i^LQ_i^L=Q_{i}^R\widetilde{Q}_{i-1}^R$. The higher order terms are Hilbert syzygies (a.k.a. relations between relations).

We can also consider the Coulomb limit
\begin{equation}
\lim_{\tau_2\to0}\lim_{\tau_3\to0}\HS(\tau_1,\tau_2,\tau_3;\mathcal{M})=\HS(T;CB)=\PE\left[2T^2\right]
\end{equation}
corresponding to the operators $\prod_{i=1}^k\Phi_{i}^L$, $\prod_{i=1}^k\Phi^R_{i}$ which indeed have dimension two and have fugacity $\prod_{i=1}^k\frac{\gamma_i}{\beta_{i-1}}=1$ and $\prod_{i=1}^k\frac{\gamma_i}{\beta_{i}}=1$ under the intrinsic symmetries.

For higher values of $k,\ell$ the computations of the Hilbert series become increasingly complex, due to the requirement of making primary decomposition in a larger number of variables using \textit{Macaulay2}. We were able to compute the Higgs-branch Hilbert series for $k=\ell=3$. Again setting $\alpha_1\alpha_2\alpha_3=1$ since that can be reintroduced by rescaling $\tau_2,\tau_3$, is
\begin{equation}
\HS(\tau_2,\tau_3;HB)=\PE[\tau_2^{3}+\tau_3^{3}]+\PE[3\tau_2^{3}]+\PE[3\tau_3^{3}]-\PE[\tau_2^{3}]-\PE[\tau_3^{3}]
\end{equation}
we note that $\gamma_i$, $\beta_i$ do not appear and $\alpha_n$ enters only via $\alpha_1\alpha_2\alpha_3$ in the Hilbert series for the Higgs Branch. Moreover, we conjecture that the form of the Hilbert series for arbitrary $k=\ell$ reads
\begin{equation}
\label{eq:hsfin}
\HS(\tau_2,\tau_3;HB)=\PE[\tau_2^{\ell}+\tau_3^{\ell}]+\PE[\ell\tau_2^{\ell}]+\PE[\ell\tau_3^{\ell}]-\PE[\tau_2^{\ell}]-\PE[\tau_3^{\ell}]\,.
\end{equation}
This form for the Hilbert series implies that the moduli space is made up of a copy $\mathbb{C}^2/(\mathbb{Z}_{\ell}\times\mathbb{Z}_{\ell})$, two copies of $(\mathbb{C}/\mathbb{Z}_{\ell})^{\ell}$ minus $\mathbb{C}/\mathbb{Z}_{\ell}$ common intersections.
The Plethystic Logarithm of the Hilbert series (\ref{eq:hsfin}) reads
\begin{equation}
\textrm{PLog}[\textrm{HS}(\tau_2,\tau_3;HB)] = \ell(\tau_2^{\ell}+\tau_3^{\ell}) -(\ell^2-1)\tau_2^{\ell}\tau_3^{\ell} + \cdots \ \, ,
\end{equation}
where, at the first order of the expansion we recognise $\ell$ generators with fugacity $\tau_2^{\ell}$ and $\ell$ generators with fugacity $\tau_3^{\ell}$. While, at the next order of the expansion, we observe the presence of $\ell^2-1$ relations between the generators. Finally the higher terms in the PLog expansion are Hilbert syzygies.


\subsubsection{Hall-Littlewood Index}
We can also compute the corresponding Hall-Littlewood index for our theories.
The general expression is given by
\begin{equation}
\label{eq:HLlk}
\begin{aligned}
&\textrm{HL}(\tau_2,\tau_3) =\\
&\oint\prod_{n=1}^{\ell}\prod_{i=1}^k\left(\frac{dz_{n,i}}{2\pi\iu z_{n,i}}\PE\left[\left(\frac{\beta_{i+n-1}}{\alpha_n}\frac{z_{n,i}}{z_{n+1,i}}\tau+\frac{\alpha_n}{\gamma_i}\frac{z_{n+1,i-1}}{z_{n,i}}\tau-\frac{\beta_{i+n-1}}{\gamma_i}\frac{z_{n+1,i-1}}{z_{n+1,i}}\tau^2\right)\right]\right)\,.
\end{aligned}
\end{equation}
For example, for $k=\ell=2$, $N=1$ we have
\begin{equation}
\textrm{HL}(\tau_2,\tau_3) = \frac{1-4\tau_2^2\tau_3^2+2\tau_2^2\tau_3^4+2\tau_2^4\tau_3^2-\tau_2^4\tau_3^4}{(1-\tau_2^2)^2(1-\tau_3^2)^2}\,.
\end{equation}

For general $\ell=k$ with $N=1$ we were propose the following conjecture, which we checked for various low values of $\ell=k$,
\begin{equation}\label{eqn:HLarbell}
\HL(\tau_2,\tau_3) = \PE[\ell \tau_2^{\ell}]+\PE[\ell \tau_3^\ell]-1\,.
\end{equation}
The ratio $\HL/\HS$ is then counting, with signs, the protected operators of the theory which have $j_1+j_2\geq\frac{1}{2}$ and $r+2j_2=\frac{4}{3}q_t$ and  are given in \eqref{eq:operatorsHL}.

It is possible compute the Hall-Littlewood index for the class $\mathcal{S}_k$ theories with $SU(N)$ gauge groups. For $k=\ell=2$ with $SU(2)^{k\ell}$ gauge group this reads
\begin{equation}
\textrm{HL}_{\alpha_1\alpha_2} =\oint\prod_{n=1}^{\ell}\prod_{i=1}^k\left(\frac{dz_{n,i}}{4\pi\iu z_{n,i}}\frac{\left(1-z_{n,i}^{\pm2}\right)\left(1-\frac{\beta_{i+n-1}}{\gamma_i}\frac{z_{n+1,i-1}^{\pm1}}{z_{n+1,i}^{\pm1}}\tau^2\right)}{\left(1-\frac{\beta_{i+n-1}}{\alpha_n}\frac{z_{n,i}^{\pm1}}{z_{n+1,i}^{\pm1}}\tau\right)\left(1-\frac{\alpha_n}{\gamma_i}\frac{z_{n+1,i-1}^{\pm1}}{z_{n,i}^{\pm1}}\tau\right)}\right)\, \ ,
\end{equation}
where $\alpha_1$ and $\alpha_2$ denote the global symmetry fugacities associated to the minimal punctures.

Expanding and then taking the PLog we have
\begin{equation}
\begin{aligned}
\PLog\left[\HL_{\alpha_1\alpha_2}\right]=&\sum_{i=1}^2\left[\sum_{n=1}^2\left(\frac{\beta_i^2}{\alpha_n ^2}+\gamma_i^2\alpha_n ^2\right)+\left(\frac{1}{\alpha_1  \alpha_2 }+ \alpha_1\alpha_2\right) \right]\tau^2\\
 &+\sum_{n=1}^2\left[\frac{\alpha_n ^2}{\alpha_{n-1} ^2}-\sum_{i=1}^2\frac{\alpha_n}{\alpha_{n-1}}\left(\beta_i^2+\frac{1}{\gamma^{2}_i}\right)-\sum_{i=1}^2\frac{\beta_{i+n-1}^2}{\gamma_{i-1}^2}\right]\tau^4+\mathcal{O}(\tau^6)\,,
\end{aligned}
\end{equation}
recall we also have $\gamma_2=\gamma_1^{-1}$, $\beta_2=\beta_1^{-1}$ and the sums over $i,n$ are taken modulo $k=2$ and $\ell=2$, respectively. The operators with $q_t=1$ correspond to `baryonic type' $\det Q_{i}^{o}$, $\det \widetilde{Q}_{i}^{o}$, with $o=L/R$ and `mesonic type' $\tr Q_{i}^{L}Q_{i}^{R}$ and $\tr \widetilde{Q}_{i}^{L}\widetilde{Q}_{i}^{R}$. At the next order we have bosonic operators $\tr\widetilde{Q}_1^{o}Q_1^{o}\widetilde{Q}_2^{o}Q_2^o$ and fermionic operators of the form $\det Q_i^R\widetilde{Q}_{i-1}^L\overline{\lambda}_j^L$, $\det Q_i^L\widetilde{Q}_{i-1}^R\overline{\lambda}_j^R$ and finally $\tr Q_i^o\widetilde{Q}_{i-1}^o\overline{\lambda}_i^o$. Here $\overline{\lambda}_i^o=\overline{\lambda}^o_{i\dot+}$ denotes fermion in the superfield expansion $\overline{\Phi}^o_i=\overline{\Phi}^o_i+\overline{\theta}^{\dot\alpha}\overline{\lambda}^o_{i\dot\alpha}+\dots$. 

The unrefined $\alpha_n=\gamma_i=\beta_i=1$ limit can be computed exactly and reads
\begin{equation}
\textrm{HL}_{\delta\alpha}=\frac{1+6\tau^2+11\tau^4-12\tau^8-4\tau^{10}}{(1-\tau^2)^6}\,.
\end{equation}

\subsection{Deconstruction Limit}

In this subsection we digress a bit from the main objective of this paper to obtain an exact check for the dimensional-deconstruction prescription of Arkani-Hamed, Cohen, Kaplan, Karch and
Motl \cite{ArkaniHamed:2001ie}. 
The last few years there
 have been a few precision tests \cite{Hayling:2017cva,Hayling:2018fmv,Hayling:2018fgy,Lambert:2012qy,Bourget:2017sxr} of the deconstruction proposal  \cite{ArkaniHamed:2001ie}.  Of most interest to us in this article is the fact that \cite{Hayling:2017cva,Bourget:2017sxr} were able to show that the $\frac{1}{2}$-BPS partition function of the $\mathcal{N}=(2,0)$ theory is equal to the Higgs branch Hilbert series of the corresponding 4d $\mathcal{N}=2$ theory in the deconstruction limit. This naturally leads one to expect that a similar story should also exist for the $\mathcal{N}=(1,1)$ Little String Theory (LST). The $\mathcal{N}=(1,1)$ LST arises as the worldvolume theory on a stack of $N$ parallel NS5-branes in type IIB string theory \cite{Giveon:1999zm,Aharony:1998ub,Seiberg:1997zk,Berkooz:1997cq,Aharony:1999ks}. The basic deconstruction proposal is that the 6d theory may be effectively described by considering the $\mathcal{N}=1$ $\mathfrak{u}(N)^{\oplus\ell k}$ toroidal quiver gauge theory realised as the $l_s\to0$ limit of the worldvolume theory on a stack of $N$ D3-branes probing a transverse $\mathbb{C}^3/(\mathbb{Z}_{\ell}\times\mathbb{Z}_k)$ singularity where the quotient acts as in \eqref{eqn:quotientactZkZl}.
We then go to the point in parameter space where the vevs $v_5=\langle Q_{(i,n)}\rangle$, $v_6=\langle \Phi_{(i,n)}\rangle$  and couplings $G=g^{\text{YM}}_{(i,n)}$ are equal for all nodes in the quiver and then take the limit\footnote{Our parameters are related to those of \cite{ArkaniHamed:2001ie} by $N_5=k$, $N_6=\ell$.}
\begin{equation}
\quad\ell\to\infty\,,\quad k\to\infty\,,\quad G\to\infty\,,\quad v_{5,6}\to\infty\,,\quad 
\end{equation}
while holding $2\pi R_5 Gv_5=\ell$ and $2\pi R_6 Gv_6=k$ fixed. The main point is that, in this limit, the transverse $\mathbb{C}^3/(\mathbb{Z}_{\ell}\times\mathbb{Z}_k)$ can be approximated by $T^2\times\mathbb{R}^4$ where the radii of the torus are $r_{5}=v_5/k$ and $r_6=v_6/\ell$. 
Performing T-duality along the two circles and then S-duality gives the rank $N$ $\mathcal{N}=(1,1)$ LST on a torus with radii $R_5,R_6$.

Representations of the $\mathcal{N}=(1,1)$ supersymmetry algebra may be decomposed into a finite sum of representations of the bosonic subalgebra given by the sum of Lorentz algebra $\mathfrak{so}(6)$ and R-symmetry algebra $\mathfrak{so}(4)\iso\mathfrak{su}(2)_{R_1}\oplus \mathfrak{su}(2)_{R_2}$. The supercharges sit in the representations $\Qsd\in[0,1,0]^{\left(\frac{1}{2},0\right)}_{\frac{1}{2}}$ and $\widetilde{\Qsd}\in[0,0,1]^{\left(0,\frac{1}{2}\right)}_{\frac{1}{2}}$ with representations labelled by $[h_1,h_2,h_3]_E^{(R_1,R_2)}$. In other words $\Qsd_{h_1h_2h_3}^a$ with $h_i=\pm\frac{1}{2}$ such that $8h_1h_2h_3=-1$ and $\widetilde{\Qsd}_{h_1h_2h_3}^{\dot a}$ with $8h_1h_2h_3=+1$. They obey
\begin{equation}
\{\Qsd^a,\Qsd^b\}=\epsilon^{ab}\eta^{\mu}p_{\mu}\,,\quad\{\widetilde{\Qsd}^{\dot a},\widetilde{\Qsd}^{\dot b}\}=\epsilon^{\dot a\dot b}\widetilde{\eta}^{\mu}p_{\mu}\,,\quad\{\Qsd^{ a},\widetilde{\Qsd}^{\dot b}\}=0\,,
\end{equation}
where $\eta^{\mu},\widetilde{\eta}^{\mu}$, $\mu=1,2,\dots,6$ denote the `t Hooft symbols which intertwine between $\mathfrak{su}(4)$ and $\mathfrak{so}(6)$. The 4d $\mathcal{N}=1$ supersymmetry algebra plus the residual global  $\mathfrak{u}(1)_t\oplus\mathfrak{u}(1)_b$ symmetry algebras can be embedded into the 6d $\mathcal{N}=(1,1)$ algebra with the relations\footnote{These can be obtained in the following way: the first two relations are simply the identifications one would make between the Cartans for $\mathfrak{su}(2)_1\oplus\mathfrak{su}(2)_2\iso\mathfrak{so}(4)\subset\mathfrak{so}(6)$. When compactifying the 6d theory on $T^2$ the spinor label $h_1$  $h_1=H=H_R+H_L$ on $T^2$ and $2H_L=q_1$, $2H_R=q_2$. Finally $2R_2=q_1-q_2$ is fixed by demanding that $R_2$ evaluates to zero on the $\mathcal{N}=1$ subsector. This then fixes $q_3=R_1$. The identifications \eqref{eqn:N4N1ides} then give \eqref{eqn:6dembedding}.}
\begin{gather}
2j_1=h_2+h_3\,,\quad 2j_2=h_2-h_3\,,\\
\frac{3}{2}r=2h_1+R_1\,,\quad 2q_t=-h_1+R_1-3R_2\,,\quad q_b=h_1-R_1-R_2\label{eqn:6dembedding}\,.
\end{gather}
The relationship between the 4d and 6d supercharges is given in Table \ref{tab:6d4dembedding}.
\begin{table}
\centering
\renewcommand{\arraystretch}{1.2}
\begin{tabular}{|c|c|c|c|c|c|c|} 
\hline
 6d $\mathcal{N}=(1,1)$&  $(h_1,h_2,h_3,R_1,R_2)$ & $(j_1,j_2)$ & $r$ &$q_t$&$q_b$& 4d $\mathcal{N}=1$\\ 
 \hline\hline
 $\Qsd^{1}_{+\pm\mp}$&$(+\frac{1}{2},\pm\frac{1}{2},\mp\frac{1}{2},+\frac{1}{2},0)$&$(0,\pm\frac{1}{2})$&$+1$&$0$&$0$&$\widetilde{\mathcal{Q}}_{\dot\pm}$\\\hline
   $\Qsd^{1}_{-\pm\pm}$&$(-\frac{1}{2},\pm\frac{1}{2},\mp\frac{1}{2},+\frac{1}{2},0)$&$(\pm\frac{1}{2},0)$&$-\frac{1}{3}$&$1$&$-1$&\\\hline
 $\Qsd^{2}_{+\pm\mp}$&$(+\frac{1}{2},\pm\frac{1}{2},\mp\frac{1}{2},-\frac{1}{2},0)$&$(0,\pm\frac{1}{2})$&$+\frac{1}{3}$&$-1$&$+1$&\\\hline
    $\Qsd^{2}_{-\pm\pm}$&$(-\frac{1}{2},\pm\frac{1}{2},\pm\frac{1}{2},-\frac{1}{2},0)$&$(\pm\frac{1}{2},0)$&$-1$&$0$&$0$&$\mathcal{Q}_{\pm}$\\\hline
$\widetilde{\Qsd}^{\dot1}_{+\pm\pm}$&$(+\frac{1}{2},\pm\frac{1}{2},\pm\frac{1}{2},0,+\frac{1}{2})$&$(\pm\frac{1}{2},0)$&$+\frac{2}{3}$&$-1$&$0$&\\\hline
$\widetilde{\Qsd}^{\dot1}_{-\pm\mp}$&$(-\frac{1}{2},\pm\frac{1}{2},\mp\frac{1}{2},0,+\frac{1}{2})$&$(0,\pm\frac{1}{2})$&$-\frac{2}{3}$&$-\frac{1}{2}$&$-1$&\\\hline
$\widetilde{\Qsd}^{\dot2}_{+\pm\pm}$&$(+\frac{1}{2},\pm\frac{1}{2},\pm\frac{1}{2},0,-\frac{1}{2})$&$(\pm\frac{1}{2},0)$&$+\frac{2}{3}$&$+\frac{1}{2}$&$+1$&\\\hline
$\widetilde{\Qsd}^{\dot2}_{-\pm\mp}$&$(-\frac{1}{2},\pm\frac{1}{2},\mp\frac{1}{2},0,-\frac{1}{2})$&$(0,\pm\frac{1}{2})$&$-\frac{2}{3}$&$+1$&$0$&\\\hline
\end{tabular}
\caption{\textit{One choice of embedding of the 4d $\mathcal{N}=1$ superalgebra into the 6d $\mathcal{N}=(1,1)$ superalgebra.}}
\label{tab:6d4dembedding}
\end{table}

Let us now move to our candidate set of operators that will reproduce the $\frac{1}{2}$-BPS scalar operators in the $(1,1)$ LST.
Let's assume that, after primary decomposition, we can always identify a irreducible mesonic branch $\mathcal{M}^{\text{mes}}$ inside the full moduli space $\mathcal{M}$. $\mathcal{M}^{\text{mes}}$ will be our candidate for reproducing the 6d $\frac{1}{2}$-BPS ring. For any $\mathcal{N}=1$ theory, the operators parametrising $\mathcal{M}$ are themselves $\frac{1}{2}$-BPS with respect to the $\mathcal{N}=1$ supersymmetry algebra, namely they are annihilated by $\widetilde{\mathcal{Q}}_{\dot\alpha}$. 

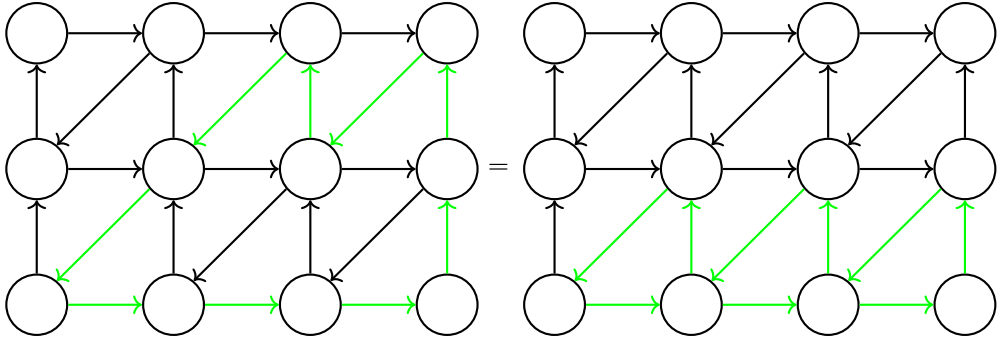
\begin{figure}
\centering
\begin{subfigure}[t]{0.4\textwidth}
\centering
  \begin{tikzpicture}[square/.style={regular polygon,regular polygon sides=4},thick,inner sep=0.1em,scale=0.9]
    \node (G11) at (0,0)[circle,draw,minimum size=0.8cm]{};
    \node (G21) at (2,0) [circle,draw,minimum size=0.8cm]{};
    \node (G31) at (4,0) [circle,draw,minimum size=0.8cm]{};
    \node (G41) at (6,0) [circle,draw,minimum size=0.8cm]{};
    \node (G12) at (0,-2)[circle,draw,minimum size=0.8cm]{};
    \node (G22) at (2,-2) [circle,draw,minimum size=0.8cm]{};
    \node (G32) at (4,-2) [circle,draw,minimum size=0.8cm]{};
    \node (G42) at (6,-2)[circle,draw,minimum size=0.8cm]{};
	\node (G13) at (0,-4)[circle,draw,minimum size=0.8cm]{};
    \node (G23) at (2,-4) [circle,draw,minimum size=0.8cm]{};
    \node (G33) at (4,-4) [circle,draw,minimum size=0.8cm]{};
    \node (G43) at (6,-4)[circle,draw,minimum size=0.8cm]{};
    
    \draw [<-] (G11.270) to (G12.90);
    \draw [<-] (G21.270) to (G22.90);
    \draw [<-,color=green] (G31.270) to (G32.90);
    \draw [<-,color=green] (G41.270) to (G42.90);
    \draw [<-] (G12.270) to (G13.90);
    \draw [<-] (G22.270) to (G23.90);
    \draw [<-] (G32.270) to (G33.90);
    \draw [<-,color=green] (G42.270) to (G43.90);
   
    \draw [->](G11.0) to (G21.180);
    \draw [->] (G21.0) to (G31.180);
    \draw [->] (G31.0) to (G41.180);
    \draw [->] (G12.0) to (G22.180);
    \draw [->] (G22.0) to (G32.180);
    \draw [->] (G32.0) to (G42.180);
    \draw [->,color=green] (G13.0) to (G23.180);
    \draw [->,color=green] (G23.0) to (G33.180);
    \draw [->,color=green] (G33.0) to (G43.180);

	\draw [<-,color=green] (G13.50) to (G22.220);
    \draw [<-] (G23.50) to (G32.220);
    \draw [<-] (G33.50) to (G42.220);
    \draw [<-,color=green] (G32.50) to (G41.220);
    \draw [<-,color=green] (G22.50) to (G31.220);
    \draw [<-] (G12.50) to (G21.220);
    \node at (6.75,-2) {$=$};
  \end{tikzpicture}
  \end{subfigure}\qquad
  \begin{subfigure}[t]{0.4\textwidth}
  \centering
    \begin{tikzpicture}[square/.style={regular polygon,regular polygon sides=4},thick,inner sep=0.1em,scale=0.9]
    \node (G11) at (0,0)[circle,draw,minimum size=0.8cm]{};
    \node (G21) at (2,0) [circle,draw,minimum size=0.8cm]{};
    \node (G31) at (4,0) [circle,draw,minimum size=0.8cm]{};
    \node (G41) at (6,0) [circle,draw,minimum size=0.8cm]{};
    \node (G12) at (0,-2)[circle,draw,minimum size=0.8cm]{};
    \node (G22) at (2,-2) [circle,draw,minimum size=0.8cm]{};
    \node (G32) at (4,-2) [circle,draw,minimum size=0.8cm]{};
    \node (G42) at (6,-2)[circle,draw,minimum size=0.8cm]{};
	\node (G13) at (0,-4)[circle,draw,minimum size=0.8cm]{};
    \node (G23) at (2,-4) [circle,draw,minimum size=0.8cm]{};
    \node (G33) at (4,-4) [circle,draw,minimum size=0.8cm]{};
    \node (G43) at (6,-4)[circle,draw,minimum size=0.8cm]{};
    
    \draw [<-] (G11.270) to (G12.90);
    \draw [<-] (G21.270) to (G22.90);
    \draw [<-] (G31.270) to (G32.90);
    \draw [<-] (G41.270) to (G42.90);
    \draw [<-] (G12.270) to (G13.90);
    \draw [<-,color=green] (G22.270) to (G23.90);
    \draw [<-,color=green] (G32.270) to (G33.90);
    \draw [<-,color=green] (G42.270) to (G43.90);
    
    \draw [->](G11.0) to (G21.180);
    \draw [->] (G21.0) to (G31.180);
    \draw [->] (G31.0) to (G41.180);
    \draw [->] (G12.0) to (G22.180);
    \draw [->] (G22.0) to (G32.180);
    \draw [->] (G32.0) to (G42.180);
    \draw [->,color=green] (G13.0) to (G23.180);
    \draw [->,color=green] (G23.0) to (G33.180);
    \draw [->,color=green] (G33.0) to (G43.180);

	\draw [<-,color=green] (G13.50) to (G22.220);
    \draw [<-,color=green] (G23.50) to (G32.220);
    \draw [<-,color=green] (G33.50) to (G42.220);
    \draw [<-] (G32.50) to (G41.220);
    \draw [<-] (G22.50) to (G31.220);
    \draw [<-] (G12.50) to (G21.220);
  \end{tikzpicture}
  \end{subfigure}
  \caption{\it Visualisation for the example of $p=3$. On the left\\ $Q_{(i,n)}Q_{(i,n+1)}Q_{(i,n+2)}\Phi_{(i,n+3)}\Phi_{(i-1,n+3)}\widetilde{Q}_{(i-1,n+2)}\Phi_{(i-1,n+2)}\widetilde{Q}_{(i-1,n+1)}\widetilde{Q}_{(i,n)}$ \\reduced to $\prod_{m=0}^2 Q_{(i,n+m)}\Phi_{(i,n+m+1)}\widetilde{Q}_{(i,n+m)}$. Again further applying the F-terms means this operator can be completely written in terms of $M$.}
  \label{fig:pathexample}
\end{figure}

One may wonder why we have picked the subvariety $\mathcal{M}^{\text{mes}}$ as opposed to the full moduli space $\mathcal{M}$. Shortly we will prove that, for the theory with $U(1)$ gauge groups, in the $k,\ell\to\infty$ limit that $\mathcal{M}^{\text{mes}}$ coincides with $\mathcal{M}$. Setting, without loss of generality in this limit $k=\ell$, the possible gauge invariant operators parametrising $\mathcal{M}$ for the theory with $U(1)$ gauge groups are in correspondence with the number of closed, directed paths that one can draw on the quiver diagram.  There are four main types of operators. There are those which involve an equal number $p\leq k=\ell$ of $Q$, $\widetilde{Q}$ and $\Phi$ fields. Such an operator enters the Hilbert series with fugacity $\tau_1^p\tau_2^p\tau_3^p$. Operators of this form correspond to picking a base node, say $(i=1,n=1)$, and drawing a closed loop involving $p$ vertical steps, $p$ horizontal steps and $p$ diagonal steps. However, recall the F-terms set
\begin{align}
\label{eqn:PhiF}&F_{\Phi_{(i,n)}}=\widetilde{Q}_{(i,n-1)}Q_{(i,n-1)}-Q_{(i-1,n)}\widetilde{Q}_{(i,n)}=0\,,\\
\label{eqn:QF}&F_{Q_{(i,n)}}=\Phi_{(i,n+1)}\widetilde{Q}_{(i,n)}-\widetilde{Q}_{(i+1,n)}\Phi_{(i+1,n)}=0\,,\\
\label{eqn:QtF}&F_{\widetilde{Q}_{(i,n)}}=Q_{(i,n)}\Phi_{(i,n+1)}-\Phi_{(i,n)}Q_{(i-1,n)}=0\,.
\end{align}
In terms of the quiver diagram Figure \ref{fig:Skquivergenus1}, this means that the operations of moving right, up or diagonally all commute. In other words this means that the associated operator is independent of the choice of base node and of the specific path chosen, it depends only on the length $3p$. See Figure \ref{fig:pathexample} for a diagrammatic example.
We can therefore write these operators as $M^p$ with, say, $M=\widetilde{Q}_{(1,\ell)}Q_{(1,\ell)}\Phi_{(1,1)}$. The other types of operators involve only $Q$'s, $\widetilde{Q}$'s or $\Phi$'s. They can be written as powers of
\begin{equation}
B_{\Phi_n}=\prod_{i=1}^{\ell}\Phi_{(i,n)}\,,\quad B_{Q_i}=\prod_{n=1}^{\ell}Q_{(i,n)}\,,\quad B_{\widetilde{Q}_i}=\prod_{j=1}^{\ell}\widetilde{Q}_{(i+j,i-j)}\,,
\end{equation}
they enter the Hilbert series with fugacity $\tau_1^{\ell}$, $\tau_2^{\ell}$ and $\tau_3^{\ell}$, respectively. In degree $\geq\min(k,\ell)$ there can be complicated relations and higher syzgies between $M$, $B_{Q_i}$, $B_{\widetilde{Q}_i}$ and $B_{\Phi_n}$, leading to a complicated structure for $\mathcal{M}$. However, the main point is that in the $\ell=k\to\infty$ limit the $B_{Q_i}$, $B_{\widetilde{Q}_i}$ and $B_{\Phi_n}$ operators all become infinity heavy and their dimensions $E=k=\ell$ tend to infinity, in particular $\lim_{\ell\to\infty}\tau_{1,2,3}^{\ell}\to0$ and their contribution to the Hilbert series vanishes. This also implies that the corresponding ring becomes freely generated, since there are no relations between the operators in degree smaller than $\min(k,\ell)\to\infty$. Only the dimension of $M$ remains finite in the limit. $M$ is a purely mesonic operator and therefore, in this limit, one indeed expects $\mathcal{M}^{\text{mes}}$ to coincide with $\mathcal{M}$. 

The above arguments can be simply extended to the case $N\geq2$ by taking traces. Each path with an equal number $p$ of $Q$, $\widetilde{Q}$ and $\Phi$'s now corresponds to $A=1,\dots,N$ operators of dimension $3pA$ corresponding to operators schematically of the form $\tr (Q^p\widetilde{Q}^p\Phi^p)^A$. As long as we consider the traces we can apply the same rules that we did for $N=1$, namely that the operations of moving right, up or diagonally commute. This again means any path of length $3p$ can be written as the loop given by $M^p$ with, say, $M=\widetilde{Q}_{(1,\ell)}Q_{(1,\ell)}\Phi_{(1,1)}$ which is now a $N\times N$ matrix. The $N$ operators corresponding to one of these closed paths can then be written  as $M_{(A)}^p$ with
\begin{equation}
M_{(A)}:=\tr M^A=\tr\left(\widetilde{Q}_{(1,\ell)}Q_{(1,\ell)}\Phi_{(1,1)}\right)^A    \quad \mbox{with} \quad A=1,\dots,N \,.
\end{equation}
As before the $M_{(A)}$ can have complicated relations with the operators which are of the form $\tr B_{Q_i}^A$, $\tr B_{\widetilde{Q}_i}^A$ and $\tr B^A_{\Phi_i}$ but the dimensions of the latter are $Ak=A\ell\to\infty$. 
We therefore arrive at the conclusion that, {\it in the deconstruction $k=\ell\to\infty$ limit $\mathcal{M}$ coincides with $\mathcal{M}^{\text{mes}}$ for all $N$}
\begin{equation}\lim_{\ell,k \to\infty}\HS(\tau_1,\tau_2,\tau_3;\mathcal{M})= \lim_{\ell,k \to\infty}\HS(\tau_1,\tau_2,\tau_3;\mathcal{M}^{\text{mes}}) \, .
\end{equation}

For the $N=1$ case we can identify $\mathcal{M}^{\text{mes}}=\mathbb{C}^3/(\mathbb{Z}_{\ell}\times\mathbb{Z}_k)$, so the Hilbert series for the $N=1$ case is given by \eqref{eqn:moilen}. Therefore, the Hilbert series for general $N$ is given by
\begin{equation}
\label{eq:hsg}
\HS(\tau_1,\tau_2,\tau_3;\mathcal{M}^{\text{mes}}) = \frac{1}{N!}\frac{\partial^{N}}{\partial\nu^{N}}\left.\PE\left[\nu M(\tau_1,\tau_2,\tau_3;\mathbb{C}^3/(\mathbb{Z}_{\ell}\times\mathbb{Z}_k))\right]\right|_{\nu=0}\, .
\end{equation}
Let us now consider the deconstruction limit. Taking the limit on the $N=1$ result, recalling that $|\tau_{1,2,3}|<1$, gives
\begin{equation}
\begin{aligned}
\lim_{k,\ell\to\infty}M(\tau_1,\tau_2,\tau_3;\mathbb{C}^3/(\mathbb{Z}_{\ell}\times\mathbb{Z}_k))&=\lim_{k,\ell\to\infty}M(\tau_1,\tau_2,\tau_3;\mathcal{M})\\
&=\lim_{k\to\infty}\PE\left[\tau_1\tau_2\tau_3+\tau_1^k+\tau_2^k\tau_3^k-\tau_1^k\tau_2^k\tau_3^k\right]\\&=\PE\left[\tau_1\tau_2\tau_3\right]=\PE[pq]\,.
\end{aligned}
\end{equation}
Therefore, for general $N$ we have 
\begin{equation}\label{eqn:decon4d}
\lim_{\ell,k \to\infty}\HS(\tau_1,\tau_2,\tau_3;\mathcal{M})= \PE\left[\sum_{p=1}^{N}(\tau_1\tau_2\tau_3)^{p}\right]=\frac{1}{(\tau_1\tau_2\tau_3;\tau_1\tau_2\tau_3)_N} \, .
\end{equation}
The coordinate-ring of $\mathcal{M}$ for this theory in the $k,\ell\to\infty$ limit is therefore simply $\mathbb{C}[M_{(1)},M_{(2)},\dots,M_{(N)}]$.
\subsubsection{Computation of the $\frac{1}{2}$-BPS Partition Function for $(1,1)$ LST}
Let us now move to the computation of the 6d quantity that we would like to match to the 4d quantity \eqref{eqn:decon4d}. We can use the fact that, at low energies, the $\mathcal{N}=(1,1)$ LST admits an effective description as 6d maximally supersymmetric SYM theory with gauge group $U(N)$.  
The on-shell degrees of freedom of the $(1,1)$ SYM theory contains a 2-form gauge field strength $F\in[0,1,1]_{2}^{(0,0)}$, scalars $X\in[0,0,0]_{2}^{(\frac{1}{2},\frac{1}{2})}$ and fermions $\lambda\in[0,1,0]_{\frac{5}{2}}^{(0,\frac{1}{2})}$, $\widetilde{\lambda}\in[0,0,1]_{\frac{5}{2}}^{(\frac{1}{2},0)}$ all in the adjoint of $\mathfrak{g}=\mathfrak{u}(N)$.
The supersymmetry transformations of interest to us are
\begin{equation}
\Qsd^{a}X^{b\dot b}=\frac{1}{\sqrt{2}}\epsilon^{ab}\lambda^{\dot b}\,,\quad \widetilde{\Qsd}^{\dot a}X^{b \dot b}=\frac{1}{\sqrt{2}}\epsilon^{\dot a \dot b}\widetilde{\lambda}^{b}\,.
\end{equation}
We can therefore construct $\frac{1}{2}$-BPS multiplets whose highest weight state is annihilated by both $\Qsd^1$ and $\widetilde{\Qsd}^{\dot1}$. The supersymmetric primary of these multiplets is given by
\begin{equation}
\mathcal{O}^{\text{h.w.}}_A:=\tr\left(X^{1\dot1}\right)^A\,, \quad \text{with $h_1=h_2=h_3=0$ and $E=4R_1=4R_2=2A$}\,,
\end{equation}
which are independent for $A=1,\dots,N$.
By acting with all possible supersymmetries $\Qsd,\widetilde{\Qsd}$ and $\mathfrak{so}(6)\oplus\mathfrak{su}(2)_{R_1}\oplus\mathfrak{su}(2)_{R_2}$ generators we can generate the entire $\frac{1}{2}$-BPS multiplet by acting on the highest weight state $\mathcal{O}^{\text{h.w.}}_A$.
We can define a $\frac{1}{2}$-BPS partition function (Hilbert series) by passing to the scalar sector of the $\Qsd^1\cap\widetilde{\Qsd}^{\dot1}$ -cohomology
\begin{equation}\label{eqn:1comma1BPS}
Z^{(1,1)}_{\text{$\frac{1}{2}$-BPS}}=\Tr_{\mathcal{H}}x^{2R_1}\,,\quad \mathcal{H}=\left\{\mathbb{C}[\mathbf{\mathcal{O}}]^{\mathfrak{g}}\middle|E=4R_1=4R_2\,,h_1=h_2=h_3=0\right\}\,.
\end{equation}
With this definition we have constructed an object which is counting only the gauge invariant words comprised of scalar component $X^{1\dot1}$ in the $(1,1)$ SYM theory. Using letter counting we can compute
\begin{equation}
Z^{(1,1)}_{\text{$\frac{1}{2}$-BPS}}=\oint d\mu_{U(N)}\PE\left[x\,\chi_{\text{adj}}(\mathbf{z})\right]=\frac{1}{(x;x)_N}\,,
\end{equation}
this is the Hilbert series for $\mathbb{C}[\mathcal{O}^{\text{h.w.}}_1,\mathcal{O}^{\text{h.w.}}_2,\dots,\mathcal{O}^{\text{h.w.}}_N]$. Identifying $x=\tau_1\tau_2\tau_3=pq$ we have matched $Z^{(1,1)}_{\text{$\frac{1}{2}$-BPS}}$ with \eqref{eqn:decon4d} and the map between the operators is just
\begin{equation}
M_{(A)}\xleftrightarrow[]{\text{4d:6d}}\mathcal{O}^{\text{h.w.}}_A\,.
\end{equation}
These operators can be expected to be found as a subset of those counted by the $\mathbb{S}^1\times \mathbb{S}^5$ partition function. By the Nahm classification \cite{Nahm:1977tg}, there is no superconformal algebra associated to $\mathcal{N}=(1,1)$ SUSY in 6d so there is no superconformal index associated to this theory, nevertheless we can define the $\mathbb{S}^1\times \mathbb{S}^5$ partition function of the theory
\begin{equation}\label{eqn:ZLST}
Z^{(1,1)}_{\mathbb{S}^1\times \mathbb{S}^5}(x,y,p_1,p_2)=\Tr_{\mathbb{S}^5}(-1)^Fe^{-\beta H}x^{E-R_1}y^{R_2}p_1^{h_1-h_3}p_2^{h_2-h_3}
\end{equation}
where $H=\{\Qsd,\Qsd^{\dagger}\}=E-h_1-h_2-h_3-4R_1$ with $\Qsd:=\Qsd_{---}^{1}$. The partition function \eqref{eqn:ZLST} can be computed using the elliptic genus method \cite{Kim:2015gha,Kim:2017xan,Kim:2018gak} or using the refined topological string \cite{Iqbal:2007ii,Lockhart:2012vp}. $Z^{(1,1)}_{\mathbb{S}^1\times \mathbb{S}^5}$ is expected to receive both perturbative contributions as well as non-perturbative contributions from 6d SYM instanton string states.

If we consider taking $xy=p_1=p_2=1$ then
\begin{equation}\label{eqn:ZLSTunref}
Z^{(1,1)}_{\mathbb{S}^1\times \mathbb{S}^5}(x,x^{-1},1,1)=\Tr_{\mathbb{S}^5}(-1)^Fe^{-\beta H}x^{E-R_1-R_2}
\end{equation}
receives extra shortening and is annihilated by $\Qsd$ and $\widetilde{\Qsd}_{+--}^{\dot1}$ ,$\widetilde{\Qsd}_{-+-}^{\dot1}$, $\widetilde{\Qsd}_{--+}^{\dot1}$. Therefore the unrefined limit \eqref{eqn:ZLSTunref} receives non-zero contributions only from states with $h_1=h_2=h_3=E-4R_2$ and $E+2R_1=6R_2$. $\mathcal{H}$ is clearly contained as a subset of those states when $E=4R_2$. 

We also note that \eqref{eqn:1comma1BPS} is equal to the $\frac{1}{2}$-BPS limit of the index of the $\mathcal{N}=(2,0)$ theory of type $\mathfrak{g}=\mathfrak{u}(N)$, which falls into the general result that, in all known examples, the $\frac{1}{2}$-BPS partition function seems to be a universal quantity in all maximally supersymmetric theories in $3$, $4$ and $6$ dimensions \cite{Kim:2016usy}. Additionally,  
the $(1,1)$ and $(2,0)$ LSTs are related by T-duality. T-duality exchanges winding and momentum modes along the temporal $\mathbb{S}^1$. Since, by definition, $Z^{(1,1)}_{\text{$\frac{1}{2}$-BPS}}$ and the analogous quantity $Z^{(2,0)}_{\text{$\frac{1}{2}$-BPS}}$ \cite{Bhattacharyya:2007sa,Kim:2015gha} defined for the $(2,0)$ theory count operators only in the zero winding and zero momentum sectors one naturally expects $Z^{(1,1)}_{\text{$\frac{1}{2}$-BPS}}=Z^{(2,0)}_{\text{$\frac{1}{2}$-BPS}}$.
It would be interesting to further investigate if it is possible to obtain the more general partition functions, such as \eqref{eqn:ZLST} and \eqref{eqn:ZLSTunref}, from 4d class $\mathcal{S}_k$ quantities using deconstruction \cite{Hayling:2017cva,Hayling:2018fmv,Hayling:2018fgy,Bourget:2017sxr}.

\section{\boldmath Interacting Trinion Theories in Class $\mathcal{S}_{k=2}$}
\label{sec:trinion}

This Section is devoted to the study of the strongly interacting  Trinion theories in Class $\mathcal{S}_{k=2}$ that correspond to
 three punctured Riemann spheres and have no  Lagrangian description.

\begin{figure}
\centering
\begin{tikzpicture}[thick]
  \draw (0,0) ellipse (1.3cm and 1.3cm);
  \draw (4,0) ellipse (1.3cm and 1.3cm);
  \node (LU) at (0.5,0.05) {};
  \node (LD) at (0.5,-0.65) {};
  \node (RU) at (3.5,0.05) {};
  \node (RD) at (3.5,-0.65) {};
  \draw [black,fill=black] (-0.1,0.6) circle (2pt) node [below]{$\delta$};
  \draw [black,fill=black] (4.1,0.6) circle (2pt) node [below]{$\alpha$};
  \draw [black,fill=green] (-0.5,-0.3) circle (6pt) node [below=4pt]{$\mathbf{u}$} node{$\ell$};
  \draw [black,fill=green] (4.5,-0.3) circle (6pt) node [below=4pt]{$\mathbf{v}$} node {$r$};
  \draw [black,fill=red] (3.3,-0.3) circle (6pt) node [below=4pt]{$\mathbf{z}$} node {${\ell}$};
  \draw [black,fill=red] (0.7,-0.3) circle (6pt) node [below=4pt]{$\mathbf{z}$} node {$r$};
 \draw [red](LU) to [bend right=15] (RU); 
 \draw [red](LD) to [bend left=15] (RD);
 \draw [<->] (0,-1.5) to (-0.5,-2);
\draw [<->] (4,-1.5) to (4.5,-2);
\end{tikzpicture}\\
\begin{tikzpicture}[thick]
  \draw (0,0) ellipse (1.3cm and 1.3cm);
  \draw (4,0) ellipse (1.3cm and 1.3cm);
  \node (LU) at (0.5,0.05) {};
  \node (LD) at (0.5,-0.65) {};
  \node (RU) at (3.5,0.05) {};
  \node (RD) at (3.5,-0.65) {};
  \draw [black,fill=black] (-0.1,0.6) circle (2pt) node [below]{$\delta$};
  \draw [black,fill=green] (4.1,0.6) circle (6pt) node [below=4pt]{$\mathbf{u}$} node {$\ell$};
  \draw [black,fill=black] (-0.5,-0.3) circle (2pt) node [below]{$\alpha$};
  \draw [black,fill=green] (4.5,-0.3) circle (6pt) node [below=4pt]{$\mathbf{v}$} node {$r$};
  \draw [black,fill=red] (3.3,-0.3) circle (6pt) node [below=4pt]{\tiny $ (\alpha\delta,z_2)$} node {$\ell$};
 \draw [red] (LU) to [bend right=15] (RU); 
 \draw [red](LD) to [bend left=15] (RD);
 \draw [<->] (5.5,0) to (6,0);
\end{tikzpicture}
\begin{tikzpicture}[thick]
  \draw (0,0) ellipse (1.3cm and 1.3cm);
  \draw (4,0) ellipse (1.3cm and 1.3cm);
  \node (LU) at (0.5,0.05) {};
  \node (LD) at (0.5,-0.65) {};
  \node (RU) at (3.5,0.05) {};
  \node (RD) at (3.5,-0.65) {};
  \draw [black,fill=black] (-0.1,0.6) circle (2pt) node [below]{$\delta$};
  \draw [black,fill=green] (4.1,0.6) circle (6pt) node [below=4pt]{$\mathbf{u}$} node {$\ell$};
  \draw [black,fill=black] (-0.5,-0.3) circle (2pt) node [below]{$\alpha$};
  \draw [black,fill=green] (4.5,-0.3) circle (6pt) node [below=4pt]{$\mathbf{v}$} node {$r$};
  \draw [black,fill=green] (3.3,-0.3) circle (6pt) node [below=4pt]{\tiny $ (\alpha\delta,z_2)$} node {$\ell$};
 \draw [green](LU) to [bend right=15] (RU); 
 \draw [green](LD) to [bend left=15] (RD);
\end{tikzpicture}
\caption{\textit{Three `S-dual' frames of the basic $A_1$ four punctured sphere in class $\mathcal{S}_2$. On the top is the canonical weakly coupled frame with fluxes for $U(1)_{\beta}\times U(1)_{\gamma}\times U(1)_t$ given by $\mathcal{F}=(0,0,1)$. On the bottom left is the S-dual frame obtained by gluing a quiver tail to the $T_B$ theory with flux $\mathcal{F}_B=(-\frac{1}{4},\frac{1}{4},1)$. On the bottom right is the S-dual frame obtained by gluing a quiver tail to the $T_A$ theory with fluxes $\mathcal{F}_A=(\frac{1}{4},\frac{1}{4},1)$. Green (red) circles indicate maximal punctures of colour $c=0$ ($c=1$). $o=l,r$ indicates the orientation of a maximal puncture. Dots indicate minimal punctures.}}
\label{fig:sduals}
\end{figure}
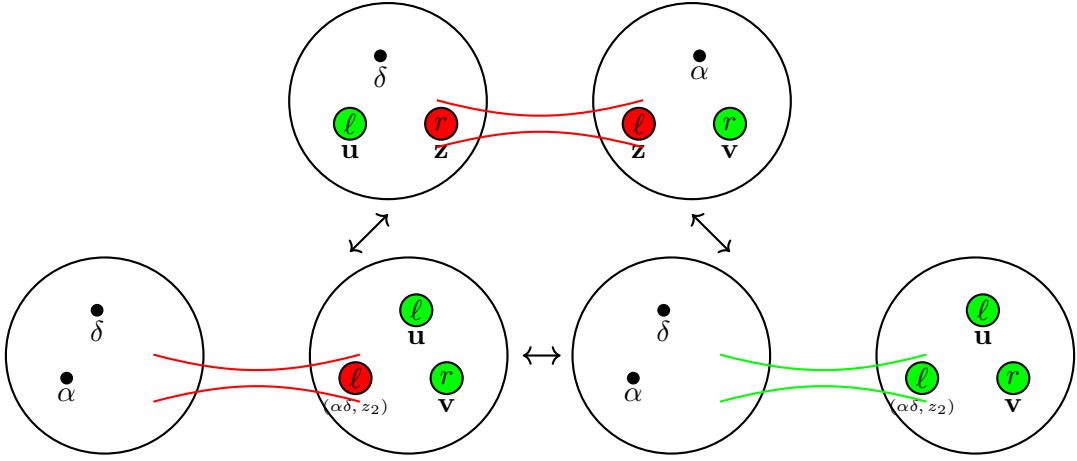
\begin{figure}
\centering
\begin{tikzpicture}[square/.style={regular polygon,regular polygon sides=4},thick,inner sep=0.1em,scale=0.9]
    \node (L1) at (0,0) [square,draw,minimum size=1.8cm]{$u_1$};
    \node (R1) at (10,0) [square,draw,minimum size=1.8cm]{$v_2$};
    \node (G1) at (5,0) [circle,draw,minimum size=1.3cm]{$b$};
    \node (G2) at (5,-3) [circle,draw,minimum size=1.3cm]{$a$};
    \node (L2) at (0,-3) [square,draw,minimum size=1.8cm]{$u_2$};
    \node (R2) at (10,-3) [square,draw,minimum size=1.8cm]{$v_1$};

    \draw [->] (L1.0) -- (G1.180) node[midway,above] {$ \scriptstyle{Q^-_{1}\,,\,\delta\gamma\sqrt{t}}$};
    \draw [->] (L2.0) -- (G2.180) node[midway,below] {$\scriptstyle{Q_2^-\,,\,\frac{\delta}{\gamma}\sqrt{t}}$};
    \draw [->] (G1.0) -- (R1.180) node[midway,above] {$\scriptstyle{{Q'_2}^+\,,\,\frac{\alpha}{\gamma}\sqrt{t}}$};
    \draw [->] (G2.0) -- (R2.180) node[midway,below] {$\scriptstyle{{Q'_1}^+\,,\,\alpha\gamma\sqrt{t}}$};
    
    \draw [->] (G1.225) -- (L2.45) node[near start,above,sloped] {$\scriptstyle{Q_2^+\,,\,\frac{1}{\delta\beta}\sqrt{t}}$};
       \draw [->] (R1.225) -- (G2.45) node[near end,below,sloped] {$\scriptstyle{{Q'_2}^-\,,\,\frac{1}{\alpha\beta}\sqrt{t}}$};
    \draw [->] (G2.135) -- (L1.315) node[near start,below,sloped] {$\scriptstyle{Q_1^+\,,\,\frac{\beta}{\delta}\sqrt{t}}$};
    \draw [->] (R2.135) -- (G1.315) node[near end,above,sloped] {$\scriptstyle{{Q'_1}^-\,,\,\frac{\beta}{\alpha}\sqrt{t}}$};
    
     \draw [<-] (G1.280) -- (G2.80) node[midway,right] {$\scriptstyle{\Phi_2\,,\,\gamma\beta\frac{pq}{t}}$};
     \draw [->] (G1.260) -- (G2.100) node[midway,left] {$\scriptstyle{\Phi_1\,,\,\frac{1}{\gamma\beta}\frac{pq}{t}}$};
         
 \end{tikzpicture}
  \caption{\it Quiver diagram of the theory associated to a sphere with two minimal punctures and two maximal punctures of colour $c_l=c_r=0$.}
  \label{fig:4puncspheretrin}
\end{figure}
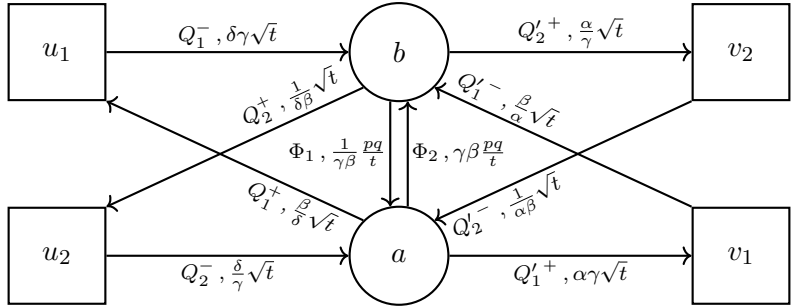
As pictured in Figure \ref{fig:sduals} the basic $A_1$ four punctured sphere in class $\mathcal{S}_2$ admits three `S-dual' decomposition. The first is the most familiar being the standard Lagrangian frame and is given in Figure \ref{fig:4puncspheretrin}. The other descriptions involve strongly interacting SCFTs, associated to spheres with three maximal punctures, with an $SU(2)$ gauging to a quiver tail. These SCFTs, denoted $T_A$ \& $T_B$ in \cite{Razamat:2016dpl} carry global symmetries of, at least $\mathfrak{su}(2)_{z}^2\oplus \mathfrak{su}(2)_{v}^2\oplus \mathfrak{su}(2)_{u}^2\oplus \mathfrak{u}(1)_{\gamma}\oplus \mathfrak{u}(1)_{\beta}\oplus \mathfrak{u}(1)_{t}$. 
Finally,
 there is a simpler Trinion for Class $\mathcal{S}_2$ and $N=2$ than the $T_A$ and $T_B$ which was studied in \cite{Razamat:2019ukg}, however, we will not study it in this paper\footnote{Our main tool in this Section is the  Spiridonov-Warnaar inversion formula \cite{2004math11044S}. The simpler Trinion of \cite{Razamat:2019ukg}
is not obtained via an inversion and it involves a new type of a puncture.}.

The field content of the $B$- and $A$-type quiver tails is listed in Table \ref{tab:tailB} and Table \ref{tab:tailA}, respectively. The superpotential  of the tail, in these cases, will include
\begin{align}\label{eqn:tailsuperpot}
W_B\supset&\sum_{a=\pm}\left(q^{(a)}\Phi'^{(+)}B_{1,a+}+q^{(a)}\Phi'^{(-)}B_{2,a-}\right)+q^{(+)}q^{(-)}T_0\,,\\\
W_A\supset&\sum_{a,b=\pm}q^{(a)}\Phi'^{(b)}B_{1,ab}+q^{(+)}q^{(-)}T_0\,.
\end{align}
\begin{table}
\centering
\begin{tabular}{|c |c|c|c|c|c|c|c|c| c|} 
 \hline
 Field(s) &$\mathfrak{su}(2)_{z_2}$&$\mathfrak{u}(1)_{\delta}$&$\mathfrak{u}(1)_{\alpha}$&$\mathfrak{u}(1)_{r}$&$\mathfrak{u}(1)_{\beta}$&$\mathfrak{u}(1)_{\gamma}$&$\mathfrak{u}(1)_{t}$&$\delta_{1\pm}$&$\widetilde{\delta}_{2\dot+}$\\[2pt] 
  \hline\hline
$q^{(\pm)}$&$\mathbf{2}$&$\pm1$&$\mp1$&$0$&$1$&$-1$&$0$&$0$&$0$\\
$\Phi'^{(\pm)}$&$\mathbf{2}$&$\mp1$&$\mp1$&$2/3$&$\mp1$&$\mp1$&$-1$&$2$&$0$\\
$B_{2,\pm-}$&$\mathbf{1}$&$-1\mp1$&$-1\pm1$&$4/3$&$-2$&$0$&$1$&$0$&$4$ \\
$B_{1,\pm+}$&$\mathbf{1}$&$1\mp1$&$1\pm1$&$4/3$&$0$&$2$&$1$&$0$ &$4$\\
$T_0$&$\mathbf{1}$&$0$&$0$&$2$&$-2$&$2$&$0$&$2$&$4$\\\hline
\end{tabular}
\caption{\textit{Field content associated to the type $B$ `quiver tail' in Class $\mathcal{S}_2$ corresponding to a sphere with two minimal punctures $\alpha,\delta$ and a `pinched' maximal puncture $(z_1=\alpha\delta,z_2)$ which can be glued to a maximal puncture to convert it to two minimal punctures. Each field is an $\mathcal{N}=1$ chiral multiplet. In the final columns we firstly list the value of $\delta_{1\pm}=r+2j_2-\frac{4}{3}q_t\pm 2j_1$ and $\widetilde{\delta}_{2\dot+}=4j_2+2r+\frac{4}{3}q_t$ of the corresponding field.}}
\label{tab:tailB}
\end{table}
\begin{table}
\centering
\begin{tabular}{|c |c|c|c|c|c|c|c|c| c|} 
 \hline
 Field(s) &$\mathfrak{su}(2)_{z_2}$&$\mathfrak{u}(1)_{\delta}$&$\mathfrak{u}(1)_{\alpha}$&$\mathfrak{u}(1)_{r}$&$\mathfrak{u}(1)_{\beta}$&$\mathfrak{u}(1)_{\gamma}$&$\mathfrak{u}(1)_{t}$&$\delta_{1\pm}$&$\widetilde{\delta}_{2\dot+}$\\[2pt] 
  \hline\hline
$q^{(\pm)}$&$\mathbf{2}$&$\pm1$&$\mp1$&$0$&$-1$&$-1$&$0$&$0$&$0$\\
$\Phi'^{(\pm)}$&$\mathbf{2}$&$\mp1$&$\mp1$&$2/3$&$\pm1$&$\mp1$&$-1$&$2$&$0$\\
$B_{1,\pm\pm}$&$\mathbf{1}$&$0$&$\pm2$&$4/3$&$1\mp1$&$1\pm1$&$1$&$0$&$4$ \\
$B_{1,\mp\pm}$&$\mathbf{1}$&$\pm2$&$0$&$4/3$&$1\mp1$&$1\pm1$&$1$&$0$ &$4$\\
$T_0$&$\mathbf{1}$&$0$&$0$&$2$&$2$&$2$&$0$&$2$&$4$\\\hline
\end{tabular}
\caption{\textit{Field content associated to the type $A$ `quiver tail' in class $\mathcal{S}_2$ corresponding to a sphere with two minimal punctures $\alpha,\delta$ and a `pinched' maximal puncture $(z_1=\alpha\delta,z_2)$ which can be glued to a maximal puncture to convert it to two minimal punctures. Each field is an $\mathcal{N}=1$ chiral multiplet. In the final columns we firstly list the value of $\delta_{1\pm}=r+2j_2-\frac{4}{3}q_t\pm 2j_1$ and $\widetilde{\delta}_{2\dot+}=4j_2+2r+\frac{4}{3}q_t$ of the corresponding field.}}
\label{tab:tailA}
\end{table}
The respective $T_B$ \& $T_A$ SCFTs have mesonic operators $M^{u}_{\pm}, M^{v}_{\pm},M^{z}_{\pm}$ which are in the bifundamental representation of the corresponding $\mathfrak{su}(2)_{c}\times \mathfrak{su}(2)_{c}$ symmetry with $c=\{z,v,u\}$ and in the $+1$ representation of $\mathfrak{u}(1)_t$. We list such operators, with the corresponding charges under the flavor symmetry group, in Table \ref{tab:HBOpsTB} and in Table \ref{tab:HBOpsTA} for the $T_B$ and $T_A$ theory respectively. In the case of the $T_B$ theory, as depicted in Table \ref{tab:HBOpsTB}, $M^{u}_{\pm}, M^{v}_{\pm}$ have charge $\pm1$ under $\mathfrak{u}(1)_{\gamma}$ and $\mp1$ $\mathfrak{u}(1)_{\beta}$ while $M^{z}_{\pm}$ has $\pm1$, $\pm1$. In the case of the $T_A$ theory $M^{u}_{\pm}, M^{v}_{\pm},M^{z}_{\pm}$ all have charge $\pm1$ under $\mathfrak{u}(1)_{\gamma}$ and $\mp1$ $\mathfrak{u}(1)_{\beta}$ as in Table \ref{tab:HBOpsTA}. The tails couple to their respective trinions through superpotentials
\begin{equation}\label{eqn:tailsuperpot2}
\Delta W_B=\sum_{a=\pm}\Phi'^{(a)}M^z_{a}\,,\quad \Delta W_A=\sum_{a=\pm}\Phi'^{(a)}M^z_{a}\,.
\end{equation}
Ultimately we have the following identity between the indices in the three different frames
\begin{equation}\label{eqn:indequivtrin}
I \indices{_{\mathbf{u}}_{\delta}_{\alpha}^{\mathbf{v}}}=I \indices{_{\mathbf{u}}_{\delta}^{\mathbf{z}}}\cdot I \indices{_{\mathbf{z}}_{\alpha}^{\mathbf{v}}}=I ^{(B)\mathbf{z}}_{\delta\alpha}\cdot I  ^{(T_B)\mathbf{v}}_{\mathbf{z}\mathbf{u}}= I  ^{(A)\mathbf{z}}_{\delta\alpha}\cdot I  ^{(T_A)\mathbf{v}}_{\mathbf{z}\mathbf{u}}
\end{equation}
and we set $\mathbf{z}=(\alpha\delta,z_2)$. The expression for $ I  \indices{_{\mathbf{u}}_{\delta}_{\alpha}^{\mathbf{v}}}$ is given in \eqref{eqn:4puncsphereindex}.
The final two  expressions read \cite{Gaiotto:2015usa,Razamat:2016dpl}
\begin{equation}\label{eqn:4puncstrong}
\begin{aligned}
& I  ^{(B)\mathbf{z}}_{\delta\alpha}\cdot I  ^{(T_B)\mathbf{v}}_{\mathbf{z}\mathbf{u}}=\kappa\oint\frac{dz_2}{4\pi\iu z_2}\delta\left(z_2,s;\frac{\beta}{\gamma}\right)\frac{\Gamma_e\left(t\frac{\gamma}{\beta}(\beta\gamma z_1)^{\pm1}s^{\pm1}\right)}{\Gamma_e\left(t(\gamma\beta z_1)^{\pm1}z_2^{\pm1}\right)} I  ^{(T_B)\mathbf{v}}_{\mathbf{z}\mathbf{u}}\\
& I  ^{(A)\mathbf{z}}_{\delta\alpha}\cdot I  ^{(T_A)\mathbf{v}}_{\mathbf{z}\mathbf{u}}=\kappa\oint\frac{dz_2}{4\pi\iu z_2}\delta\left(z_2,s;\frac{1}{\beta\gamma}\right)\frac{\Gamma_e\left(t\beta\gamma\left(\frac{z_1\gamma}{\beta}\right)^{\pm1}s^{\pm1}\right)}{\Gamma_e\left(t\left(\frac{z_1\gamma}{\beta}\right)^{\pm1}z_2^{\pm1}\right)} I  ^{(T_A)\mathbf{v}}_{\mathbf{z}\mathbf{u}}\,,
\end{aligned}
\end{equation}
where we defined $s=\frac{\delta}{\alpha}$ and $z_1=\alpha\delta$ and the function $\delta(x,y;T)$ is defined in \eqref{eqn:deltafac}. 
We now need to apply the Spiridonov-Warnaar inversion formula \cite{2004math11044S} to invert the above integrals. The inverse formula is reviewed and explained in Appendix \ref{sec:appB} and in the form that we will use it is presented in equation (\ref{eqn:SWinversion}).

\subsection{$T_B$ Theory}
Applying \eqref{eqn:SWinversion} the index for the $T_B$ theory is
\begin{equation}\label{eqn:TBindex}
 I  ^{(T_B)\mathbf{v}}_{\mathbf{z}\mathbf{u}}=\Gamma_e\left(t(\gamma\beta z_1)^{\pm1}z_2^{\pm1}\right)\kappa\oint_{C_{z_2}}\frac{ds}{4\pi\iu s}\delta\left(s,z_2;\frac{\gamma}{\beta}\right)\frac{ I  \indices{_{\mathbf{u}\sqrt{sz_1}\sqrt{\frac{z_1}{s}}}^{\mathbf{v}}}}{\Gamma_e\left(t\frac{\gamma}{\beta}(\beta\gamma z_1)^{\pm1}s^{\pm1}\right)}\,,
\end{equation}
where $C_{z_2}$ is a deformation of the unit circle which encloses $s=\frac{\gamma}{\beta}z_2^{\pm1}$ and excludes $s=\frac{\beta}{\gamma}z_2^{\pm1}$.
The same expression was also given in (A.1) and (5.20) of \cite{Razamat:2016dpl} so we will not analyse the full expression further. But rather let us focus on the Hall-Littlewood limit $\frac{p}{\sqrt{t}},\frac{q}{\sqrt{t}}\to0$ of \eqref{eqn:TBindex}. We know that this limit is well defined for the original Lagrangian four punctured sphere theory. We can also see from Table \ref{tab:tailB} that each letter for the tail has $\delta_{1\pm}\geq0$, meaning that the limit is also well defined for the $B$-type tail theory; implying that the Hall-Littlewood limit of the index for the $T_B$ theory is also expected to be well defined. In this limit the three punctured sphere index becomes
\begin{equation}\label{eqn:HLTBindex}
\HL^{(T_B)\mathbf{v}}_{\mathbf{z}\mathbf{u}}=\frac{\left(1-\frac{\gamma^2}{\beta^2}\right)}{\left(1-\tau^2(\gamma\beta z_1)^{\pm1}z_2^{\pm1}\right)}\oint_{C_{z_2}}\frac{ds}{4\pi\iu s}\left(1-s^{\pm2}\right)\frac{\left(1-\tau^2\frac{\gamma}{\beta}(\beta\gamma z_1)^{\pm1}s^{\pm1}\right)}{\left(1-\frac{\gamma}{\beta}s^{\pm1}z_2^{\pm1}\right)}\HL\indices{_{{\mathbf{u}}{\sqrt{sz_1}}{\sqrt{\frac{z_1}{s}}}}^{\mathbf{v}}}
\end{equation}
where $\HL\indices{_{{\mathbf{u}}{\sqrt{sz_1}}{\sqrt{\frac{z_1}{s}}}}^{\mathbf{v}}}$ is given in \eqref{eqn:4puncHL}. The global symmetry of this theory does not enhance from the expected one and is given by $\mathfrak{g}^{(T_B)}=\mathfrak{su}(2)_{\mathbf{z}}^2\oplus \mathfrak{su}(2)_{\mathbf{v}}^2\oplus \mathfrak{su}(2)_{\mathbf{u}}^2\oplus \mathfrak{u}(1)_{\gamma}\oplus \mathfrak{u}(1)_{\beta}\oplus \mathfrak{u}(1)_t$. In expansion around $\tau=0$ and $\frac{\gamma}{\beta}=0$
\begin{equation}
\HL^{(T_B)\mathbf{v}}_{\mathbf{z}\mathbf{u}}=1+P_1\tau^2+P_2\tau^4+\mathcal{O}(\tau^6,\frac{\gamma^6}{\beta^6})
\end{equation}
with the coefficient of $\tau^2$ being
\begin{equation}
\begin{aligned}\label{eqn:P1TB}
P_1=&f_1^-([0;0;0;0;1;1]+[0;0;1;1;0;0])+f_1^+ [1;1;0;0;0;0]\\
&+\gamma ^2[0;1;1;0;1;0]+[1;0;0;1;1;0]+[1;0;1;0;0;1]+\frac{1}{\beta^2}[0;1;0;1;0;1] \, \ ,
\end{aligned}
\end{equation}
where we write the character of $SU(2)_{z_1}\otimes SU(2)_{z_2}\otimes SU(2)_{v_1}\otimes SU(2)_{v_2}\otimes SU(2)_{u_1}\otimes SU(2)_{u_2}$ as $[n_1;n_2;n_3;n_4;n_5;n_6]$. In order to condense the notation we also defined $f^{\pm}_n\equiv\sum_{i=0}^{n}\gamma^{n-2i}\beta^{\pm (n-2i)}$. While the coefficient of $\tau^4$ being
\begingroup
\allowdisplaybreaks
\begin{equation}\label{eqn:P2TB}
\begin{aligned}
P_2=&1+\frac{\beta ^2}{\gamma ^2}+f^-_2([0;0;0;0;2;2]+[0;0;2;2;0;0])+f^+_2[2;2;0;0;0;0]\\
&+\left(f^-_2+1\right)
   [0;0;1;1;1;1]+f_1^+f_1^-([1;1;1;1;0;0]+[1;1;0;0;1;1])\\
 &+f_1^-([1;0;1;0;1;0] +[1;0;0;1;0;1])+f_1^+([0;1;1;0;0;1]+red[0;1;0;1;1;0])\\
   &+[2;0;1;1;1;1]+\frac{\gamma ^2}{\beta ^2}[0;2;1;1;1;1]+\frac{1}{\beta ^2}[1;1;1;1;0;2]+\gamma ^2[1;1;1;1;2;0]\\
   &+\gamma ^2[1;1;2;0;1;1]+\frac{1}{\beta ^2}[1;1;0;2;1;1]+\gamma^2f_1^-([0;1;2;1;1;0]+[0;1;1;0;2;1])\\
   &+f_1^- ([1;0;2;1;0;1]+[1;0;1;0;1;2]+[1;0;0;1;2;1]+[1;0;1;2;1;0])\\
   &+\frac{1}{\beta^2}f_1^-([0;1;1;2;0;1]+[0;1;0;1;1;2])+f_1^+ ([2;1;1;0;0;1]+[2;1;0;1;1;0])\\
&+\gamma^2f_1^+[1;2;1;0;1;0]+\frac{1}{\beta^2}f_1^+[1;2;0;1;0;1]+\gamma ^4[0;2;2;0;2;0]+[2;0;2;0;0;2]\\
   &+\frac{1}{\beta ^4}[0;2;0;2;0;2]+\left([2;0;0;0;0;0]+\text{permutations}\right)
\end{aligned}
\end{equation}
\endgroup
 Note the appearance of the Mesonic generators $M^{u}_{\pm},M^{v}_{\pm},M^{z}_{\pm}$ in the first line of \eqref{eqn:P1TB}. We also have additional generators, in the second line of \eqref{eqn:P1TB}, in trifundamental representations of the corresponding $\mathfrak{su}(2)^{\oplus3}$'s. These are the $B$-operators reported in the last four rows of Table \ref{tab:HBOpsTB}. 
The conformal R-symmetry of the $T_B$ theory is \cite{Razamat:2016dpl}
\begin{equation}
r_c=r+0.0985(q_{\gamma}-q_{\beta})-0.043523q_t\,,
\end{equation}
where $r$ denotes the free R-symmetry, for the Hall-Littlewood index this can be easily computed by using $r=-2j_2+\frac{4}{3}q_t$ from \eqref{eqn:HL}.
Operators contributing to the Hall-Littlewood limit of the index therefore have conformal dimension
\begin{equation}
E_c=2j_2+\frac{3}{2}r_c=-j_2+0.14775(q_{\gamma}-q_{\beta})+1.93472q_t\,.
\end{equation}
Recall for unitary that $E_c\geq1$, this implies that the (bosonic) operators with $q_t=1$ appearing in \eqref{eqn:P1TB} must have $j_2=0$. This in particular means that they all are the top components of $\overline{\mathcal{B}}_{r_c,(0,0)}$ multiplets and therefore are Higgs branch operators, in the definition of Section \ref{sec:hcbc}. Therefore, the Higgs branch of the $T_B$ theory contains at least the operators listed in Table \ref{tab:HBOpsTB}. It is tempting to conjecture that Table \ref{tab:HBOpsTB} is actually the complete list of single trace operators on the Higgs branch. Expansion of the Plethystic Logarithm $\PLog\HL^{(T_B)}=1+L_1\tau^2+L_2\tau^4+\dots$, $P_1=L_1$ shows that every coefficient in $L_2$ is negative implying those operators are all fermionic and there are no new bosonic operators appearing at order $\tau^4$. This of course does not rule out the possibility of the existence of additional bosonic operators that are cancelled by fermionic operators due to the factor $(-1)^F$. We can perform a similar analysis for $P_2$ coefficient. It turns out that all the operators, at this order of the expansion, are multitrace bosonic operators.


\begin{table}
\centering
\begin{tabular}{|c |c|c|c|c|c|c|c|c|c| c|} 
 \hline
 &$E_c$&$\mathfrak{su}(2)_{z_1}$&$\mathfrak{su}(2)_{z_2}$&$\mathfrak{su}(2)_{v_1}$&$\mathfrak{su}(2)_{v_2}$&$\mathfrak{su}(2)_{u_1}$&$\mathfrak{su}(2)_{u_2}$&$q_{\gamma}$&$q_{\beta}$&$q_{t}$\\
  \hline\hline
$M^{u}_{+}$&$2.23022$&$\mathbf{1}$&$\mathbf{1}$&$\mathbf{1}$&$\mathbf{1}$&$\mathbf{2}$&$\mathbf{2}$&$+1$&$-1$&$1$\\
$M^{u}_{-}$&$1.63922$&$\mathbf{1}$&$\mathbf{1}$&$\mathbf{1}$&$\mathbf{1}$&$\mathbf{2}$&$\mathbf{2}$&$-1$&$+1$&$1$\\
$M^{v}_{+}$&$2.23022$&$\mathbf{1}$&$\mathbf{1}$&$\mathbf{2}$&$\mathbf{2}$&$\mathbf{1}$&$\mathbf{1}$&$+1$&$-1$&$1$\\
$M^{v}_{-}$&$1.63922$&$\mathbf{1}$&$\mathbf{1}$&$\mathbf{2}$&$\mathbf{2}$&$\mathbf{1}$&$\mathbf{1}$&$-1$&$+1$&$1$\\
$M^{z}_{\pm}$&$1.93472$&$\mathbf{2}$&$\mathbf{2}$&$\mathbf{1}$&$\mathbf{1}$&$\mathbf{1}$&$\mathbf{1}$&$\pm1$&$\pm1$&$1$\\
$B_{211}$&$2.23022$&$\mathbf{1}$&$\mathbf{2}$&$\mathbf{2}$&$\mathbf{1}$&$\mathbf{2}$&$\mathbf{1}$&$2$&$0$&$1$\\
$B_{121}$&$1.93472$&$\mathbf{2}$&$\mathbf{1}$&$\mathbf{1}$&$\mathbf{2}$&$\mathbf{2}$&$\mathbf{1}$&$0$&$0$&$1$\\
$B_{112}$&$1.93472$&$\mathbf{2}$&$\mathbf{1}$&$\mathbf{2}$&$\mathbf{1}$&$\mathbf{1}$&$\mathbf{2}$&$0$&$0$&$1$\\
$B_{222}$&$2.23022$&$\mathbf{1}$&$\mathbf{2}$&$\mathbf{1}$&$\mathbf{2}$&$\mathbf{1}$&$\mathbf{2}$&$0$&$-2$&$1$\\\hline
\end{tabular}
\caption{\textit{Higgs branch operators of the $T_B$ SCFT that appear at the order $\tau^2$ of the expansion.}}
\label{tab:HBOpsTB}
\end{table}
 
We can consider an unrefined $\mathbf{z}=\mathbf{u}=\mathbf{v}=\mathbf{1}$, $\gamma\beta=1$ limit. In this limit we were able to cast $\HL\indices{_{{\mathbf{1}}{\sqrt{s}}{\sqrt{\frac{1}{s}}}}^{\mathbf{1}}}$ into a relatively closed form and the three punctured sphere index becomes
\begin{equation}
\left.\HL^{(T_B)\mathbf{1}}_{\mathbf{1}\mathbf{1}}\right|_{\gamma\beta=1}=\oint_{C_1}\frac{ds}{2\pi\iu}\frac{P(s,\gamma\beta^{-1},\tau)}{\left(\frac{\beta}{\gamma}-s\right)^2 \left( \frac{s\beta}{\gamma}-1\right)^2 \left(s-\tau ^2\right)^6 \left(s \tau ^2-1\right)^6 \left( \frac{\tau ^2\beta}{\gamma}-s\right)^2 \left(\frac{s \tau ^2\beta}{\gamma}-1\right)^2 }
\end{equation}
where $P(s,\gamma\beta^{-1},\tau)$ is a polynomial in $s$ of degree $17$. We then have to collect the residues of the poles in the interior of $C_1$; located at $s=\frac{\gamma}{\beta},\tau^2,\frac{\tau^2\beta}{\gamma}$. The final result is given by
\begin{equation}
\left.\HL^{(T_B)\mathbf{1}}_{\mathbf{1}\mathbf{1}}\right|_{\gamma\beta=1}=\frac{\tau^{19}\gamma^2}{\beta^2}\frac{Q_B\left(\frac{\beta}{\gamma},\frac{1}{\tau}\right)-Q_B\left(\frac{\gamma}{\beta},\tau\right)}{\left(1-\frac{\beta}{\gamma}\tau^2\right)^{7}\left(1-\frac{\gamma}{\beta}\tau^2\right)^{9}\left(1-\tau^2\right)^{11}}\,,
\end{equation}
where $Q_B(\gamma\beta^{-1},\tau)$ is a degree $19$ polynomial in $\tau$. For brevity we relegate the full expression of $Q_B(\gamma\beta^{-1},\tau)$ in  \eqref{eqn:Qpolyn} to Appendix \ref{app:3puncHL} and quote only the result for $\gamma=\beta=1$ 
\begin{equation}
\begin{aligned}\label{eqn:gb1limHL}
\left.\HL^{(T_B)\mathbf{1}}_{\mathbf{1}\mathbf{1}}\right|_{\gamma=\beta=1}=\frac{1}{(1-\tau^2)^{18}}&\left(\tau ^{20}+38 \tau ^{18}+474 \tau ^{16}+2582 \tau ^{14}+6895 \tau ^{12}\right.\\
&\left.+9516 \tau ^{10}+6895 \tau ^8+2582 \tau ^6+474 \tau ^4+38 \tau ^2+1\right)\,.
\end{aligned}
\end{equation}

\smallskip

Finally we conclude this Section by computing the Coulomb index for the $T_B$ theory, i.e. we consider the limit ($p,q,t\to0$, $\frac{pq}{t}\to T$), which was originally given in \cite{Razamat:2018zus}. We have that
\begin{equation}\label{eqn:TBindexCoulomb}
 {I^C}  ^{(T_B)\mathbf{v}}_{\mathbf{z}\mathbf{u}}=\left(1-T(\gamma\beta z_1)^{\pm1}z_2^{\pm1}\right)\oint_{C_{z_2}}\frac{ds}{4\pi\iu s}\frac{\left(1-\frac{\gamma^2}{\beta^2}\right)(1-s^{\pm2})}{\left(1-\frac{\gamma}{\beta}z_2^{\pm1}s^{\pm1}\right)}\frac{ {I^C}  \indices{_{\mathbf{u}\sqrt{sz_1}\sqrt{\frac{z_1}{s}}}^{\mathbf{v}}}}{\left(1-T\frac{\beta}{\gamma}(\beta\gamma z_1)^{\pm1}s^{\pm1}\right)}\,,
\end{equation}
The Coulomb index for the 4-punctured sphere was given in \eqref{eqn:hscbhfn} and is given by
\begin{equation}
{I^C}  \indices{_{\mathbf{u}\sqrt{sz_1}\sqrt{\frac{z_1}{s}}}^{\mathbf{v}}}=\PE\left[\left(\gamma^2\beta^2+\frac{1}{\gamma^2\beta^2}+1\right)T^2\right]\,,
\end{equation}
notice that it is independent of $s$ and therefore we can easily compute the above integral by collecting the residues of the poles inside $C_{z_2}$ at $s=T\beta^2 z_1,\frac{T}{\gamma^2 z_1},\frac{\gamma}{\beta}z_2^{\pm1}$. The final result is
\begin{equation}
 {I^C}  ^{(T_B)\mathbf{v}}_{\mathbf{z}\mathbf{u}}=\PE\left[\left(\gamma^2\beta^2+\frac{1}{\gamma^2\beta^2}+\frac{\beta^2}{\gamma^2}\right)T^2\right]\,.
 \end{equation}
This result is in agreement with the row number six of Table 1 in \cite{Razamat:2018zus}. We conclude that the Coulomb branch for the $T_B$ theory is freely generated and it is generated by three bosonic operators. In particular the first operator has charge $+2$ under both $\mathfrak{u}(1)_{\beta }$ and $\mathfrak{u}(1)_{\gamma}$, the second has charge $-2$ under both $\mathfrak{u}(1)_{\beta }$ and $\mathfrak{u}(1)_{\gamma}$, while the third has charge $+2$ under $\mathfrak{u}(1)_{\beta}$ and charge $-2$ under $\mathfrak{u}(1)_{\gamma}$.   

\subsection{$T_A$ Theory}
We now consider the $T_A$ theory. Applying the Spiridonov-Warnaar inversion formula \eqref{eqn:SWinversion} the index for the $T_A$ theory is
\begin{equation}\label{eqn:TAindex}
 I  ^{(T_A)\mathbf{v}}_{\mathbf{z}\mathbf{u}}=\Gamma_e\left(t \left(\frac{\gamma z_1}{\beta}\right)^{\pm1}z_2^{\pm1}\right)\kappa\oint_{C_{z_2}}\frac{ds}{4\pi\iu s}\delta\left(s,z_2;\beta\gamma\right)\frac{ I  \indices{_{\mathbf{u}\sqrt{sz_1}\sqrt{\frac{z_1}{s}}}^{\mathbf{v}}}}{\Gamma_e\left(t\beta\gamma\left(\frac{z_1\gamma}{\beta}\right)^{\pm1}s^{\pm1}\right)}\,,
\end{equation}
where now $C_{z_2}$ encloses $s=\beta\gamma z_2^{\pm1}$ and excludes $s=\frac{1}{\beta\gamma}z_2^{\pm1}$.
The same expression was also given in (B.3) and (5.17) of \cite{Razamat:2016dpl}. As pointed out in \cite{Razamat:2016dpl} the flavour symmetry enhances $\mathfrak{su}(2)^2_{z}\oplus \mathfrak{su}(2)^2_{v}\oplus \mathfrak{su}(2)^2_{u}\oplus \mathfrak{u}(1)_{\gamma}\oplus \mathfrak{u}(1)_{\beta}\oplus \mathfrak{u}(1)_{t}\to \mathfrak{g}^{(T_A)}=\mathfrak{so}(8)_{w}\oplus \mathfrak{su}(2)_{z_2}\oplus \mathfrak{su}(2)_{v_1}\oplus \mathfrak{su}(2)_{u_1}\oplus \mathfrak{u}(1)_{\gamma\beta}\oplus \mathfrak{u}(1)_{t}$. The enhancement to $\mathfrak{so}(8)$ is made manifest by identifying $\mathbf{w}=(\frac{\gamma}{\beta}u_2 , \frac{\beta}{\gamma}\frac{1}{u_2}, \frac{\beta}{\gamma}u_2 , \frac{\gamma}{\beta}\frac{1}{u_2} , v_2z_1, \frac{1}{v_2z_1}, \frac{v_2}{z_1},\frac{z_1}{v_2})$ as the $SO(8)$ fugacities. We can again take the Hall-Littlewood limit. 
In an expansion around $\tau=0$, $\beta\gamma=0$ the Hall-Littlewood index reads
\begin{equation}
\begin{aligned}\label{eqn:HLTA}
\HL^{(T_A)\mathbf{v}}_{\mathbf{z}\mathbf{u}}=&1+\Big[[1,0,0,0;0;0;1]+[0,0,1,0;0;1;0]+[0,0,0,1;1;0;0]\\
&+\gamma^2\beta^2[0,0,0,0;1;1;1]\Big]\tau^2+\Big[\gamma^4\beta^4[0,0,0,0;2;2;2]\\
&+\gamma^2\beta^2([1,0,0,0;1;1;2]+[0,0,1,0;1;2;1]+[0,0,0,1;2;1;1]-1)\\
&+[2,0,0,0;0;0;2]+[0,0,2,0;0;2;0]+[0,0,0,2;2,0,0]\\
&+[1,0,1,0;0;1;1]+[1,0,0,1;1;0;1]+[0,0,1,1;1;1;0]\\
&+[0,1,0,0;0;0;0]\Big]\tau^4+\mathcal{O}(\tau^6,\beta^6\gamma^6)
\end{aligned}
\end{equation} 
where we denote the character of $SO(8)_{\mathbf{w}}\otimes SU(2)_{z_2}\otimes SU(2)_{v_1}\otimes SU(2)_{u_1}$ \\ by $[d_1,d_2,d_3,d_4;d_5;d_6;d_7]$.
We observe that all the coefficients appearing at the order $\tau^4$ of the expansion correspond to bosonic multitrace operators, that is to say to product of single trace operators, with the only exception of the term $-\gamma^2\beta^2\tau^4$.
We can see from Table \ref{tab:shorts}, Table \ref{tab:Conserved} and equations \eqref{eqn:HLmultiind1}-\eqref{eqn:HLmultiind2} that the corresponding multiplet (since the $T_A$ theory is interacting it can not have higher spin free fields) must be a $\overline{\mathcal{C}}_{\frac{2}{3}(0,0)}$ multiplet with $\mathfrak{u}(1)_t$ charge $q_t=2$ and $q_{\beta}=q_{\gamma}=2$. In particular the negative coefficient appearing in the expansion of the Hall-Littlewood index means that the Hall-Littlewood index for generic genus zero class $\mathcal{S}_k$ theories is, in general, \textit{not} equal to the Hilbert series for its Higgs branch (as per the definition of Section \ref{sec:hcbc}) whose expansion coefficients must all be positive.  

Let us explain why we get this minus sign from multiplets recombination. The expansion of the Hall-Littlewood limit of the index of the four punctured sphere theory has all positive coefficients, in particular it does not contain a $\overline{\mathcal{C}}_{\frac{2}{3}(0,0)}$ multiplet with $q_t=q_{\beta}=q_{\gamma}=2$. This could mean that the $\overline{\mathcal{C}}_{\frac{2}{3}(0,0)}$ multiplet may be ``recombined'' in a long multiplet $\mathcal{A}$, whose Hall-Litllewood index $\mathcal{I}^{HL}(\mathcal{A})$ is equal to zero. As we approach the complete decoupling limit where the $SU(2)_{z_2}$ is ungauged the four punctured sphere theory decomposes into the $T_A$ SCFT plus the $A$-type quiver tail theory. As we are taking this limit a long multiplet $\mathcal{A}^{3}_{\frac{2}{3},(0,0)}$ of the four-punctured sphere theory decomposes via the recombination rule \eqref{eqn:recomb1}
\begin{equation}
\label{eq:rec}
\mathcal{A}^{3}_{\frac{2}{3},(0,0)}\iso\overline{\mathcal{C}}_{\frac{2}{3},(0,0)}\oplus\overline{\mathcal{B}}_{\frac{8}{3},(0,0)} \, \ .
\end{equation}
Using the relations \eqref{eqn:HLmultiind1}-\eqref{eqn:HLmultiind2} we observe that the Hall-Littlewood indices of the $\overline{\mathcal{C}}_{\frac{2}{3},(0,0)}$ multiplet and of the $\overline{\mathcal{B}}_{\frac{8}{3},(0,0)}$ multiplet satisfy
\begin{equation}
\mathcal{I}^{HL}\left(\overline{\mathcal{C}}_{\frac{2}{3},(0,0)}\right) = - \mathcal{I}^{HL}\left(\overline{\mathcal{B}}_{\frac{8}{3},(0,0)}\right)  \, \ ,
\end{equation}
therefore considering the Hall-Littlewood index of the recombination rule (\ref{eq:rec}) we get
\begin{equation}
\mathcal{I}^{HL}\left(\mathcal{A}^{3}_{\frac{2}{3},(0,0)}\right) = \mathcal{I}^{HL}\left(\overline{\mathcal{C}}_{\frac{2}{3},(0,0)}\right) + \mathcal{I}^{HL}\left(\overline{\mathcal{B}}_{\frac{8}{3},(0,0)}\right) =0 \, \ ,
\end{equation}
as it must be since $\mathcal{A}^{3}_{\frac{2}{3},(0,0)}$ is a long multiplet.
Therefore as long as the $\overline{\mathcal{C}}_{\frac{2}{3},(0,0)}$ and the $\overline{\mathcal{B}}_{\frac{8}{3},(0,0)}$ multiplet recombine their contribution to the HL index mutually cancels.

However, as we have seen, the $\overline{\mathcal{C}}_{\frac{2}{3}(0,0)}$ multiplet appear in the HL index of the $T_A$ theory. On the other hand, we are then led to expect that, the $\overline{\mathcal{B}}_{\frac{8}{3},(0,0)}$ multiplet with $q_t=q_{\beta}=q_{\gamma}=2$ lives in the $A$-type quiver tail theory. As a matter of fact, going through the list of Hall-Littlewood ($\delta_{1\pm}=0$) operators with $(-1)^F=+1$, $q_t=q_{\beta}=q_{\gamma}=2$ in the tail theory Table \ref{tab:tailA}, we obtain an operator in the $\mathbf{2}\otimes\mathbf{2}\iso\mathbf{3}\oplus\mathbf{1}$ of the enhanced $SU(2)_{s=\delta/\alpha}$ 
\begin{equation}
B_{IJ}:=\begin{pmatrix}
B_{1,++}B_{1,+-}&B_{1,++}B_{1,--}\\
B_{1,-+}B_{1,+-}&B_{1,--}B_{1,+-}
\end{pmatrix}_{IJ}=B_{(IJ)}+B_{[IJ]}
\end{equation}
with $I,J=1,2$ $SU(2)_{s}$ indices. Since the $T_A$ theory contains only $SU(2)_s$ singlets we are instructed to take the $\mathbf{1}$ in the decomposition and we conclude that the operator 
\begin{equation}
B_{[IJ]}=\frac{1}{2}\left(B_{1,++}B_{1,--}- B_{1,-+}B_{1,+-}\right)\epsilon_{IJ}
\end{equation}
is the top component of the $\overline{\mathcal{B}}_{\frac{8}{3},(0,0)}$ multiplet in the above recombination rule.

Therefore we conclude that the origin of the minus sign in the expansion (\ref{eqn:HLTA}) is due to fact that the long multiplet $\mathcal{A}^{3}_{\frac{2}{3},(0,0)}$ splits into two short multiplets that do not recombine anymore. Due to this, the negative contribution coming from the $\overline{\mathcal{C}}_{\frac{2}{3}(0,0)}$ multiplet, i.e. the term $-\gamma^2\beta^2\tau^4$, is not cancelled by the positive contribution arising from the index of the $\overline{\mathcal{B}}_{\frac{8}{3}(0,0)}$ multiplet of the $A$-type theory. It is then natural to expect that, also at higher orders of the expansions, new negative coefficients will appear. A similar feature has been observed in a 4d $\mathcal{N}=2$ context in \cite{Gadde:2009dj}. 

The conformal R-symmetry of the $T_A$ theory is \cite{Razamat:2016dpl}
\begin{equation}
r_c=r+0.0689(q_{\gamma}+q_{\beta})-0.044777q_t\,.
\end{equation}
Therefore, operators contributing to the Hall-Littlewood limit of the index have conformal energy
\begin{equation}
E_c=2j_2+\frac{3}{2}r_c=-j_2 + 1.93283 q_t+ 0.10335 (q_{\gamma}+q_{\beta})\,.
\end{equation}
\begin{table}
\centering
\begin{tabular}{|c |c|c|c|c|c|c|c|} 
 \hline
 &$E_c$&$\mathfrak{so}(8)_{\mathbf{w}}$&$\mathfrak{su}(2)_{z_2}$&$\mathfrak{su}(2)_{v_1}$&$\mathfrak{su}(2)_{u_1}$&$q_{\gamma}+q_{\beta}$&$q_{t}$\\\hline\hline
$M^{v}$&$1.93283$&$\mathbf{8}_v$&$\mathbf{1}$&$\mathbf{1}$&$\mathbf{2}$&$0$&$1$\\
$M^{c}$&$1.93283$&$\mathbf{8}_c$&$\mathbf{1}$&$\mathbf{2}$&$\mathbf{1}$&$0$&$1$\\
$M^{s}$&$1.93283$&$\mathbf{8}_s$&$\mathbf{2}$&$\mathbf{1}$&$\mathbf{1}$&$0$&$1$\\
$B$&$2.34623$&$\mathbf{1}$&$\mathbf{2}$&$\mathbf{2}$&$\mathbf{2}$&$4$&$1$\\\hline
\end{tabular}
\caption{\textit{Higgs branch operators of the $T_A$ SCFT that appear at the order $\tau^2$ of the expansion.}}
\label{tab:HBOpsTA}
\end{table}
By the same arguments as the ones made for the $T_B$ theory the operators with $q_t=1$ appearing in the expansion \eqref{eqn:HLTA} have $j_2=0$. Therefore the operators of Table \ref{tab:HBOpsTA} comprise, at least a subset of, the possible operators on the Higgs branch of the $T_A$ SCFT. In the notation of Table \ref{tab:HBOpsTB} we have combined $M^v=\{M_+^u,M_-^u,B_{121}\}$, $M^c=\{M_+^v,M_-^v,B_{112}\}$, $M^s=\{M_+^z,M_-^z,B_{222}\}$ and $B=\{B_{211}\}$. The expansion of $\PLog\HL^{(T_A)}=\tilde{L}_1\tau^2+\tilde{L}_2\tau^4+\dots$ with the coefficients of $\tilde{L}_2$ all being negative again does not rule out the possibility that Table \ref{tab:HBOpsTA} is in fact the complete list of Higgs branch generators for the $T_A$ theory and that new generators do not appear at higher orders.
We were able to obtain a closed expression in the unrefined $\mathbf{z}=\mathbf{v}=\mathbf{u}=\mathbf{1}$, $\gamma\beta^{-1}=1$ limit:
\begin{equation}\label{eqn:HLTAunref}
\begin{aligned}
\left.\HL^{(T_A)\mathbf{1}}_{\mathbf{1}\mathbf{1}}\right|_{\frac{\gamma}{\beta}=1}=&\oint_{C_1}\frac{\widetilde{P}(s,\gamma\beta,\tau)ds}{(s-\beta  \gamma )^2 (\beta  \gamma  s-1)^2 \left(\beta  \gamma  s-\tau ^2\right)^5 \left(s \tau ^2-\beta  \gamma \right)^5 \left(s-\beta  \gamma  \tau ^2\right)^3 \left(\beta  \gamma  s \tau^2-1\right)^3}\\
=&\beta^5\gamma^5\tau^{15}\frac{Q_A(\tau,\beta\gamma)-Q_A(\tau^{-1},\beta^{-1}\gamma^{-1})}{(1-\tau^2)^{19}(1-\gamma^2\beta^2\tau^2)^4}
\end{aligned}
\end{equation}
where $\widetilde{P}(s,\gamma\beta,\tau)$ is a polynomial in $s$ and in the second line we took the residues at $s=\beta\gamma,(\beta\gamma)^{\pm1}\tau^2$ and $Q_A(\tau,\beta\gamma)$ is a polynomial of degree $15$ in $\tau$. We list the full expression for it in \eqref{eqn:QApolyn} and quote here only the result for $\gamma=\beta=1$ which reads
\begin{equation}
\begin{aligned}
\left.\HL^{(T_A)\mathbf{1}}_{\mathbf{1}\mathbf{1}}\right|_{\gamma=\beta=1}=&\frac{1}{(1 -\tau^2)^{18}}\left(\tau ^{20}+38 \tau ^{18}+474 \tau ^{16}+2582 \tau ^{14}+6895 \tau ^{12}\right.\\&\quad\quad\quad\quad\left.+9516 \tau ^{10}+6895 \tau ^8+2582 \tau ^6+474 \tau ^4+38 \tau ^2+1\right)\,,
\end{aligned}
\end{equation}
again, note the palindromic structure of the numerator.
Also note that in the fully unrefined limit $\left.\HL^{(T_A)\mathbf{1}}_{\mathbf{1}\mathbf{1}}\right|_{\gamma=\beta=1}=\left.\HL^{(T_B)\mathbf{1}}_{\mathbf{1}\mathbf{1}}\right|_{\gamma=\beta=1}$. This is expected since the $T_A$ \& $T_B$ SCFTs differ only by choices for fluxes for $U(1)_{\gamma}$ and $U(1)_{\beta}$. This equality no longer holds with refinement turned on, since $T_A$ and $T_B$ have different global symmetry algebras.

The Coulomb index for the $T_A$ theory can also be computed, the computation is identical to the $T_B$ case and it reads
\begin{equation}
{I^C}  ^{(T_A)\mathbf{v}}_{\mathbf{z}\mathbf{u}}=\PE\left[\left(\beta^2\gamma^2+\frac{2}{\beta^2\gamma^2}\right)T^2\right] \, \ .
\end{equation}
The above result is in agreement with row number five of table 1 in \cite{Razamat:2018zus}. We observe that the Coulomb branch is generated by three bosonic operators without relations among them. In particular the first operator has charge $+2$ under both $\mathfrak{u}(1)_{\beta }$ and $\mathfrak{u}(1)_{\gamma}$, while the  second and the third have charge $-2$ under both $\mathfrak{u}(1)_{\beta }$ and $\mathfrak{u}(1)_{\gamma}$.

\section{Outlook and Open Problems}
\label{sec:conclusions}

In this paper we studied several aspects of Class $\mathcal{S}_k$ SCFTs. We review the notion of moduli space of vacua for this class of theories and we provide a consistent definition of its Coulomb branch and its Higgs branch. We then introduce the mathematical tools that we use for the characterization of these theories, namely the Higgs branch and Coulomb branch Hilbert series and the Superconformal index along with some of its limits.

First, we analyse Lagrangian Class $S_k$ theories  at genus $g=0$. For this subclass of theories we prove the equality between the Hall-Littlewood limit of the index and the corresponding Higgs branch Hilbert series.
Then, we consider Lagrangian Class $S_k$ theories at genus $g=1$. We compute the Higgs branch Hilbert series and the Hall-Littlewood limit of the index for some of these theories. As in the case of $\mathcal{N}=2$ class $\mathcal{S}$ theories at genus $g=1$ the Hall-Littlewood limit of the index and the Higgs branch Hilbert series are different. The reason is that the Higgs branch Hilbert series counts only  $\mathcal{\overline{B}}_{r,(0,0)}$ and $\mathcal{\overline{D}}_{(0,0)}$ multiplets, while the Hall-Littlewood limit of the index, as discussed in Appendix \ref{app:SCAreps}, counts also $\mathcal{\overline{C}}_{r,(0,j_2)}$, $\mathcal{\hat{C}}_{(0,j_2)}$ and $\mathcal{D}^{(j_2+1)}_{(j_2,0)}$ multipltes.

We show that the Hilbert series of the mesonic branch of the theory arising on a stack of $N$-D3 branes probing a $\mathbb{C}^3/(\mathbb{Z}_k \times \mathbb{Z}_l)$ singularity in the limit $l,k \rightarrow \infty$ reproduces the 1/2-BPS partition function of the $(1,1)$ LST. This equality provides an important check of the deconstruction proposal for the (1,1) LST.

Finally, we consider the $A_1$ four punctured sphere in class $\mathcal{S}_2$. This theory admits three pants decompositions corresponding to the three distinct $S$-duality frames. In two of these, the $T_A$ and $T_B$ theories,
which are strongly coupled SCFTs appear. Using the Spiridonov-Warnaar inversion formula we compute the Hall-Littlewood and the Coulomb limit of the index for these two theories. We observe that all the coefficients arising in the expansion of the HL index of the $T_B$ theory are positive. On the other hand in the expansion of the HL index of the $T_A$ theory are present negative coefficients. From an index perspective  these negative coefficients arise because a long multiplet hits a unitary bound and splits into short multiplets. Moreover this also implies that in general, for non-Langrangian class $S_k$ theories, the HL limit of the index is not equal to the Higgs branch Hilbert series.

In the future it  would be interesting to study whether it is possible to define a Coulomb branch and a Higgs branch for theories in the Class $\mathcal{S}_{\Gamma}$ \cite{Heckman:2016xdl}.
For $\mathcal{S}_{\Gamma=D,E}$, a direct generalisation of our work is not immediate since there is no analogue of $U(1)_t$. Naively, the  reason is that $A_N$ type singularity has a $U(1)$ isometry whereas $D_N$, $E_N$ do not. For example $\mathbb{C}^2/A_{N}$ has defining equations $xy=w^{N+1}$, we can give $U(1)$ charges $+1,-1,0$ to $x,y,w$, but $\mathbb{C}^2/D_{N+2}$ has $x^2+y^2w=w^{N+1}$ and has empty isometry group.

Finally, an important  question that we didn't address in this paper, is  whether a two dimensional chiral algebra \cite{Beem:2013sza} for the $4d$ $\mathcal{N}=1$ theories discussed in this paper exist. In this correspondence a central role is played by the Schur limit $I^{S}$ of the $\mathcal{N}=1$ index \eqref{eqn:schur}. In Section \ref{sec:sci} we observed that the current $j^{\mu}_{11}$, whose top component is identified with the stress-energy tensor of the corresponding chiral algebra, is not present in $I^{S}$. This fact seems
to lead to the conclusion that such chiral algebra does not have a stress-energy tensor. It has been observed in \cite{Cordova:2017mhb,Cordova:2017ohl,Cordova:2016uwk} that in the context of 4d $\mathcal{N}=2$  SCFT with defects it is possible to construct a chiral algebra without a  stress-energy tensor. Therefore, it would be interesting to  clarify if a similar phenomenon could take place also for the $4d$ $\mathcal{N}=1$ theories considered in this paper.
As a possibly smaller step in this direction it would be interesting  to clarify which of the properties, valid for $\mathcal{N}=2$ 4d Higgs branches, discussed \cite{Niarchos:2018mvl} have a counterpart for the $\mathcal{N}=1$ theories discussed in this paper.

\section*{Acknowledgments}
It is a pleasure to thank Antoine Bourget  for useful discussions, and Shlomo Razamat for not only important discussions but also for reading and commenting on our draft in its final stages.
The work of the authors was partially supported by the DFG via the Emmy
Noether program ``Exact results in Gauge theories'' and the GIF Research Grant I-1515-
303./2019.

\appendix
\section{The Algebra-Geometry Correspondence}
\label{app:ag}
In this Appendix we review  the correspondence between complex algebraic geometry and affine varieties (we refer the interested reader to \cite{Cox:2007} for a more broad introduction to this topic). Given an affine complex variety $\mathcal{V} \subset \mathbb{C}^{n}$, we can associate to it an ideal $\mathbf{I}(\mathcal{V}) \subset \mathbb{C}[x_1,...,x_n]$ using the following map
\begin{equation}
\label{eq:mapv}
\mathbf{I}(\mathcal{V}) = \{ f \in \mathbb{C}[x_1,...,x_n] \mid f(x) = 0 \ \forall \ x \in \mathcal{V} \}\, ,
\end{equation}
on the other hand given an ideal $\mathcal{I} \subset \mathbb{C}[x_1,...x_n]$ we can associate to it a complex variety $\mathbf{V}(\mathcal{I})$ in the following way 
\begin{equation}
\label{eq:mapI}
\mathbf{V}(\mathcal{I}) = \{ x \in \mathbb{C}^{n} \mid f(x) = 0 \ \forall \ f \in \mathcal{I} \}\, .
\end{equation}
Therefore these two maps provide a way to relate ideals and complex affine varieties. However in general this correspondence (more precisely the map \eqref{eq:mapv}) is not one to one. \footnote{For example let's consider the family of distinct ideals $\langle x^n \rangle$ (with $n \in \mathbb{N}$) in $\mathbb{C}$, then it's easy to see that to the map \eqref{eq:mapv} associates to each of them the same variety, namely $\mathbf{V}(x^n) = \{0\}$.} In order understand under which conditions we can get a bijection let us introduce the notion of the \textit{radical} of a given ideal $\mathcal{I} \subset \mathbb{C}[x_1,...,x_n]$, that will be denoted by $\sqrt{\mathcal{I}}$, and it is defined as
\begin{equation}
\sqrt{\mathcal{I}} := \{ f \mid f^m \in \mathcal{I} \ \textrm{for some} \ m \geq 1  \}\, .
\end{equation}
From its definition it follows $\mathcal{I} \subset \sqrt{\mathcal{I}}$ (since $f \in \mathcal{I}$ implies that $f^{1} \in \mathcal{I}$). Moreover the radical of an ideal $\mathcal{I}$ is always a \textit{radical ideal}\footnote{An ideal $\mathcal{J} \subset \mathbb{C}[x_1,...,x_n]$ is \textit{radical} if $f^m \in \mathcal{J}$ for some integer $m \geq 1$ implies that $f \in \mathcal{J}$.}\cite{Cox:2007}. The Hilbert's \textit{Nullstellensatz} theorem states that if $\mathcal{I}$ is an ideal in $\mathbb{C}[x_1,...x_n]$ then
\begin{equation}
\mathbf{I}(\mathbf{V}(\mathcal{I})) = \sqrt{\mathcal{I}}\, .
\end{equation} 
Therefore this theorem and the maps \eqref{eq:mapI}-\eqref{eq:mapv} provide us a one to one correspondence between complex varietes and algebraic quantities given by radical ideals.
This dictionary can be extended reformulating in algebraic terms geometrical problems, see Table \ref{tab:ag}. A particular useful class of ideals is provided by the so called \textit{prime ideals}\footnote{An ideal $\mathcal{I} \subset \mathbb{C}[x_1,...,x_n]$ is \textit{prime} whenever $f,g \in \mathbb{C}[x_1,...x_n]$ and $fg \in \mathcal{I}$, then either $f \in \mathcal{I}$ or $g \in \mathcal{I}$}. It's easy to prove that every prime ideal is also a radical ideal, and that moreover there is a one-to-one correspondence between \textit{irreducible varieties}\footnote{An affine variety $\mathcal{V} \subset \mathbb{C}^{n}$ is said to be \textit{irreducible} if whenever $\mathcal{V}$ is written in the form $\mathcal{V} =\mathcal{V}_1 \cup \mathcal{V}_2$, where $\mathcal{V}_1$ and $V_2$ are affine varietes, then either $\mathcal{V}_1=\mathcal{V}$ or $\mathcal{V}_2=\mathcal{V}$.} and prime ideals \cite{Cox:2007}.

\begin{table}
\center{
\begin{tabular}{ccc}
Algebra & $\leftrightarrow$ & Geometry\\
\hline
Radical ideal $\mathcal{I}$ &  & Affine variety $\mathbf{V}(\mathcal{I})$\\
\hline
Addition of ideals &  & Intersection of varietes\\
$\mathcal{I}+\mathcal{J}$             &   & $\mathbf{V}(\mathcal{I}) \cap \mathbf{V}(\mathcal{J})$ \\

\hline
Product of ideals &  & Union of varietes\\
$\mathcal{I}\mathcal{J}$             &   & $\mathbf{V}(\mathcal{I}) \cup \mathbf{V}(\mathcal{J})$ \\

\hline
Prime ideal $\mathcal{I}$ &  & Affine irreducible variety $\mathbf{V}(\mathcal{I})$\\
\hline
Regular sequence & & Complete intersection\\
\hline
\end{tabular}
}
\caption{\textit{Summary of the algebra-geometry correspondence} \label{tab:ag}.}
\end{table}

\subsection{The F-flat Moduli Space}
\label{app:fterms}
Therefore, using the dictionary summarized in the previous subsection, the problem concerning the characterization of the different branches of the master space $\mathcal{F}$ can be recast into an algebraic problem. As a matter of fact the full information contained in $\mathcal{F}$ considered as a variety can be equivalently encoded in the quotient ring between the ring $\mathcal{R}:=\mathbb{C}[Q_i]$ of all the  polynomials with complex coefficients that can be written starting from the scalar fields $Q_i$ that are taking a VEV  and the ideal $\mathcal{I}$, that is enconding the F-terms constraints. Schematically we can write
\begin{equation}
\mathcal{F} \ \xleftrightarrow{1:1} \  \mathcal{R}/\mathcal{I} \, ,
\end{equation}
Moreover since $\mathcal{R}$ is a \textit{Noetherian ring}\footnote{A ring $\mathcal{R}$ is called \textit{Noetherian} if there is no infinite ascending sequence of left (or right) ideals. Therefore, given any chain of left (or right) ideals,
\begin{equation*}
\mathcal{I}_1 \ \subseteq \ \dotsc \ \mathcal{I}_{k-1} \ \subseteq \ \mathcal{I}_{k} \ \subseteq \ \mathcal{I}_{k+1} \ \dotsc \, ,
\end{equation*}
there exists an $n$ such that
\begin{equation*}
\mathcal{I}_n = \mathcal{I}_{n+1} = \dotsc
\end{equation*}
 }, using the \textit{Lasker-Noether theorem}, we can always decompose and ideal of $\mathcal{R}$ as an irredundant intersection of a finite set $\{\mathcal{J}_i\}$ of primary ideals \cite{Cox:2007}, this procedure is called \textit{primary decomposition}. In particular we can take into account the above ideal $\mathcal{I}$ and we get
\begin{equation}
\label{eq:primarydec}
	\mathcal{I} = \bigcap\limits_{i=1}^{d}\mathcal{J}_{i}\, \ . 
\end{equation}
We can then consider the different radical ideals $\sqrt{\mathcal{J}_i}$ associated to each of the primary ideals in \eqref{eq:primarydec} that 
turn out to be a prime ideals. Therefore using the algebra-geometry dictionary, we have a one to one correspondence  between affine irreducible complex variates and the radical ideals $\sqrt{\mathcal{J}_i}$. This implies that the master space $\mathcal{F}$, considered as a complex variety, can be written as 
\begin{equation}
\label{eq:fdec}
\mathcal{F} = \bigcup_{i=1}^{d} \mathbf{V}(\sqrt{\mathcal{J}_i})\, ,
\end{equation}
in this way the algebraic approach turns out to be very powerful since it provides a systematic way to decompose the F-terms moduli space of a theory into different irreducible branches.

Remarkably we can also establish when the space $\mathcal{F}$ is a complete intersection. It holds the following \cite{stanley1978hilbert}

\begin{theorem}
Given the ring of polynomials $\mathcal{R}=\mathbb{C}[x_1,...x_n]$ and the ideal $\mathcal{I} \subset \mathcal{R}$ then the algebraic variety associated to the quotient ring $\mathcal{R}/\mathcal{I}$ is a complete intersection if and only if $\mathcal{I}$ is generated by a regular sequence of homogeneous polynomials.
\end{theorem}

Therefore if the ideal is generated by a \textit{regular sequence}\footnote{A sequence of non-costant polynomials $P_1,P_2,...P_r$ is said to be \textit{regular} if for all $i=1,...,r$, $P_i$ is not a zero divisor modulo the partial ideal $(P_1,...,P_{i-i})$.} of polynomials then we can use letter counting for the computation of the corresponding Hilbert series. For application of the above theorem in a different context see \cite{Bourget:2017sxr}, we also refer the interested reader to Appendix A of that paper for a more detailed discussion related to this issue. 

\section{Identities and Special Functions}
\label{sec:appB}

The Plethystic exponential of a function $f(x)$, with $f(0)=0$ is defined to be
\begin{equation}\label{eqn:PE}
\PE\left[f(x)\right]:=\exp\left({\sum_{n=1}^{\infty}\frac{f(x^n)}{n}}\right)\,,\quad \PLog[f(x)]=\PE^{-1}[f(x)]=\sum_{n=1}^{\infty}\frac{\mu(n)}{n}\log f(x^n)\,,
\end{equation}
and $\mu(n)$ is the M\"obius $\mu$ function. 
Some basic identities are
\begin{equation}
\PE[t]=\frac{1}{1-t}\,,\quad \PE[-t]=1-t\,,\quad \sum_{n=0}^{N-1}t^n=\frac{1-t^N}{1-t}=\PE[t-t^N]\,.
\end{equation}
The Elliptic Gamma function is defined as
\begin{equation}
\label{eqn:EllGamma}
\Gamma_e(z):=\Gamma(z;p,q)=\prod_{i,j=0}^{\infty}\frac{1-z^{-1}p^{i+1}q^{j+1}}{1-zp^iq^j}=\PE\left[\frac{z-\frac{pq}{z}}{(1-p)(1-q)}\right]\,.
\end{equation}
An obvious, yet important, identity is 
\begin{equation}\label{eqn:gammaid}
\Gamma_e(z)\Gamma_e(pq/z)=1\,.
\end{equation}
We will often use the shorthand notation
\begin{equation}
f(z^{\pm n})\equiv f(z^n)f(z^{-n})\,.
\end{equation}
We use the following notation for the $q$-Pochammer symbols
\begin{equation}
(a;q)_N=\prod_{n=0}^{N}(1-aq^n)\,,\quad (a;q):=(a;q)_{\infty}=\PE\left[\frac{a}{1-q}\right]\,.
\end{equation}
\paragraph{$SU(N)$ characters}
Highest weight, irreducible representations $\mathcal{R}_{(d_1,d_2,\dots,d_{N-1})}$ of $SU(N)$ may be labelled by a Young diagram $\lambda$ of length $\ell(\lambda)=N$ 
\begin{equation}
\lambda=\{\lambda_1,\lambda_2,\dots,\lambda_A,\dots\lambda_{N}|\lambda_A\in\mathbb{N}\cup\{0\}\,,\lambda_1\geq\lambda_2\geq\dots\lambda_A,\dots\geq\lambda_{N-1}\geq\lambda_N=0\}\,,
\end{equation}
where the relations between the Dynkin labels of the representation and the Young diagram is
\begin{equation}\label{eqn:youngdynkin}
d_A=\lambda_A-\lambda_{A+1}\,,\quad \lambda_A=\sum_{i=A}^{N-1}d_i\,.
\end{equation}
The conjugate representation $\overline{\mathcal{R}_{(d_1,d_2,\dots,d_{N-1})}}=\mathcal{R}_{(d_{N-1},d_{N-2},\dots,d_1)}$ is therefore associated to the Young diagram $\overline{\lambda}$ with $\overline{\lambda}_A=\sum_{r=A}^{N-1}d_{N-r}=\sum_{r=A}^{N-1}(\lambda_{N-r}-\lambda_{N-r+1})=\lambda_{1}-\lambda_{N-A+1}$.

The characters for the representation $\mathcal{R}_{(d_1,d_2,\dots,d_{N-1})}$ are given by Schur polynomials
\begin{equation}\label{eqn:SUNChar}
\chi_{(d_1,d_2,\dots,d_{N-1})}(\mathbf{x})=s_{\lambda}(\mathbf{x})=\frac{\det_{AB} x_A^{\lambda_B+N-B}}{\det_{AB} x_A^{N-B}}\,,
\end{equation}
with $\lambda_N=0$ and $\prod_{A=1}^Nx_A=1$. We also often abuse notation and denote these characters simply by their Dynkin labels $\chi_{(d_1,d_2,\dots,d_{N-1})}(\mathbf{x})\equiv[d_1,d_2,\dots,d_{N-1}]$. The representation labelled by $\lambda$ has dimension
\begin{equation}
|\mathcal{R}_{(d_1,d_2,\dots,d_{N-1})}|=s_{\lambda}(\mathbf{1})=\prod_{1\leq A<B\leq N}\frac{\lambda_A-\lambda_B-A+B}{-A+B}\,.
\end{equation}

Schur polynomials are orthogonal with respect to the Haar measure of $SU(N)$ 
\begin{equation}
\langle s_{\lambda},s_{\mu}\rangle:=\oint d\mu(\mathbf{x})s_{\lambda}(\mathbf{x})s_{\mu}(\mathbf{x})=\delta_{\lambda,\mu}\,,
\end{equation}
\begin{equation}\label{eqn:Haarmeasure}
\oint d\mu(\mathbf{z})=\frac{1}{N!}\oint_{|z_A|=1}\prod_{A=1}^N\frac{dz_A}{2\pi\iu z_A}\delta\left(\prod_{A=1}^Nz_A-1\right)\prod_{A\neq B}\left(1-\frac{z_A}{z_B}\right)\,.
\end{equation}
One fact that we will often use is, that for any class function $f:SU(N)\to\mathbb{C}$ that is also invariant under the Weyl group of $SU(N)$ we can write
\begin{equation}
\oint d\mu(\mathbf{z})f(\mathbf{z})=\oint_{|z_A|=1}\prod_{A=1}^{N-1}\frac{dz_A}{2\pi\iu z_A}\prod_{1\leq A<B\leq N}\left(1-\frac{z_A}{z_B}\right)f(\mathbf{z})\,.
\end{equation}
Using
\begin{equation}
s_{\mu}s_{\nu}=\sum_{\eta}c_{\mu\nu}^{\eta}s_{\eta}\,,
\end{equation}
where $c^{\mu}_{\nu\eta}$ are Littlewood-Richardson coefficients, we can write
\begin{equation}
\langle s_{\lambda},s_{\mu}s_{\nu}\rangle=\sum_{\eta}c^{\eta}_{\mu\nu}\langle s_{\lambda},s_{\eta}\rangle=c^{\lambda}_{\mu\nu}\,.
\end{equation}
For Schur polynomials of type $A_1$ the Littlewood-Richardson coefficients are, of course, simply
\begin{equation}
s_{2j}s_{2j'}=\sum_{2j''=2|j-j'|}^{2j+2j'}s_{2j''}\,,
\end{equation}
with $2j,2j',2j''\in\mathbb{N}\cup\{0\}$.
\paragraph{$SO(2N)$ characters}
We will also sporadically make use of $SO(2N)$ characters. These are given by
\begin{equation}
\hat{s}_{\lambda}(\mathbf{x})=\frac{\det_{AB} \left(x_A^{\lambda_B+N-B}-x_A^{-\lambda_B-N+B}\right)}{\det_{AB} (x_A-x_A^{-1})^{N-B}}\,.
\end{equation}
where the $\lambda_1\geq\lambda_2\geq,\dots,\geq|\lambda_N|\geq0$ is related to the Dynkin labels $[d_1,d_2,\dots,d_N]$ by
\begin{equation}
\lambda_A=-d_{N-1}\delta_{A,N}+\frac{1}{2}(d_N+d_{N-1})+\sum_{n=A}^{N-2}d_n\,.
\end{equation}

\paragraph{Spiridonov-Warnaar Inversion Formula}
Define
\begin{equation}\label{eqn:deltafac}
\delta(x,y;T):=\frac{\Gamma_e(Tx^{\pm1}y^{\pm1})}{\Gamma_e(T^2)\Gamma_e(x^{\pm2})}\,,
\end{equation}
and consider the integral
\begin{equation}
f(z)=\kappa\oint\frac{dw}{4\pi\iu w}\delta(w,z;T)\hat{f}(w)\,,
\end{equation}
such that $\max\{|p|,|q|\}<|T|^2<1$. A consequence of the Spiridonov-Warnaar theorem is that \cite{2004math11044S,Gadde:2010te}
\begin{equation}\label{eqn:SWinversion}
\hat{f}(w)=\kappa\oint_{C_w}\frac{dz}{4\pi\iu z}\delta(z,w;T^{-1})f(z)\,,
\end{equation}
where $C_w$ is a deformation of the unit circle that includes the poles at $z=T^{-1}w^{\pm1}$ but excludes those at $Tw^{\pm1}$. Note that, if $\lim_{p,q\to0}(f,\hat{f},T):=(\tilde{f},\tilde{\hat{f}},T)$ is smooth, \eqref{eqn:SWinversion} implies
\begin{equation}
\tilde{f}(z)=\oint\frac{dw}{4\pi\iu w}\frac{(1-T^2)(1-w^{\pm2})}{(1-Tw^{\pm1}z^{\pm1})}\tilde{\hat{f}}(w)\, \implies\,
\tilde{\hat{f}}(w)=\oint_{C_w}\frac{dz}{4\pi\iu z}\frac{(1-T^{-2})(1-z^{\pm2})}{(1-T^{-1}w^{\pm1}z^{\pm1})}\tilde{f}(z)\,.
\end{equation}

\section{\boldmath Representation Theory of $\mathfrak{su}(2,2|1)$}
\label{app:SCAreps}
In this Appendix we discuss the representation theory of (the complexification of) the $\mathfrak{su}(2,2|1)$ superalgebra. Unitary representations of $\mathfrak{su}(2,2|1)$ can be decomposed into a finite sum of representations $[j_1,j_2]_E^{(\rN1)}$ of the maximal compact bosonic subalgebra
\begin{equation}
\mathfrak{u}(1)_E\oplus \mathfrak{su}(2)_1\oplus\mathfrak{su}(2)_2\oplus \mathfrak{u}(1)_{\rN1}\subset\mathfrak{su}(2,2|1)\,.
\end{equation}
The $\mathfrak{su}(2,2|1)$ superalgebra has four Poincar\'e supercharges 
\begin{equation}
\mathcal{Q}\in[1/2,0]_{\frac{1}{2}}^{(-1)}\,,\quad \widetilde{\mathcal{Q}}\in[0,1/2]^{(1)}_{\frac{1}{2}}\,.
\end{equation}
Note that with respect to \cite{Cordova:2016emh} we use $j=2j_1$, $\bar{j}=2j_2$. The $\mathcal{Q}$ and $\widetilde{\mathcal{Q}}$ shortening conditions are listed in Table \ref{tab:Qshort} and Table \ref{tab:Qtilshort} respectively.
\begin{table}
\centering
\begin{tabular}{|c|c|c|c|}
\hline
Name&Primary&Unitarity bound& $\mathcal{Q}$-Null state\\\hline
$L$&$[j_1,j_2]_{E}^{(r)}$&$E>2+2j_1-\frac{3}{2}r$&None\\\hline
$A_1$&$[j_1\geq1/2,j_2]_{E}^{(r)}$&$E=2+2j_1-\frac{3}{2}r$&$[j_1-1/2,j_2]_{E+1/2}^{(r-1)}$\\\hline
$A_2$&$[0,j_2]_{E}^{(r)}$&$E=2-\frac{3}{2}r$&$[0,j_2]_{E+1}^{(r-2)}$\\\hline
$B_1$&$[0,j_2]_{E}^{(r)}$&$E=-\frac{3}{2}r$&$[1/2,j_2]_{E+1/2}^{(r-1)}$\\\hline
\end{tabular}
\caption{\textit{$\mathcal{Q}$-shortening conditions.}}
\label{tab:Qshort}
\end{table}
\begin{table}
\centering
\begin{tabular}{|c|c|c|c|}
\hline
Name&Primary&Unitarity bound& $\widetilde{\mathcal{Q}}$-Null state\\\hline
$\overline{L}$&$[j_1,j_2]_{E}^{(r)}$&$E>2+2j_1+\frac{3}{2}\rN1$&None\\\hline
$\overline{A}_1$&$[j_1,j_2\geq1/2]_{E}^{(r)}$&$E=2+2j_1+\frac{3}{2}r$&$[j_1,j_2-1/2]_{E+1/2}^{(r+1)}$\\\hline
$\overline{A}_2$&$[j_1,0]_{E}^{(r)}$&$E=2+\frac{3}{2}r$&$[j_1,0]_{E+1}^{(r+2)}$\\\hline
$\overline{B}_1$&$[j_1,0]_{E}^{(r)}$&$E=\frac{3}{2}r$&$[j_1,1/2]_{E+1/2}^{(r+1)}$\\\hline
\end{tabular}
\caption{\textit{$\widetilde{\mathcal{Q}}$-shortening conditions.}}
\label{tab:Qtilshort}
\end{table}
We list all possible short unitary multiplets of the $\mathfrak{su}(2,2|1)$ superconformal algebra in Table \ref{tab:shorts}.
\begin{table}
\centering
\begin{tabular}{|c|c|c|}
\hline
CDI-notation&Quantum number relations& DO-notation\\\hline\hline
$L\overline{L}[j_1,j_2]_{E}^{(r)}$&$E>2+\max\left\{2j_1-\frac{3}{2}r,2j_2+\frac{3}{2}r\right\}$&$\mathcal{A}^E_{r,(j_1,j_2)}$\\\hline\hline
$B_1\overline{L}[0,j_2]_E^{(r)}$&$r<-\frac{2}{3}(j_2+1)$, $E=-\frac{3}{2}r$&$\mathcal{B}_{r,(0,j_2)}$\\\hline
$L\overline{B}_1[j_1,0]_E^{(r)}$&$r>\frac{2}{3}(j_1+1)$, $E=\frac{3}{2}r$&$\overline{\mathcal{B}}_{r,(j_1,0)}$\\\hline
$B_1\overline{B}_1[0,0]_E^{(r)}$&$E=r=0$&$\hat{\mathcal{B}}$\\\hline\hline
$A_{\ell}\overline{L}[j_1,j_2]_E^{(r)}$&$r<\frac{2}{3}(j_1-j_2)$, $E=2+2j_1-\frac{3}{2}r$&$\mathcal{C}_{r,(j_1,j_2)}$\\\hline
$L\overline{A}_{\overline{\ell}}[j_1,j_2]_E^{(r)}$&$r>\frac{2}{3}(j_1-j_2)$, $E=2+2j_2+\frac{3}{2}r$&$\overline{\mathcal{C}}_{r,(j_1,j_2)}$\\\hline
$A_{\ell}\overline{A}_{\overline{\ell}}[j_1,j_2]_E^{(r)}$&$r=\frac{2}{3}(j_1-j_2)$, $E=2+j_1+j_2$&$\hat{\mathcal{C}}_{(j_1,j_2)}$\\\hline\hline
$B_1\overline{A}_{\overline{\ell}}[0,j_2]_E^{(r)}$&$E=-\frac{3}{2}r=1+j_2$&$\mathcal{D}_{(0,j_2)}$\\\hline
$A_{\ell}\overline{B}_1[j_1,0]_E^{(r)}$&$E=\frac{3}{2}r=1+j_1$&$\overline{\mathcal{D}}_{(j_1,0)}$\\\hline
\end{tabular}
\caption{\textit{Unitary representations of the $\mathfrak{su}(2,2|1)$ superconformal algebra. In the above we have $\ell=1$ if $j_1\geq\frac{1}{2}$, $\ell=2$ if $j_1=0$, $\overline{\ell}=1$ if $j_2\geq\frac{1}{2}$ and $\overline{\ell}=2$ if $j_2=0$. In the first column we list the notation of \cite{Cordova:2016emh} and in third column we list the corresponding Dolan \& Osborn style notation \cite{Dolan:2002zh} which was also used in \cite{Liendo:2011wc,Beem:2012yn,Rastelli:2016tbz,Gadde:2010en}.}}
\label{tab:shorts}
\end{table}
\begin{table}
\renewcommand{\arraystretch}{1.1}
\centering
\begin{tabular}{|c|c|}
\hline
Conserved current multiplet(s)&Comment(s)\\\hline
$\hat{\mathcal{C}}_{(0,0)}$&Flavour current\\\hline
$\hat{\mathcal{C}}_{(\frac{1}{2},0)}$, $\hat{\mathcal{C}}_{(0,\frac{1}{2})}$&Supersymmetric currents\\\hline
$\hat{\mathcal{C}}_{(\frac{1}{2},\frac{1}{2})}$&Stress tensor, contains $\mathfrak{u}(1)_{r}$ current\\\hline
$\hat{\mathcal{C}}_{(j_1,j_2)}|_{j_1+j_2>1}$&Higher spin currents\\\hline
$\overline{\mathcal{D}}_{(0,0)}$ $(\mathcal{D}_{(0,0)})$&Free (anti-)Chiral field $\Phi$ ($\overline{\Phi}$)\\\hline
$\overline{\mathcal{D}}_{(\frac{1}{2},0)}$ ($\mathcal{D}_{(0,\frac{1}{2})}$)&Free (anti-)vector superfield $W_{\alpha}$ $(\overline{W}_{\dot\alpha})$\\\hline
$\overline{\mathcal{D}}_{(j_1\geq1,0)}$, $\mathcal{D}_{(0,j_2\geq1)}$&Higher spin free fields.\\\hline
\end{tabular}
\caption{\textit{Conserved current multiplets of $\mathfrak{su}(2,2|1)$ superconformal algebra.}}
\label{tab:Conserved}
\end{table}
We list all multiplets that contain conserved currents in Table \ref{tab:Conserved}. Using
\begin{equation}
\mathcal{C}_{r,(-\frac{1}{2},j_2)}\iso\mathcal{B}_{r-1,(0,j_2)}\,,\quad\overline{\mathcal{C}}_{r,(j_1,-\frac{1}{2})}\iso\overline{\mathcal{B}}_{r+1,(j_1,0)}\,,
\end{equation}
the $\mathcal{N}=1$ recombination rules can be written as
\begin{gather}
\label{eqn:recomb1}\mathcal{A}^{2+2j_1-\frac{3}{2}r}_{r<\frac{2}{3}(j_1-j_2),(j_1,j_2)}\iso\mathcal{C}_{r,(j_1,j_2)}\oplus\mathcal{C}_{r-1,(j_1-\frac{1}{2},j_2)}\,,\quad\mathcal{A}^{2+2j_2+\frac{3}{2}r}_{r>\frac{2}{3}(j_1-j_2),(j_1,j_2)}\iso\overline{\mathcal{C}}_{r,(j_1,j_2)}\oplus\overline{\mathcal{C}}_{r+1,(j_1,j_2-\frac{1}{2})}\,,\\
\mathcal{A}^{2+j_1+j_2}_{\frac{2}{3}(j_1-j_2),(j_1,j_2)}\iso\hat{\mathcal{C}}_{(j_1,j_2)}\oplus\mathcal{C}_{\frac{2}{3}(j_1-j_2)-1,(j_1-\frac{1}{2},j_2)}\oplus\overline{\mathcal{C}}_{\frac{2}{3}(j_1-j_2)+1,(j_1,j_2-\frac{1}{2})}\,.\label{eqn:recomb2}
\end{gather} 
The multiplets $\overline{\mathcal{D}}_{(j_1,0)}$, $\overline{\mathcal{B}}_{r<\frac{2}{3}j_1+2,(j_1,0)}$ and their complex conjugates have no recombination rules and are therefore absolutely protected.
We list below the contribution of each multiplet to the right-handed index \eqref{eqn:SCI2}
\begin{align}
& I_{\mathcal{A}^E_{r,(j_1,j_2)}}= I_{\mathcal{B}_{r,(0,j_2)}}= I_{\mathcal{C}_{r,(j_1,j_2)}}=0\,,\label{eqn:multiind1}
\quad I_{\hat{\mathcal{C}}_{(j_1,j_2)}}=\frac{(-1)^{2j_1+2j_2+1}(pq)^{\frac{2}{3}j_2+\frac{1}{3}j_1+1}}{(1-p)(1-q)}\chi_{2j_1}\,,\\
& I_{\overline{\mathcal{C}}_{r,(j_1,j_2)}}=\frac{(-1)^{2j_1+2j_2+1}(pq)^{\frac{r}{2}+j_2+1}}{(1-p)(1-q)}\chi_{2j_1}\,,\quad I_{\overline{\mathcal{B}}_{r,(j_1,0)}}=\frac{(-1)^{2j_1}(pq)^{\frac{r}{2}}}{(1-p)(1-q)}\chi_{2j_1}\\
& I_{\mathcal{D}_{(0,j_2)}}=\frac{(-1)^{2j_2+1}(pq)^{\frac{2}{3}j_2+\frac{2}{3}}}{(1-p)(1-q)}\,,\quad  I_{\overline{\mathcal{D}}_{(j_1,0)}}=\frac{(-1)^{2j_1}(pq)^{\frac{j_1+1}{3}}\left(\chi_{2j_1}-\sqrt{pq}\chi_{2j_1-1}\right)}{(1-p)(1-q)}\,,\label{eqn:multiind2}
\end{align}
where the character of the spin-$\frac{s}{2}$ representation of $SU(2)$ is $\chi_{s}(y)=\frac{y^{s+1}-y^{-s-1}}{y-y^{-1}}$ and it should be understood in the above that, $\chi_{2j_1}\equiv\chi_{2j_1}\left(\sqrt{p/q}\right)=\chi_{2j_1}\left(\sqrt{\sigma/\rho}\right)$.

By construction the right-handed index of all multiplets of the type $X\overline{L}[j_1,j_2]_E^{(r)}$ is zero. In computing the above one must carefully deal with equations of motion. If any given $\mathfrak{so}(4,2)$ representation appearing in a multiplet saturates the unitarity bound then there will be a corresponding equation of motion which must enter the index with opposite statistics. We list the the possible null states of $\mathfrak{so}(4,2)$ in Table \ref{tab:so42nulls}.
\begin{table}
\centering
\begin{tabular}{|c|c|c|}
\hline
Primary&Unitarity bound& Null state\\\hline
$[j_1\geq1/2,j_2\geq1/2]_{E}$&$E\geq2+j_1+j_2$&$[j_1-1/2,j_2-1/2]_{E+1}$\\\hline
$[j_1\geq1/2,0]_{E}$&$E\geq j_1+1$&$[j_1-1/2,1/2]_{E+1}$\\\hline
$[0,j_2\geq1/2]_{E}$&$E\geq j_2+1$&$[1/2,j_2-1/2]_{E+1}$\\\hline
$[0,0]_{E}$&$E\geq1$&$[0,0]_{E+2}$\\\hline
$[0,0]_{E}$&$E=0$&$[1/2,1/2]_{E+1}$\\\hline
\end{tabular}
\caption{\textit{Unitary representations of $\mathfrak{so}(4,2)$. In the final column we list the associated null state when the unitarity bound is saturated. The null states correspond to equations of motion and as such their contribution must be subtracted from the index.}}
\label{tab:so42nulls}
\end{table}

\subsection{$\mathcal{N}=1$ Index Equivalence Classes}
By either examining the recombination rules \eqref{eqn:recomb1} \eqref{eqn:recomb2}, or, by directly observing the contribution to the index from each multiplet \eqref{eqn:multiind1}-\eqref{eqn:multiind2} we can immediately read off the (left)right-handed index equivalence classes \cite{Gadde:2009dj,Beem:2012yn,Evtikhiev:2017heo}. That is, the set of multiplets with equal contribution to the (left)right-handed index. Here we focus only on the right-handed index. The equivalence classes (leaving implicit the quantum number inequalities of Table \ref{tab:shorts}) are
\begin{align}
&[\widetilde{r},j_1]_+:=\{\overline{\mathcal{C}}_{r,(j_1,\frac{\widetilde{r}-r}{2})}|\widetilde{r}-r\in2\mathbb{Z}_{\geq0}\}\cup\{\hat{\mathcal{C}}_{(j_1,\frac{3\widetilde{r}-2j_1}{4})}|3\widetilde{r}-2j_1\in4\mathbb{Z}_{\geq0}\}\\
&[\widetilde{r},j_1]_-:=\{\overline{\mathcal{C}}_{r,(j_1,\frac{\widetilde{r}-r}{2})}|\widetilde{r}-r\in-1+2\mathbb{Z}_{\geq0}\}\cup\{\hat{\mathcal{C}}_{(j_1,\frac{3\widetilde{r}-2j_1}{4})}|3\widetilde{r}-2j_1\in2+4\mathbb{Z}_{\geq0}\}
\end{align}
and their contributions to the index are
\begin{equation}\label{eqn:indequiv}
I_{[\widetilde{r},j_1]_+}=-I_{[\widetilde{r},j_1]_-}=\frac{(-1)^{2j_1+1}(pq)^{\frac{\widetilde{r}}{2}+1}}{(1-p)(1-q)}\chi_{2j_1}\,.
\end{equation} 
The cases in which we can extract the most information regarding the spectrum from the index are those in which the equivalence class contains a small number of representatives. For example, for example if $\widetilde{r}\in(2j_1/3,4/3+2j_1/3]$ then $[\widetilde{r},j_1]_+$ is empty and $[\widetilde{r},j_1]_-$ can contain only $\overline{\mathcal{B}}_{\widetilde{r}-1,(j_1,0)}$. The multiplets $\mathcal{D}_{(0,j_2)}$ and $\overline{\mathcal{D}}_{(j_1,0)}$ are free fields and sit in their own equivalence classes. Finally the multiplets $\hat{\mathcal{C}}_{(\frac{1}{2},0)}$ and $\hat{\mathcal{C}}_{(0,\frac{1}{2})}$ are the only representatives within the classes $[\frac{1}{3},\frac{1}{2}]_+$ and $[\frac{2}{3},0]_-$, respectively. $[\frac{1}{3},\frac{1}{2}]_-$ and $[\frac{2}{3},0]_+$ also contain only a single representatives, being $\overline{\mathcal{B}}_{\frac{7}{3}(\frac{1}{2},0)}$ and $\overline{\mathcal{C}}_{\frac{2}{3}(0,0)}$ respectively. 
We also defined the \textit{net degeneracy}
\begin{equation}\label{eqn:netdegen}
\text{ND}[\widetilde{r},j_2]:=\#[\widetilde{r},j_2]_+-\#[\widetilde{r},j_2]_-\,.
\end{equation}

\subsection{Hall-Littlewood Limit}
The indices \eqref{eqn:multiind1}-\eqref{eqn:multiind2} can of course enter into the character expansion of \eqref{eqn:SCI} with factors of $\left(\tau/\rho\sigma\right)^{2q_t/3}=(t/(pq)^{2/3})^{q_t}$. By construction the Hall-Littlewood limit of the index \eqref{eqn:HL} counts only those operators with $2j_2=-r+\frac{4}{3}q_t$ and $j_1=0$. Assuming that this limit always exists for theories in class $\mathcal{S}_k$ we can extract bounds for the value of $\mathfrak{u}(1)_t$ charges for given multiplets appearing the character expansion of the index. In particular, using the fact that
\begin{equation}
\lim_{\sigma\to0}\lim_{\rho\to0}\,(\sigma\rho)^a\chi_{2j_1}\left(\sqrt{\sigma/\rho}\right)=\begin{cases}
\delta_{j_1,0}&a=j_1\,,\\
0&a> j_1\,,\\
\infty&a<j_1\,,
\end{cases}
\end{equation}
with $j_1\geq0$ so that the limit exists only if $a\geq j_1$. Therefore, from \eqref{eqn:multiind1}-\eqref{eqn:multiind2} one can see that the the $\mathfrak{u}(1)_t$ charges of the multiplets contributing to the right handed index must obey the constraints of Table \ref{tab:HLconstraint}. By appying conjugation it is possible to find similar bounds for multiplets appearing in the character expansion of the left-handed index. One may also repeat such an exercise with the Macdonald limit ($\sigma\to0$) of the index. 
\begin{table}
\centering
\begin{tabular}{|c|c|}
\hline
Multiplet&$q_t$ bound\\\hline\hline
$\overline{\mathcal{B}}_{r,(j_1,0)}$&$\frac{r}{2}-\frac{2}{3}q_t\geq j_1$\\\hline
$\overline{\mathcal{C}}_{r,(j_1,j_2)}$&$\frac{r}{2}+j_2+1-\frac{2}{3}q_t\geq j_1$\\\hline
$\hat{\mathcal{C}}_{(j_1,j_2)}$&$1-\frac{2}{3}q_t\geq\frac{2}{3}(j_1-j_2)	$\\\hline
$\mathcal{D}_{(0,j_2)}$&$j_2+1-q_t\geq0$\\\hline
$\overline{\mathcal{D}}_{(j_1,0)}$&$-2j_1+1-q_t\geq0$\\\hline
\end{tabular}
\caption{\it Restrictions imposed on the $\mathfrak{u}(1)_t$ representations implied by the existence of the Hall-Littlewood limit of the index. In order that a multiplet contributes to the Hall-Littlewood index it must have $j_1=0$ and saturate the bound.}
\label{tab:HLconstraint}
\end{table}
Defining $\HL_{\mathcal{M}^{(q_t)}}=\lim_{\sigma,\rho\to0}\left(\frac{\tau}{\sigma\rho}\right)^{2q_t/3} I_{\mathcal{M}}$ and assuming that the bounds of Table \ref{tab:HLconstraint} are satisfied (such that the limit always always exists) we have
\begin{align}
&\HL_{{\mathcal{A}^{E,(q_t)}_{r,(j_1,j_2)}}}=\HL_{\mathcal{B}^{(q_t)}_{r,(0,j_2)}}=\HL_{\mathcal{C}^{(q_t)}_{r,(j_1,j_2)}}=0\,,\quad\label{eqn:HLmultiind1}\HL_{\overline{\mathcal{B}}^{(q_t)}_{r,(j_1,0)}}=\tau^{2q_t}\delta_{j_1,0}\delta_{\frac{3r}{4},q_t}\,,\\
&\HL_{\overline{\mathcal{C}}^{(q_t)}_{r,(j_1,j_2)}}=(-1)^{2j_2+1}\tau^{2q_t}\delta_{j_1,0}\delta_{q_t,\frac{3}{2}(j_2+1)+\frac{3}{4}r}\,,\quad\HL_{\hat{\mathcal{C}}^{(q_t)}_{(j_1,j_2)}}=(-1)^{2j_2+1}\tau^{2q_t}\delta_{j_1,0}\delta_{q_t,j_2+\frac{3}{2}}\,,\\
&\HL_{\mathcal{D}^{(q_t)}_{(0,j_2)}}=(-1)^{2j_2+1}\tau^{2q_t}\delta_{q_t,j_2+1}\,,\quad\HL_{\overline{\mathcal{D}}^{(q_t)}_{(j_1,0)}}=\tau\delta_{j_1,0}\delta_{q_t,\frac{1}{2}}\,.\label{eqn:HLmultiind2}
\end{align}
so the only multiplets that can contribute to the right-handed index in the Hall-Littlewood limit are
\begin{equation}
\label{eq:operatorsHL}
\overline{\mathcal{B}}^{(\frac{4r}{3})}_{r,(0,0)}\,,\quad \overline{\mathcal{C}}^{(\frac{3}{2}(j_2+1)+\frac{3}{4}r)}_{r,(0,j_2)}\,,\quad \hat{\mathcal{C}}^{(j_2+\frac{3}{2})}_{(0,j_2)}\,,\quad \mathcal{D}^{(j_2+1)}_{(0,j_2)}\,,\quad \overline{\mathcal{D}}^{(\frac{1}{2})}_{(0,0)}\,.
\end{equation}
Equality of the Hall-Littlewood limit of the index with the Hilbert series of the Higgs branch at would imply that in those theories the Hall-Littlewood index receives contribution only from $\overline{\mathcal{D}}^{(\frac{1}{2})}_{(0,0)}$ and $\overline{\mathcal{B}}^{(\frac{4r}{3})}_{r,(0,0)}$ multiplets.

\section{Superconformal Index Notation}
\label{sec:appD}

The right-handed superconformal index computed with respect to $\widetilde{\mathcal{Q}}_{\dot-}$ is given by
\begin{equation}\label{eqn:SCI2}
I(p,q)=\textrm{Tr}(-1)^Fp^{j_1+j_2+\frac{r}{2}}q^{-j_1+j_2+\frac{r}{2}}e^{-\beta\widetilde{\delta}_{\dot-}}\,.
\end{equation}
The superconformal index \eqref{eqn:SCI2} receives contributions only from those states satisfying
\begin{equation}
\widetilde{\delta}_{\dot-}=2\{\widetilde{\mathcal{Q}}_{\dot-},\widetilde{\mathcal{S}}^{\dot-}\}=E-2j_2-\frac{3}{2}r=0\,.
\end{equation}
The single letter contributions of $\mathcal{N}=1$ Chiral multiplets and Vector multiplets may be computed by enumerating all letters with $\widetilde{\delta}_{\dot-}=0$. These are listed in Table \ref{tab:letters}. 
\begin{table}
\renewcommand{\arraystretch}{1.2}
\centering
\begin{tabular}{|c||c|c|c|c|c|c|} 
\hline
 & $E_{\text{UV}}$ & $j_1$ & $j_2$ & $r_{\text{UV}}$ & $r_{\text{IR}}$& Index\\ 
 \hline\hline
  $x$ & $1$ & $0$ & $0$ & $\frac{2}{3}$ & $r_{X}$ &$(pq)^{\frac{r_{X}}{2}}$\\ \hline
 $\overline{\psi}_{\dot+}$ & $3/2$ & $0$ & $+\frac{1}{2}$ & $\frac{1}{3}$ & $-r_{X}+1$ & $-(pq)^{\frac{2-r_{X}}{2}}$\\ 
 \hline\hline
  $\widetilde{F}_{\dot+\dot+}$ & $2$ & $0$ & $1$ & $0$ & $0$ & $pq$\\ 
 \hline
   $\lambda_{\pm}$ & $3/2$ & $\pm\frac{1}{2}$ & $0$ & $1$ & $1$ & $-p$, $-q$\\ 
 \hline
    $\partial\lambda=0$ & $5/2$ & $0$ & $+\frac{1}{2}$ & $1$ & $1$ & $pq$\\ 
 \hline\hline
    $\partial_{\pm\dot+}$ & $1$ & $\pm\frac{1}{2}$ & $+\frac{1}{2}$ & $0$ & $0$ & $p$, $q$\\ 
 \hline
\end{tabular}
\caption{\textit{Letters of a free chiral multiplet $X=\{x,\psi,F\}$, its conjugate $\overline{X}=\{\overline{x},\overline{\psi},\overline{F}\}$ and free vector multiplet $V=\{\lambda_{\alpha},\overline{\lambda}_{\dot\alpha},F_{\alpha\beta},\widetilde{F}_{\dot\alpha\dot\beta},D,\overline{D}\}$. For the vector multiplet we must take into account the equation of motion $\partial\lambda=\partial_{+\dot+}\lambda_{-}+\partial_{-\dot+}\lambda_{+}=0$.}}
\label{tab:letters}
\end{table}
The single letter contribution of a chiral multiplet $X$ and its conjugate $\overline{X}$ is given by
\begin{equation}
i_{X}(p,q)=\frac{(pq)^{\frac{r_{X}}{2}}-(pq)^{\frac{2-r_{X}}{2}}}{(1-p)(1-q)}\,,
\end{equation}
here, $r_X:=r_{\text{IR}}[X]$. Taking the PE yields \cite{Dolan:2008qi}
\begin{equation}
I_{X}(p,q):=\PE[i_{X}(p,q)]=\Gamma_e\left((pq)^{\frac{r_{X}}{2}}\right)\,,
\end{equation}
where the Elliptic Gamma function is defined in \eqref{eqn:EllGamma}.
Similarly, the single letter contribution for a vector multiplet $V$ in the adjoint representation of $SU(N)$ is given by
\begin{equation}
i_{V}(p,q,\mathbf{z})=\left(-\frac{p}{1-p}-\frac{q}{1-q}\right)\chi_{(1,0,\dots,0,1)}(\mathbf{z})\,.
\end{equation} 
Taking the PE yields
\begin{equation}\label{eqn:veccont}
I_{V}(p,q,\mathbf{z})=\frac{\kappa}{\Delta(\mathbf{z})}\prod_{A\neq B}\frac{1}{\Gamma_e\left(\frac{z_A}{z_B}\right)}\,,\quad \Delta(\mathbf{z})=\prod_{A\neq B}\left(1-\frac{z_A}{z_B}\right)\,,\quad \kappa:=(p;p)^{N-1}(q;q)^{N-1}\,.
\end{equation}
The vector multiplet also comes with an integral over the gauge group \eqref{eqn:Haarmeasure}.

\section{\boldmath Unrefined Hall-Littlewood Index for Class $\mathcal{S}_2$ Interacting Trinions}
\label{app:3puncHL}
\subsection{$T_B$ Theory}
The Hall-Littlewood index for the three-punctured sphere $T_B$ theory in the unrefined $\mathbf{z}=\mathbf{u}=\mathbf{v}=\mathbf{1}$, $\gamma\beta=1$ limit is given by
\begin{equation}
\label{eq:HLTB111}
\left.\HL^{(T_B)\mathbf{1}}_{\mathbf{1}\mathbf{1}}\right|_{\gamma\beta=1}=\frac{\gamma^2}{\beta^2}\frac{\tau^{19}\left(Q_B(\gamma^{-1}\beta,\tau^{-1})-Q_B(\gamma\beta^{-1},\tau)\right)}{\left(1-\frac{\beta}{\gamma}\tau^2\right)^{7}\left(1-\frac{\gamma}{\beta}\tau^2\right)^{9}\left(1-\tau^2\right)^{11}}\,.
\end{equation}
The polynomial $Q_B(\gamma\beta^{-1},\tau)$ is given by
\begin{equation}
\begin{aligned}
&Q_B(\gamma\beta^{-1},\tau)=-\frac{3 \gamma ^5 \tau ^9}{\beta ^5}-\frac{15 \gamma ^5 \tau ^7}{\beta ^5}-\frac{17 \gamma ^5 \tau ^5}{\beta ^5}-\frac{5 \gamma ^5 \tau ^3}{\beta ^5}+\frac{21 \gamma ^4 \tau ^{11}}{\beta ^4}+\frac{73 \gamma ^4
   \tau ^9}{\beta ^4}\\
   &-\frac{14 \gamma ^4 \tau ^7}{\beta ^4}-\frac{6 \gamma ^4 \tau ^5}{\beta ^4}+\frac{201 \gamma ^4 \tau ^3}{\beta ^4}+\frac{45 \gamma ^4 \tau }{\beta ^4}+\frac{\gamma ^3 \tau ^{17}}{\beta
   ^3}-\frac{3 \gamma ^3 \tau ^{15}}{\beta ^3}-\frac{67 \gamma ^3 \tau ^{13}}{\beta ^3}-\frac{55 \gamma ^3 \tau ^{11}}{\beta ^3}\\
 &+\frac{305 \gamma ^3 \tau ^9}{\beta ^3}+\frac{\beta ^3 \tau ^9}{\gamma
   ^3}-\frac{387 \gamma ^3 \tau ^7}{\beta ^3}-\frac{11 \beta ^3 \tau ^7}{\gamma ^3}-\frac{885 \gamma ^3 \tau ^5}{\beta ^3}+\frac{67 \beta ^3 \tau ^5}{\gamma ^3}+\frac{919 \gamma ^3 \tau ^3}{\beta ^3}\\
   &-\frac{265\beta ^3 \tau ^3}{\gamma ^3}-\frac{298 \gamma ^3 \tau }{\beta ^3}+\frac{818 \beta ^3 \tau }{\gamma ^3}+\frac{\gamma ^2 \tau ^{19}}{\beta ^2}+\frac{13 \gamma ^2 \tau ^{17}}{\beta ^2}+\frac{36 \gamma ^2 \tau
   ^{15}}{\beta ^2}-\frac{156 \gamma ^2 \tau ^{13}}{\beta ^2}\\
   &-\frac{231 \gamma ^2 \tau ^{11}}{\beta ^2}+\frac{15 \beta ^2 \tau ^{11}}{\gamma ^2}+\frac{1421 \gamma ^2 \tau ^9}{\beta ^2}-\frac{149 \beta ^2 \tau
   ^9}{\gamma ^2}-\frac{132 \gamma ^2 \tau ^7}{\beta ^2}+\frac{582 \beta ^2 \tau ^7}{\gamma ^2}-\frac{3340 \gamma ^2 \tau ^5}{\beta ^2}\\
 &-\frac{714 \beta ^2 \tau ^5}{\gamma ^2}+\frac{4332 \gamma ^2 \tau ^3}{\beta
   ^2}-\frac{2327 \beta ^2 \tau ^3}{\gamma ^2}+\frac{1484 \gamma ^2 \tau }{\beta ^2}+\frac{4101 \beta ^2 \tau }{\gamma ^2}+\frac{15 \gamma  \tau ^{17}}{\beta }+\frac{83 \gamma  \tau ^{15}}{\beta }\\
   &-\frac{193
   \gamma  \tau ^{13}}{\beta }+\frac{52 \beta  \tau ^{13}}{\gamma }-\frac{805 \gamma  \tau ^{11}}{\beta }-\frac{308 \beta  \tau ^{11}}{\gamma }+\frac{2208 \gamma  \tau ^9}{\beta }+\frac{81 \beta  \tau ^9}{\gamma
   }+\frac{1224 \gamma  \tau ^7}{\beta }\\
   &+\frac{2933 \beta  \tau ^7}{\gamma }-\frac{8621 \gamma  \tau ^5}{\beta }-\frac{4192 \beta  \tau ^5}{\gamma }+\frac{5807 \gamma  \tau ^3}{\beta }-\frac{2504 \beta  \tau
   ^3}{\gamma }+\frac{7024 \gamma  \tau }{\beta }+\frac{12360 \beta  \tau }{\gamma }\\
   &+52 \tau ^{15}-36 \tau ^{13}-982 \tau ^{11}+1322 \tau ^9+3423 \tau ^7-9605 \tau ^5+1849 \tau ^3+14421 \tau\,.\label{eqn:Qpolyn}
\end{aligned}
\end{equation}

\subsection{$T_A$ Theory}
The Hall-Littlewood index for the $T_A$ theory in the unrefined limit with $\gamma=\beta$ reads
\begin{equation}
\left.\HL^{(T_A)\mathbf{1}}_{\mathbf{1}\mathbf{1}}\right|_{\frac{\gamma}{\beta}=1}=\frac{\beta^5\gamma^5\tau^{15}\left(Q_A(\tau,\beta\gamma)-Q_A(\tau^{-1},\beta^{-1}\gamma^{-1})\right)}{(1-\tau^2)^{19}(1-\gamma^4\beta^4\tau^2)^4}
\end{equation}
where 
\begin{equation}
\begin{aligned}\label{eqn:QApolyn}
&Q_A(\tau,\beta\gamma)=-\beta ^5 \gamma ^5 \tau ^{15}-29 \beta ^5 \gamma ^5 \tau ^{13}-274 \beta ^5 \gamma ^5 \tau ^{11}-1122 \beta ^5 \gamma ^5 \tau ^9-2222 \beta ^5 \gamma ^5 \tau ^7\\
&-2222 \beta ^5 \gamma ^5 \tau ^5-1122 \beta ^5
   \gamma ^5 \tau ^3+\frac{\tau ^3}{\beta ^5 \gamma ^5}-274 \beta ^5 \gamma ^5 \tau +\frac{29 \tau }{\beta ^5 \gamma ^5}-4 \beta ^3 \gamma ^3 \tau ^{13}-19 \beta ^3 \gamma ^3 \tau ^{11}\\
 &+525 \beta ^3 \gamma ^3
   \tau ^9+4216 \beta ^3 \gamma ^3 \tau ^7+11048 \beta ^3 \gamma ^3 \tau ^5-\frac{4 \tau ^5}{\beta ^3 \gamma ^3}+12523 \beta ^3 \gamma ^3 \tau ^3-\frac{144 \tau ^3}{\beta ^3 \gamma ^3}\\
   &+6519 \beta ^3 \gamma ^3
   \tau -\frac{1524 \tau }{\beta ^3 \gamma ^3}-\beta  \gamma  \tau ^{11}+15 \beta  \gamma  \tau ^9-350 \beta  \gamma  \tau ^7+\frac{6 \tau ^7}{\beta  \gamma }-5418 \beta  \gamma  \tau ^5\\
   &+\frac{286 \tau
   ^5}{\beta  \gamma }-19741 \beta  \gamma  \tau ^3+\frac{3706 \tau ^3}{\beta  \gamma }-27485 \beta  \gamma  \tau +\frac{15982 \tau }{\beta  \gamma }\,.
\end{aligned}
\end{equation}

\bibliography{biblio}

\providecommand{\href}[2]{#2}\begingroup\raggedright\begin{thebibliography}{10}

\bibitem{Gaiotto:2009we}
D.~Gaiotto, \emph{{N=2 dualities}},
  \href{http://dx.doi.org/10.1007/JHEP08(2012)034}{\emph{JHEP} {\bfseries 08}
  (2012) 034}, [\href{https://arxiv.org/abs/0904.2715}{{\ttfamily 0904.2715}}].

\bibitem{Alday:2009aq}
L.~F. Alday, D.~Gaiotto and Y.~Tachikawa, \emph{{Liouville Correlation
  Functions from Four-dimensional Gauge Theories}},
  \href{http://dx.doi.org/10.1007/s11005-010-0369-5}{\emph{Lett. Math. Phys.}
  {\bfseries 91} (2010) 167--197},
  [\href{https://arxiv.org/abs/0906.3219}{{\ttfamily 0906.3219}}].

\bibitem{Gadde:2009kb}
A.~Gadde, E.~Pomoni, L.~Rastelli and S.~S. Razamat, \emph{{S-duality and 2d
  Topological QFT}},
  \href{http://dx.doi.org/10.1007/JHEP03(2010)032}{\emph{JHEP} {\bfseries 03}
  (2010) 032}, [\href{https://arxiv.org/abs/0910.2225}{{\ttfamily 0910.2225}}].

\bibitem{Gadde:2011ik}
A.~Gadde, L.~Rastelli, S.~S. Razamat and W.~Yan, \emph{{The 4d Superconformal
  Index from q-deformed 2d Yang-Mills}},
  \href{http://dx.doi.org/10.1103/PhysRevLett.106.241602}{\emph{Phys. Rev.
  Lett.} {\bfseries 106} (2011) 241602},
  [\href{https://arxiv.org/abs/1104.3850}{{\ttfamily 1104.3850}}].

\bibitem{Bah:2012dg}
I.~Bah, C.~Beem, N.~Bobev and B.~Wecht, \emph{{Four-Dimensional SCFTs from
  M5-Branes}}, \href{http://dx.doi.org/10.1007/JHEP06(2012)005}{\emph{JHEP}
  {\bfseries 06} (2012) 005},
  [\href{https://arxiv.org/abs/1203.0303}{{\ttfamily 1203.0303}}].

\bibitem{Gaiotto:2015usa}
D.~Gaiotto and S.~S. Razamat, \emph{{$ \mathcal{N}=1 $ theories of class $
  {\mathcal{S}}_k $}},
  \href{http://dx.doi.org/10.1007/JHEP07(2015)073}{\emph{JHEP} {\bfseries 07}
  (2015) 073}, [\href{https://arxiv.org/abs/1503.05159}{{\ttfamily
  1503.05159}}].

\bibitem{Coman:2015bqq}
I.~Coman, E.~Pomoni, M.~Taki and F.~Yagi, \emph{{Spectral curves of $
  \mathcal{N} $ = 1 theories of class $ {\mathcal{S}}_k $}},
  \href{http://dx.doi.org/10.1007/JHEP06(2017)136}{\emph{JHEP} {\bfseries 06}
  (2017) 136}, [\href{https://arxiv.org/abs/1512.06079}{{\ttfamily
  1512.06079}}].

\bibitem{Mitev:2017jqj}
V.~Mitev and E.~Pomoni, \emph{{2D CFT blocks for the 4D class $\mathcal{S}_k$
  theories}}, \href{http://dx.doi.org/10.1007/JHEP08(2017)009}{\emph{JHEP}
  {\bfseries 08} (2017) 009},
  [\href{https://arxiv.org/abs/1703.00736}{{\ttfamily 1703.00736}}].

\bibitem{Bourton:2017pee}
T.~Bourton and E.~Pomoni, \emph{{Instanton counting in Class $\mathcal{S}_k$}},
   \href{https://arxiv.org/abs/1712.01288}{{\ttfamily 1712.01288}}.

\bibitem{Razamat:2018zus}
S.~S. Razamat and E.~Sabag, \emph{{A freely generated ring for N=1 models in
  class $S_k$}},  \href{https://arxiv.org/abs/1804.00680}{{\ttfamily
  1804.00680}}.

\bibitem{Heckman:2016xdl}
J.~J. Heckman, P.~Jefferson, T.~Rudelius and C.~Vafa, \emph{{Punctures for
  theories of class $ {\mathcal{S}}_{\varGamma } $}},
  \href{http://dx.doi.org/10.1007/JHEP03(2017)171}{\emph{JHEP} {\bfseries 03}
  (2017) 171}, [\href{https://arxiv.org/abs/1609.01281}{{\ttfamily
  1609.01281}}].

\bibitem{Kim:2018lfo}
H.-C. Kim, S.~S. Razamat, C.~Vafa and G.~Zafrir, \emph{{Compactifications of
  ADE conformal matter on a torus}},
  \href{http://dx.doi.org/10.1007/JHEP09(2018)110}{\emph{JHEP} {\bfseries 09}
  (2018) 110}, [\href{https://arxiv.org/abs/1806.07620}{{\ttfamily
  1806.07620}}].

\bibitem{Kim:2018bpg}
H.-C. Kim, S.~S. Razamat, C.~Vafa and G.~Zafrir, \emph{{D-type Conformal Matter
  and SU/USp Quivers}},
  \href{http://dx.doi.org/10.1007/JHEP06(2018)058}{\emph{JHEP} {\bfseries 06}
  (2018) 058}, [\href{https://arxiv.org/abs/1802.00620}{{\ttfamily
  1802.00620}}].

\bibitem{Kim:2017toz}
H.-C. Kim, S.~S. Razamat, C.~Vafa and G.~Zafrir, \emph{{E-String Theory on
  Riemann Surfaces}},
  \href{http://dx.doi.org/10.1002/prop.201700074}{\emph{Fortsch. Phys.}
  {\bfseries 66} (2018) 1700074},
  [\href{https://arxiv.org/abs/1709.02496}{{\ttfamily 1709.02496}}].

\bibitem{Bah:2017gph}
I.~Bah, A.~Hanany, K.~Maruyoshi, S.~S. Razamat, Y.~Tachikawa and G.~Zafrir,
  \emph{{4d $ \mathcal{N}=1 $ from 6d $ \mathcal{N}=\left(1,0\right) $ on a
  torus with fluxes}},
  \href{http://dx.doi.org/10.1007/JHEP06(2017)022}{\emph{JHEP} {\bfseries 06}
  (2017) 022}, [\href{https://arxiv.org/abs/1702.04740}{{\ttfamily
  1702.04740}}].

\bibitem{Razamat:2020bix}
S.~S. Razamat and E.~Sabag, \emph{{SQCD and pairs of pants}},
  \href{http://dx.doi.org/10.1007/JHEP09(2020)028}{\emph{JHEP} {\bfseries 09}
  (2020) 028}, [\href{https://arxiv.org/abs/2006.03480}{{\ttfamily
  2006.03480}}].

\bibitem{Seiberg:1994rs}
N.~Seiberg and E.~Witten, \emph{{Electric - magnetic duality, monopole
  condensation, and confinement in N=2 supersymmetric Yang-Mills theory}},
  \href{http://dx.doi.org/10.1016/0550-3213(94)90124-4,
  10.1016/0550-3213(94)00449-8}{\emph{Nucl. Phys.} {\bfseries B426} (1994)
  19--52}, [\href{https://arxiv.org/abs/hep-th/9407087}{{\ttfamily
  hep-th/9407087}}].

\bibitem{Seiberg:1994aj}
N.~Seiberg and E.~Witten, \emph{{Monopoles, duality and chiral symmetry
  breaking in N=2 supersymmetric QCD}},
  \href{http://dx.doi.org/10.1016/0550-3213(94)90214-3}{\emph{Nucl. Phys.}
  {\bfseries B431} (1994) 484--550},
  [\href{https://arxiv.org/abs/hep-th/9408099}{{\ttfamily hep-th/9408099}}].

\bibitem{Nekrasov:2002qd}
N.~A. Nekrasov, \emph{{Seiberg-Witten prepotential from instanton counting}},
  \href{http://dx.doi.org/10.4310/ATMP.2003.v7.n5.a4}{\emph{Adv. Theor. Math.
  Phys.} {\bfseries 7} (2003) 831--864},
  [\href{https://arxiv.org/abs/hep-th/0206161}{{\ttfamily hep-th/0206161}}].

\bibitem{Pestun:2007rz}
V.~Pestun, \emph{{Localization of gauge theory on a four-sphere and
  supersymmetric Wilson loops}},
  \href{http://dx.doi.org/10.1007/s00220-012-1485-0}{\emph{Commun. Math. Phys.}
  {\bfseries 313} (2012) 71--129},
  [\href{https://arxiv.org/abs/0712.2824}{{\ttfamily 0712.2824}}].

\bibitem{Hitchin:1986ea}
N.~J. Hitchin, A.~Karlhede, U.~Lindstrom and M.~Rocek, \emph{{Hyperkahler
  Metrics and Supersymmetry}},
  \href{http://dx.doi.org/10.1007/BF01214418}{\emph{Commun. Math. Phys.}
  {\bfseries 108} (1987) 535}.

\bibitem{Antoniadis:1996ra}
I.~Antoniadis and B.~Pioline, \emph{{Higgs branch, hyperKahler quotient and
  duality in SUSY N=2 Yang-Mills theories}},
  \href{http://dx.doi.org/10.1142/S0217751X97002620}{\emph{Int. J. Mod. Phys.}
  {\bfseries A12} (1997) 4907--4932},
  [\href{https://arxiv.org/abs/hep-th/9607058}{{\ttfamily hep-th/9607058}}].

\bibitem{hyper}
O.~Biquard and P.~Gauduchon, \emph{{Hyper-Kahler metrics on cotangent bundles
  of Hermitian symmetric spaces}}, {\emph{Lecture Notes in Pure and Appl.
  Math.} {\bfseries 128} (1997) 287--298}.

\bibitem{Baggio:2014ioa}
M.~Baggio, V.~Niarchos and K.~Papadodimas, \emph{{tt$^{*}$ equations,
  localization and exact chiral rings in 4d $ \mathcal{N} $ =2 SCFTs}},
  \href{http://dx.doi.org/10.1007/JHEP02(2015)122}{\emph{JHEP} {\bfseries 02}
  (2015) 122}, [\href{https://arxiv.org/abs/1409.4212}{{\ttfamily 1409.4212}}].

\bibitem{Baggio:2014sna}
M.~Baggio, V.~Niarchos and K.~Papadodimas, \emph{{Exact correlation functions
  in $SU(2) \mathcal N=2$ superconformal QCD}},
  \href{http://dx.doi.org/10.1103/PhysRevLett.113.251601}{\emph{Phys. Rev.
  Lett.} {\bfseries 113} (2014) 251601},
  [\href{https://arxiv.org/abs/1409.4217}{{\ttfamily 1409.4217}}].

\bibitem{Baggio:2015vxa}
M.~Baggio, V.~Niarchos and K.~Papadodimas, \emph{{On exact correlation
  functions in SU(N) $ \mathcal{N}=2 $ superconformal QCD}},
  \href{http://dx.doi.org/10.1007/JHEP11(2015)198}{\emph{JHEP} {\bfseries 11}
  (2015) 198}, [\href{https://arxiv.org/abs/1508.03077}{{\ttfamily
  1508.03077}}].

\bibitem{Gerchkovitz:2016gxx}
E.~Gerchkovitz, J.~Gomis, N.~Ishtiaque, A.~Karasik, Z.~Komargodski and S.~S.
  Pufu, \emph{{Correlation Functions of Coulomb Branch Operators}},
  \href{http://dx.doi.org/10.1007/JHEP01(2017)103}{\emph{JHEP} {\bfseries 01}
  (2017) 103}, [\href{https://arxiv.org/abs/1602.05971}{{\ttfamily
  1602.05971}}].

\bibitem{Grassi:2019txd}
A.~Grassi, Z.~Komargodski and L.~Tizzano, \emph{{Extremal Correlators and
  Random Matrix Theory}},  \href{https://arxiv.org/abs/1908.10306}{{\ttfamily
  1908.10306}}.

\bibitem{Beem:2013sza}
C.~Beem, M.~Lemos, P.~Liendo, W.~Peelaers, L.~Rastelli and B.~C. van Rees,
  \emph{{Infinite Chiral Symmetry in Four Dimensions}},
  \href{http://dx.doi.org/10.1007/s00220-014-2272-x}{\emph{Commun. Math. Phys.}
  {\bfseries 336} (2015) 1359--1433},
  [\href{https://arxiv.org/abs/1312.5344}{{\ttfamily 1312.5344}}].

\bibitem{Niarchos:2018mvl}
V.~Niarchos, \emph{{Geometry of Higgs-branch superconformal primary bundles}},
  \href{http://dx.doi.org/10.1103/PhysRevD.98.065012}{\emph{Phys. Rev.}
  {\bfseries D98} (2018) 065012},
  [\href{https://arxiv.org/abs/1807.04296}{{\ttfamily 1807.04296}}].

\bibitem{Gadde:2011uv}
A.~Gadde, L.~Rastelli, S.~S. Razamat and W.~Yan, \emph{{Gauge Theories and
  Macdonald Polynomials}},
  \href{http://dx.doi.org/10.1007/s00220-012-1607-8}{\emph{Commun. Math. Phys.}
  {\bfseries 319} (2013) 147--193},
  [\href{https://arxiv.org/abs/1110.3740}{{\ttfamily 1110.3740}}].

\bibitem{ArkaniHamed:2001ie}
N.~Arkani-Hamed, A.~G. Cohen, D.~B. Kaplan, A.~Karch and L.~Motl,
  \emph{{Deconstructing (2,0) and little string theories}},
  \href{http://dx.doi.org/10.1088/1126-6708/2003/01/083}{\emph{JHEP} {\bfseries
  01} (2003) 083}, [\href{https://arxiv.org/abs/hep-th/0110146}{{\ttfamily
  hep-th/0110146}}].

\bibitem{Hassler:2017arf}
F.~Hassler and J.~J. Heckman, \emph{{Punctures and Dynamical Systems}},
  \href{https://arxiv.org/abs/1711.03973}{{\ttfamily 1711.03973}}.

\bibitem{Hanany:2015pfa}
A.~Hanany and K.~Maruyoshi, \emph{{Chiral theories of class $ \mathcal{S} $}},
  \href{http://dx.doi.org/10.1007/JHEP12(2015)080}{\emph{JHEP} {\bfseries 12}
  (2015) 080}, [\href{https://arxiv.org/abs/1505.05053}{{\ttfamily
  1505.05053}}].

\bibitem{Luty:1995sd}
M.~A. Luty and W.~Taylor, \emph{{Varieties of vacua in classical supersymmetric
  gauge theories}},
  \href{http://dx.doi.org/10.1103/PhysRevD.53.3399}{\emph{Phys. Rev.}
  {\bfseries D53} (1996) 3399--3405},
  [\href{https://arxiv.org/abs/hep-th/9506098}{{\ttfamily hep-th/9506098}}].

\bibitem{1993alg.geom..4001B}
S.~{Bradlow}, G.~{Daskalopoulos} and R.~{Wentworth}, \emph{{Birational
  Equivalences of Vortex Moduli}}, {\emph{arXiv e-prints} (Apr, 1993)
  alg--geom/9304001}, [\href{https://arxiv.org/abs/alg-geom/9304001}{{\ttfamily
  alg-geom/9304001}}].

\bibitem{1993alg.geom..6005B}
A.~{Bertram}, G.~{Daskalopoulos} and R.~{Wentworth}, \emph{{Gromov Invariants
  for Holomorphic Maps from Riemann Surfaces to Grassmannians}}, {\emph{arXiv
  e-prints} (Jun, 1993) alg--geom/9306005},
  [\href{https://arxiv.org/abs/alg-geom/9306005}{{\ttfamily
  alg-geom/9306005}}].

\bibitem{Intriligator:1995au}
K.~A. Intriligator and N.~Seiberg, \emph{{Lectures on supersymmetric gauge
  theories and electric-magnetic duality}},
  \href{http://dx.doi.org/10.1016/0920-5632(95)00626-5}{\emph{Nucl. Phys. Proc.
  Suppl.} {\bfseries 45BC} (1996) 1--28},
  [\href{https://arxiv.org/abs/hep-th/9509066}{{\ttfamily hep-th/9509066}}].

\bibitem{Dolan:2002zh}
F.~A. Dolan and H.~Osborn, \emph{{On short and semi-short representations for
  four-dimensional superconformal symmetry}},
  \href{http://dx.doi.org/10.1016/S0003-4916(03)00074-5}{\emph{Annals Phys.}
  {\bfseries 307} (2003) 41--89},
  [\href{https://arxiv.org/abs/hep-th/0209056}{{\ttfamily hep-th/0209056}}].

\bibitem{curves}
T.~Bourton and E.~Pomoni, \emph{{To appear}}, .

\bibitem{Cordova:2016emh}
C.~Cordova, T.~T. Dumitrescu and K.~Intriligator, \emph{{Multiplets of
  Superconformal Symmetry in Diverse Dimensions}},
  \href{https://arxiv.org/abs/1612.00809}{{\ttfamily 1612.00809}}.

\bibitem{Seiberg:1994bz}
N.~Seiberg, \emph{{Exact results on the space of vacua of four-dimensional SUSY
  gauge theories}},
  \href{http://dx.doi.org/10.1103/PhysRevD.49.6857}{\emph{Phys. Rev.}
  {\bfseries D49} (1994) 6857--6863},
  [\href{https://arxiv.org/abs/hep-th/9402044}{{\ttfamily hep-th/9402044}}].

\bibitem{Feng:2007ur}
B.~Feng, A.~Hanany and Y.-H. He, \emph{{Counting gauge invariants: The
  Plethystic program}},
  \href{http://dx.doi.org/10.1088/1126-6708/2007/03/090}{\emph{JHEP} {\bfseries
  03} (2007) 090}, [\href{https://arxiv.org/abs/hep-th/0701063}{{\ttfamily
  hep-th/0701063}}].

\bibitem{Benvenuti:2006qr}
S.~Benvenuti, B.~Feng, A.~Hanany and Y.-H. He, \emph{{Counting BPS Operators in
  Gauge Theories: Quivers, Syzygies and Plethystics}},
  \href{http://dx.doi.org/10.1088/1126-6708/2007/11/050}{\emph{JHEP} {\bfseries
  11} (2007) 050}, [\href{https://arxiv.org/abs/hep-th/0608050}{{\ttfamily
  hep-th/0608050}}].

\bibitem{Gray:2008yu}
J.~Gray, A.~Hanany, Y.-H. He, V.~Jejjala and N.~Mekareeya, \emph{{SQCD: A
  Geometric Apercu}},
  \href{http://dx.doi.org/10.1088/1126-6708/2008/05/099}{\emph{JHEP} {\bfseries
  05} (2008) 099}, [\href{https://arxiv.org/abs/0803.4257}{{\ttfamily
  0803.4257}}].

\bibitem{Hanany:2008kn}
A.~Hanany and N.~Mekareeya, \emph{{Counting Gauge Invariant Operators in SQCD
  with Classical Gauge Groups}},
  \href{http://dx.doi.org/10.1088/1126-6708/2008/10/012}{\emph{JHEP} {\bfseries
  10} (2008) 012}, [\href{https://arxiv.org/abs/0805.3728}{{\ttfamily
  0805.3728}}].

\bibitem{Kinney:2005ej}
J.~Kinney, J.~M. Maldacena, S.~Minwalla and S.~Raju, \emph{{An Index for 4
  dimensional super conformal theories}},
  \href{http://dx.doi.org/10.1007/s00220-007-0258-7}{\emph{Commun. Math. Phys.}
  {\bfseries 275} (2007) 209--254},
  [\href{https://arxiv.org/abs/hep-th/0510251}{{\ttfamily hep-th/0510251}}].

\bibitem{Romelsberger:2005eg}
C.~Romelsberger, \emph{{Counting chiral primaries in N = 1, d=4 superconformal
  field theories}},
  \href{http://dx.doi.org/10.1016/j.nuclphysb.2006.03.037}{\emph{Nucl. Phys.}
  {\bfseries B747} (2006) 329--353},
  [\href{https://arxiv.org/abs/hep-th/0510060}{{\ttfamily hep-th/0510060}}].

\bibitem{Rastelli:2016tbz}
L.~Rastelli and S.~S. Razamat, \emph{{The supersymmetric index in four
  dimensions}},  2016.
\newblock \href{https://arxiv.org/abs/1608.02965}{{\ttfamily 1608.02965}}.

\bibitem{Gadde:2009dj}
A.~Gadde, E.~Pomoni and L.~Rastelli, \emph{{The Veneziano Limit of N = 2
  Superconformal QCD: Towards the String Dual of N = 2 SU(N(c)) SYM with N(f) =
  2 N(c)}},  \href{https://arxiv.org/abs/0912.4918}{{\ttfamily 0912.4918}}.

\bibitem{Beem:2012yn}
C.~Beem and A.~Gadde, \emph{{The $N=1$ superconformal index for class $S$ fixed
  points}}, \href{http://dx.doi.org/10.1007/JHEP04(2014)036}{\emph{JHEP}
  {\bfseries 04} (2014) 036},
  [\href{https://arxiv.org/abs/1212.1467}{{\ttfamily 1212.1467}}].

\bibitem{Evtikhiev:2017heo}
M.~Evtikhiev, \emph{{Studying superconformal symmetry enhancement through
  indices}}, \href{http://dx.doi.org/10.1007/JHEP04(2018)120}{\emph{JHEP}
  {\bfseries 04} (2018) 120},
  [\href{https://arxiv.org/abs/1708.08307}{{\ttfamily 1708.08307}}].

\bibitem{Spiridonov:2009za}
V.~P. Spiridonov and G.~S. Vartanov, \emph{{Elliptic Hypergeometry of
  Supersymmetric Dualities}},
  \href{http://dx.doi.org/10.1007/s00220-011-1218-9}{\emph{Commun. Math. Phys.}
  {\bfseries 304} (2011) 797--874},
  [\href{https://arxiv.org/abs/0910.5944}{{\ttfamily 0910.5944}}].

\bibitem{Hanany:2012dm}
A.~Hanany, N.~Mekareeya and S.~S. Razamat, \emph{{Hilbert Series for Moduli
  Spaces of Two Instantons}},
  \href{http://dx.doi.org/10.1007/JHEP01(2013)070}{\emph{JHEP} {\bfseries 01}
  (2013) 070}, [\href{https://arxiv.org/abs/1205.4741}{{\ttfamily 1205.4741}}].

\bibitem{Gaiotto:2012uq}
D.~Gaiotto and S.~S. Razamat, \emph{{Exceptional Indices}},
  \href{http://dx.doi.org/10.1007/JHEP05(2012)145}{\emph{JHEP} {\bfseries 05}
  (2012) 145}, [\href{https://arxiv.org/abs/1203.5517}{{\ttfamily 1203.5517}}].

\bibitem{Hanany:2014dia}
A.~Hanany and R.~Kalveks, \emph{{Highest Weight Generating Functions for
  Hilbert Series}},
  \href{http://dx.doi.org/10.1007/JHEP10(2014)152}{\emph{JHEP} {\bfseries 10}
  (2014) 152}, [\href{https://arxiv.org/abs/1408.4690}{{\ttfamily 1408.4690}}].

\bibitem{Forcella:2008bb}
D.~Forcella, A.~Hanany, Y.-H. He and A.~Zaffaroni, \emph{{The Master Space of
  N=1 Gauge Theories}},
  \href{http://dx.doi.org/10.1088/1126-6708/2008/08/012}{\emph{JHEP} {\bfseries
  08} (2008) 012}, [\href{https://arxiv.org/abs/0801.1585}{{\ttfamily
  0801.1585}}].

\bibitem{Seiberg:1997ax}
N.~Seiberg, \emph{{Notes on theories with 16 supercharges}},
  \href{http://dx.doi.org/10.1016/S0920-5632(98)00128-5}{\emph{Nucl. Phys.
  Proc. Suppl.} {\bfseries 67} (1998) 158--171},
  [\href{https://arxiv.org/abs/hep-th/9705117}{{\ttfamily hep-th/9705117}}].

\bibitem{Gang:2011xp}
D.~Gang, E.~Koh, K.~Lee and J.~Park, \emph{{ABCD of 3d ${\cal N}=8$ and 4
  Superconformal Field Theories}},
  \href{https://arxiv.org/abs/1108.3647}{{\ttfamily 1108.3647}}.

\bibitem{M2}
D.~R. Grayson and M.~E. Stillman, ``Macaulay2, a software system for research
  in algebraic geometry.'' Available at
  \url{https://faculty.math.illinois.edu/Macaulay2/}.

\bibitem{Hayling:2017cva}
J.~Hayling, C.~Papageorgakis, E.~Pomoni and D.~Rodríguez-Gómez, \emph{{Exact
  Deconstruction of the 6D (2,0) Theory}},
  \href{http://dx.doi.org/10.1007/JHEP06(2017)072}{\emph{JHEP} {\bfseries 06}
  (2017) 072}, [\href{https://arxiv.org/abs/1704.02986}{{\ttfamily
  1704.02986}}].

\bibitem{Hayling:2018fmv}
J.~Hayling, R.~Panerai and C.~Papageorgakis, \emph{{Deconstructing Little
  Strings with $\mathcal{N}=1$ Gauge Theories on Ellipsoids}},
  \href{http://dx.doi.org/10.21468/SciPostPhys.4.6.042}{\emph{SciPost Phys.}
  {\bfseries 4} (2018) 042},
  [\href{https://arxiv.org/abs/1803.06177}{{\ttfamily 1803.06177}}].

\bibitem{Hayling:2018fgy}
J.~Hayling, V.~Niarchos and C.~Papageorgakis, \emph{{Deconstructing Defects}},
  \href{https://arxiv.org/abs/1809.10485}{{\ttfamily 1809.10485}}.

\bibitem{Lambert:2012qy}
N.~Lambert, C.~Papageorgakis and M.~Schmidt-Sommerfeld, \emph{{Deconstructing
  (2,0) Proposals}},
  \href{http://dx.doi.org/10.1103/PhysRevD.88.026007}{\emph{Phys. Rev.}
  {\bfseries D88} (2013) 026007},
  [\href{https://arxiv.org/abs/1212.3337}{{\ttfamily 1212.3337}}].

\bibitem{Bourget:2017sxr}
A.~Bourget, A.~Pini and D.~Rodriguez-Gomez, \emph{{Towards Deconstruction of
  the Type D (2,0) Theory}},
  \href{http://dx.doi.org/10.1007/JHEP12(2017)146}{\emph{JHEP} {\bfseries 12}
  (2017) 146}, [\href{https://arxiv.org/abs/1710.10247}{{\ttfamily
  1710.10247}}].

\bibitem{Giveon:1999zm}
A.~Giveon, D.~Kutasov and O.~Pelc, \emph{{Holography for noncritical
  superstrings}},
  \href{http://dx.doi.org/10.1088/1126-6708/1999/10/035}{\emph{JHEP} {\bfseries
  10} (1999) 035}, [\href{https://arxiv.org/abs/hep-th/9907178}{{\ttfamily
  hep-th/9907178}}].

\bibitem{Aharony:1998ub}
O.~Aharony, M.~Berkooz, D.~Kutasov and N.~Seiberg, \emph{{Linear dilatons, NS
  five-branes and holography}},
  \href{http://dx.doi.org/10.1088/1126-6708/1998/10/004}{\emph{JHEP} {\bfseries
  10} (1998) 004}, [\href{https://arxiv.org/abs/hep-th/9808149}{{\ttfamily
  hep-th/9808149}}].

\bibitem{Seiberg:1997zk}
N.~Seiberg, \emph{{New theories in six-dimensions and matrix description of M
  theory on T**5 and T**5 / Z(2)}},
  \href{http://dx.doi.org/10.1016/S0370-2693(97)00805-8}{\emph{Phys. Lett.}
  {\bfseries B408} (1997) 98--104},
  [\href{https://arxiv.org/abs/hep-th/9705221}{{\ttfamily hep-th/9705221}}].

\bibitem{Berkooz:1997cq}
M.~Berkooz, M.~Rozali and N.~Seiberg, \emph{{Matrix description of M theory on
  T**4 and T**5}},
  \href{http://dx.doi.org/10.1016/S0370-2693(97)00800-9}{\emph{Phys. Lett.}
  {\bfseries B408} (1997) 105--110},
  [\href{https://arxiv.org/abs/hep-th/9704089}{{\ttfamily hep-th/9704089}}].

\bibitem{Aharony:1999ks}
O.~Aharony, \emph{{A Brief review of 'little string theories'}},
  \href{http://dx.doi.org/10.1088/0264-9381/17/5/302}{\emph{Class. Quant.
  Grav.} {\bfseries 17} (2000) 929--938},
  [\href{https://arxiv.org/abs/hep-th/9911147}{{\ttfamily hep-th/9911147}}].

\bibitem{Nahm:1977tg}
W.~Nahm, \emph{{Supersymmetries and their Representations}},
  \href{http://dx.doi.org/10.1016/0550-3213(78)90218-3}{\emph{Nucl. Phys.}
  {\bfseries B135} (1978) 149}.

\bibitem{Kim:2015gha}
J.~Kim, S.~Kim and K.~Lee, \emph{{Little strings and T-duality}},
  \href{http://dx.doi.org/10.1007/JHEP02(2016)170}{\emph{JHEP} {\bfseries 02}
  (2016) 170}, [\href{https://arxiv.org/abs/1503.07277}{{\ttfamily
  1503.07277}}].

\bibitem{Kim:2017xan}
J.~Kim and K.~Lee, \emph{{Little strings on D$_{n}$ orbifolds}},
  \href{http://dx.doi.org/10.1007/JHEP10(2017)045}{\emph{JHEP} {\bfseries 10}
  (2017) 045}, [\href{https://arxiv.org/abs/1702.03116}{{\ttfamily
  1702.03116}}].

\bibitem{Kim:2018gak}
J.~Kim, K.~Lee and J.~Park, \emph{{On elliptic genera of 6d string theories}},
  \href{http://dx.doi.org/10.1007/JHEP10(2018)100}{\emph{JHEP} {\bfseries 10}
  (2018) 100}, [\href{https://arxiv.org/abs/1801.01631}{{\ttfamily
  1801.01631}}].

\bibitem{Iqbal:2007ii}
A.~Iqbal, C.~Kozcaz and C.~Vafa, \emph{{The Refined topological vertex}},
  \href{http://dx.doi.org/10.1088/1126-6708/2009/10/069}{\emph{JHEP} {\bfseries
  10} (2009) 069}, [\href{https://arxiv.org/abs/hep-th/0701156}{{\ttfamily
  hep-th/0701156}}].

\bibitem{Lockhart:2012vp}
G.~Lockhart and C.~Vafa, \emph{{Superconformal Partition Functions and
  Non-perturbative Topological Strings}},
  \href{http://dx.doi.org/10.1007/JHEP10(2018)051}{\emph{JHEP} {\bfseries 10}
  (2018) 051}, [\href{https://arxiv.org/abs/1210.5909}{{\ttfamily 1210.5909}}].

\bibitem{Kim:2016usy}
S.~Kim and K.~Lee, \emph{{Indices for 6 dimensional superconformal field
  theories}}, \href{http://dx.doi.org/10.1088/1751-8121/aa5cbf}{\emph{J. Phys.}
  {\bfseries A50} (2017) 443017},
  [\href{https://arxiv.org/abs/1608.02969}{{\ttfamily 1608.02969}}].

\bibitem{Bhattacharyya:2007sa}
S.~Bhattacharyya and S.~Minwalla, \emph{{Supersymmetric states in M5/M2 CFTs}},
  \href{http://dx.doi.org/10.1088/1126-6708/2007/12/004}{\emph{JHEP} {\bfseries
  12} (2007) 004}, [\href{https://arxiv.org/abs/hep-th/0702069}{{\ttfamily
  hep-th/0702069}}].

\bibitem{Razamat:2016dpl}
S.~S. Razamat, C.~Vafa and G.~Zafrir, \emph{{4d $ \mathcal{N}=1 $ from 6d (1,
  0)}}, \href{http://dx.doi.org/10.1007/JHEP04(2017)064}{\emph{JHEP} {\bfseries
  04} (2017) 064}, [\href{https://arxiv.org/abs/1610.09178}{{\ttfamily
  1610.09178}}].

\bibitem{Razamat:2019ukg}
S.~S. Razamat and E.~Sabag, \emph{{Sequences of $6d$ SCFTs on generic Riemann
  surfaces}}, \href{http://dx.doi.org/10.1007/JHEP01(2020)086}{\emph{JHEP}
  {\bfseries 01} (2020) 086},
  [\href{https://arxiv.org/abs/1910.03603}{{\ttfamily 1910.03603}}].

\bibitem{2004math11044S}
V.~P. Spiridonov and S.~O. Warnaar, \emph{Inversions of integral operators and
  elliptic beta integrals on root systems},
  \href{http://dx.doi.org/https://doi.org/10.1016/j.aim.2005.11.007}{\emph{Advances
  in Mathematics} {\bfseries 207} (2006) 91 -- 132}.

\bibitem{Cordova:2017mhb}
C.~Cordova, D.~Gaiotto and S.-H. Shao, \emph{{Surface Defects and Chiral
  Algebras}}, \href{http://dx.doi.org/10.1007/JHEP05(2017)140}{\emph{JHEP}
  {\bfseries 05} (2017) 140},
  [\href{https://arxiv.org/abs/1704.01955}{{\ttfamily 1704.01955}}].

\bibitem{Cordova:2017ohl}
C.~Cordova, D.~Gaiotto and S.-H. Shao, \emph{{Surface Defect Indices and 2d-4d
  BPS States}}, \href{http://dx.doi.org/10.1007/JHEP12(2017)078}{\emph{JHEP}
  {\bfseries 12} (2017) 078},
  [\href{https://arxiv.org/abs/1703.02525}{{\ttfamily 1703.02525}}].

\bibitem{Cordova:2016uwk}
C.~Cordova, D.~Gaiotto and S.-H. Shao, \emph{{Infrared Computations of Defect
  Schur Indices}}, \href{http://dx.doi.org/10.1007/JHEP11(2016)106}{\emph{JHEP}
  {\bfseries 11} (2016) 106},
  [\href{https://arxiv.org/abs/1606.08429}{{\ttfamily 1606.08429}}].

\bibitem{Cox:2007}
J.~L. Cox, David and D.~O'shea., \emph{{Ideals, varieties, and algorithms}}.
\newblock 2007.

\bibitem{stanley1978hilbert}
R.~P. Stanley, \emph{Hilbert functions of graded algebras}, {\emph{Advances in
  Mathematics} {\bfseries 28} (1978) 57--83}.

\bibitem{Gadde:2010te}
A.~Gadde, L.~Rastelli, S.~S. Razamat and W.~Yan, \emph{{The Superconformal
  Index of the $E_6$ SCFT}},
  \href{http://dx.doi.org/10.1007/JHEP08(2010)107}{\emph{JHEP} {\bfseries 08}
  (2010) 107}, [\href{https://arxiv.org/abs/1003.4244}{{\ttfamily 1003.4244}}].

\bibitem{Liendo:2011wc}
P.~Liendo and L.~Rastelli, \emph{{The Complete One-loop Spin Chain of N =1
  SQCD}}, \href{http://dx.doi.org/10.1007/JHEP10(2012)117}{\emph{JHEP}
  {\bfseries 10} (2012) 117},
  [\href{https://arxiv.org/abs/1111.5290}{{\ttfamily 1111.5290}}].

\bibitem{Gadde:2010en}
A.~Gadde, L.~Rastelli, S.~S. Razamat and W.~Yan, \emph{{On the Superconformal
  Index of N=1 IR Fixed Points: A Holographic Check}},
  \href{http://dx.doi.org/10.1007/JHEP03(2011)041}{\emph{JHEP} {\bfseries 03}
  (2011) 041}, [\href{https://arxiv.org/abs/1011.5278}{{\ttfamily 1011.5278}}].

\bibitem{Dolan:2008qi}
F.~A. Dolan and H.~Osborn, \emph{{Applications of the Superconformal Index for
  Protected Operators and q-Hypergeometric Identities to N=1 Dual Theories}},
  \href{http://dx.doi.org/10.1016/j.nuclphysb.2009.01.028}{\emph{Nucl. Phys.}
  {\bfseries B818} (2009) 137--178},
  [\href{https://arxiv.org/abs/0801.4947}{{\ttfamily 0801.4947}}].

\end{thebibliography}\endgroup
\bibliographystyle{JHEP}

\end{document}